\documentclass[%
 reprint, 
 amsmath,amssymb,
 aps,
 prb
]{revtex4-2}

\usepackage{cancel} 
\usepackage{braket} 
\usepackage{mathdots}
\usepackage[usenames, dvipsnames]{xcolor}

\usepackage{graphicx}
\usepackage{dcolumn}
\usepackage{bm}
\usepackage{hyperref}
\usepackage[mathlines]{lineno}

\newcommand{\ed}[1]
{{#1}}
\newcommand{\ee}{{e}}
\newcommand{\ii}{{i}}
\newcommand{\dd}{\partial}

\newcommand{\myabs}[1]{\lvert #1 \rvert}

\begin{document}
\definecolor{persianred}{rgb}{0.8, 0.2, 0.2}
\definecolor{persianblue}{rgb}{0.11, 0.22, 0.73}
\definecolor{amber}{rgb}{.9, 0.5, 0.0}
\preprint{APS/123-QED}

\title{Classical analogue to the Kitaev model and Majorana-like topological bound states}%

\author{Ting-Wei Liu}
\author{Fabio Semperlotti}
 \email{fsemperl@purdue.edu}
\affiliation{Ray W. Herrick Laboratories, School of Mechanical Engineering, Purdue University, West Lafayette, Indiana 47907, USA.}%

\date{\today}%

\begin{abstract}
This study explores the possibility and presents a methodology to synthesize a classical mechanical analogue to the quantum mechanical 1D Kitaev model. While being fundamentally different, we will identify significant conceptual similarities between the two models that culminate in the occurrence, in the classical analogue system, of topologically non-trivial bound states that are akin to Majorana zero modes. By reformulating the Hamiltonian of the classical system in a form reminiscent of second quantization, we show that a 1D staggered classical mechanical chain can exhibit dynamic characteristics analogous to the Kitaev's 1D superconducting model, as well as its characteristic bound states. 
The non-trivial topological nature of the bound states is further confirmed by the topological band structure analysis and by the topological invariant. While the non-Abelian nature of these states remains an open question, these results allow envisioning the possibility to achieve topological braiding in classical mechanical systems.

\end{abstract}

\maketitle

\section{\label{sec:intro}
introduction
}
Several studies in the general area of metamaterials have shown the many conceptual similarities between photonic, phononic, and mechanical systems. As an example, acoustic and mechanical systems can be devised to simulate digital electronic circuits \cite{liang2009acoustic,fleury2014sound,li2014granular,yu2018elastic,ma2019valley,el2022mechanical} in which logic gates, switches, and other components are realized via acoustic or mechanical components. More recently, this same trend was observed also with respect to quantum mechanical 
\cite{zivari2022non} and even topological materials %
\cite{chen2019mechanical,barlas2020topological,qian2022observation,allein2022duality,allein2022duality,qian2023observation}, although the correspondence with classical systems becomes more elusive and often hidden in details of the mathematical structure describing the high level dynamics.
As an example, in quantum information processing,
the Majorana zero mode (i.e. the quasiparticle that represents the solid state electronic counterpart of the Majorana fermion \cite{majorana1937teoria}) has been shown to be a potential candidate to serve as a quantum bit (qubit) in future quantum computers due to its non-Abelian braiding statistics and to its topologically-protected fault-tolerant nature \cite{kitaev2001unpaired,kitaev2003fault,lutchyn2010majorana,alicea2010majorana,sau2010generic,alicea2011non, chien2018topological, chien2017thermal, greiter20141d, nadj2014observation, kim2018toward, attig2019topological, zhang2017two}; a critical aspect to control computational errors.
\label{ParaC1}
{
Only in very recent times, a handful of studies investigated the possibility to synthesize classical
\ed{
electrical \cite{ezawa2019braiding,ezawa2020non} and
}
mechanical \cite{chen2019mechanical,gao2019majorana,barlas2020topological,qian2022observation,allein2022duality,qian2023observation} analogues of the Majorana zero modes.
}

During the past decade, various concepts and experimental investigations on classical systems (including photonic and phononic) have shown 
\label{ParaA1}
{
the ability to reproduce analogue mechanisms to quantum topological mechanisms at the basis of topological insulators and other topological materials in classical electromagnetic, acoustic, and mechanical systems
\ed{\cite{raghu2008analogs,yang2015topological,ni2015topologically,khanikaev2015topologically,wang2015topological,nash2015topological,mousavi2015topologically,miniaci2018experimental,he2016acoustic,susstrunk2015observation,wu2015scheme,yang2018visualization,xia2017topological,deng2017observation,chaunsali2018subwavelength,liu2020robust,liu2021synthetic,lu2017observation,pal2017edge,vila2017observation,liu2018tunable,zhu2018design,liu2019experimental,ganti2020weyl,ganti2020topological,xiao2015geometric,xiao2017topological,chaunsali2017demonstrating,chen2018study,chen2019mechanical,barlas2020topological,qian2022observation,allein2022duality,ding2016emergence,tang2020exceptional,dominguez2020environmentally,liao2022engineering,ye2022topological}.}
A common trait of these different implementations was the synthesis of dynamical matrices (\ed{that describe the dynamics of the system based on the classical equations of motion or perturbative coupled mode methods}) resembling the Hamiltonian operator (i.e. the matrix representation) of the target topological quantum system.}
Such analogy is possible thanks to similarities in the underlying mathematical representation, that however might not lead to a direct correlation of certain physical properties.
A simple example of this discrepancy is seen in the comparison between quantum and classical plane waves. A free quantum mechanical particle having only kinetic energy in the Hamiltonian has the wavefunction of a \ed{plane wave}, while an acoustic \ed{plane wave} involves an exchange between kinetic and potential energies via the medium supporting the wave, so the Hamiltonian includes both kinetic and potential energy terms.
Although the existing approach has been shown to be successful in creating analogue systems, the dynamical matrix is not the only representation of the classical system, therefore not the only way to connect quantum and classical systems.

\label{ParaA2}
{\ed{
This study presents a first attempt to synthesize classical mechanical analogs to quantum topological systems at the Hamiltonian level, rather than at the dynamical matrix level.
This goal was achieved by developing a second-quantization-like formalism applicable to classical systems. 
By means of this method, the classical Hamiltonian is expressed in terms of on-site and hopping energy terms. These terms show a highly correlated mathematical structure with the second quantized form of a solid state system, hence offering an alternative and powerful tool to analyze differences and similarities between classical and quantum mechanical systems.
}
Especially, we focus on replicating the 1D Kitaev superconducting chain} \cite{kitaev2001unpaired} and the Majorana zero modes with classical mechanical elements (e.g. springs and mass particles) and show that a dimerized mechanical chain (having staggered particle masses or spring constants) has a similar Hamiltonian to the Kitaev's model. Also, we will show that topological bound states described by a Hamiltonian analogous to the one underlying Majorana zero modes appear at 
the ends of
topologically nontrivial chain.
The dynamical behavior and the topological invariant of the classical mechanical chain can be also obtained by substituting the Hamiltonian into the classical Hamilton's equations.

The 1D superconducting chain model proposed by \textcite{kitaev2001unpaired} is described by the following Hamiltonian in second quantization formalism,
\begin{align}\begin{split}\label{eq:HK}
    \hat{H}_\text{Kitaev}=-\mu \sum_{j=1}^N{\hat{c}_j^\dagger \hat{c}_j}
    -t &\sum_{j=1}^{N-1}{\left(\hat{c}_{j+1}^\dagger\hat{c}_j+\mathrm{H.c.}\right)}\\
    + &\sum_{j=1}^{N-1}{\left(\Delta \hat{c}_{j+1}^\dagger\hat{c}_j^\dagger+\mathrm{H.c.}\right)},
\end{split}\end{align}
where $\hat{c}_j^\dagger$ and $\hat{c}_j$ are the fermion creation and annihilation operators, and $\mu$, $t$ and $\Delta$ represent the on-site energy, hopping, and superconducting coefficients, respectively.
The system has a symmetric spectrum about zero energy that is protected by particle-hole symmetry.
Concerning the proposed classical mechanical chain that is the object of this study, a classical ``second quantized'' notation will be derived based on time-reversal eigenmodes of each building-block oscillator, and an analogue particle-hole symmetry will be identified and found to be responsible for a symmetric spectrum.
\textcite{kitaev2001unpaired} also demonstrated in the same model that under two extreme cases 1) $\mu\gg t=\lvert\Delta\rvert$, and 2) $\mu\ll t=\lvert\Delta\rvert$,
the Hamiltonians can be written as
\begin{subequations}
\begin{align}
    \label{eq:HK1}
    \hat{H}_1 &= \ii \frac{\mu}{2}\sum_{j=1}^N \left(\hat{\gamma}_1 \hat{\gamma}_2\right)_j,\\
    \label{eq:HK2}
    \hat{H}_2 &= \ii t\sum_{j=1}^{N-1} \left(\hat{\gamma}_1\right)_{j+1} \left(\hat{\gamma}_2\right)_j,
\end{align}
\end{subequations}
 with the self-conjugate Majorana operators $\hat\gamma_{1,2}$ following $\hat{c}^\dagger =\frac{1}{2}\left(\hat{\gamma}_1+\ii \hat{\gamma}_2\right)$, and
$\hat{c} =\frac{1}{2}\left(\hat{\gamma}_1-\ii \hat{\gamma}_2\right)$.
These equations indicate two types of pairing of $\hat{\gamma}_{1,2}$ and
Majorana zero modes (as unpaired Majorana operators) appear at the ends of the topologically nontrivial ($\mu<t=\lvert\Delta\rvert$) chain.
The same result is also found in the classical mechanical chain where Majorana-like bound states manifest at the terminals of the chain.

In literature, 1D and quasi-1D periodic classical mechanical systems following the conventional dynamical matrix approach usually fall into the category of classical analogues to the Su-Schrieffer–Heeger \cite{su1979solitons} (SSH) model \cite{xiao2015geometric,xiao2017topological,chaunsali2017demonstrating,yin2018band,chen2018study,vila2019role,shi2021disorder}, while there are also recent studies focusing on creating Majorana-like bound states by involving complex structures \cite{barlas2020topological,qian2022observation,allein2022duality}.
The major difference between Kitaev's model and the SSH model
lies in the superconducting pairing terms $\Delta \hat{c}_j^\dagger\hat{c}_{j+1}^\dagger + \Delta^\ast \hat{c}_j\hat{c}_{j+1}$  in Eq.~(\ref{eq:HK}).
It will be shown that the superconducting terms are inherent in the Hamiltonian of a mechanical chain under the classical second quantized notation.

With the proposed second quantized notation and the Hamiltonian analogy approach, we are able to construct a classical system analogue to the Kitaev chain and exhibiting Majorana-like bound states. This unique approach also provides a new perspective on possible strategies to link classical and quantum systems.

\section{Classical analogue Kitaev chain}\label{sec:dikitaev} 

\subsection{Hamiltonian of a 1D classical mechanical chain}
\begin{figure}[!htb]
\centering
\includegraphics[width=.85\columnwidth]{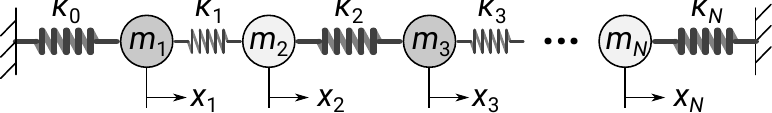}
\caption{\label{fig:chain}
A sketch of an arbitrary 1D mechanical chain. Mass of the $j^\mathrm{th}$ particle is labeled $m_j$, while $\kappa_j$ indicates the constant of the spring next to the $j^\mathrm{th}$ particle. The conjugate momenta and coordinates $(\dots,p_j,\dots;\dots,x_j,\dots)$ form the canonical coordinates.}
\end{figure}

Consider a 1D classical mechanical chain composed of $N$ particles with mass $m_j$ connected by springs with constants $\kappa_j$, where $j=1,\dots, N$, with terminals connected to the ground by springs, as shown in Fig.~\ref{fig:chain}.

Assuming only longitudinal motions are allowed (particles follow a frictionless slide), the Hamiltonian of such classical mechanical chain is the total energy,
\begin{equation}\label{eq:ham}
    H=\sum_{j=1}^N{\frac{p_j^2}{2m_j}}
    +\sum_{j=0}^N{\frac{\kappa_j}{2}(x_{j+1}-x_j)^2},
\end{equation}
where $(p_j,x_j)$ are momentum and spatial coordinate of the $j^\mathrm{th}$ particle, and the $2N$-tuple $(\mathbf{p};\mathbf{x})$ forms a set of canonical coordinates;
it can be easily verified that
\ed{
$\{x_j,x_l\}=\{p_j,p_l\}=0$},
and
$\{x_j,p_l\}=-\{p_j,x_l\}=\delta_{jl}$,
where the curly brackets represent Poisson brackets, and $\delta_{jl}$ is the Kronecker delta.
For convenience, in Eq.~(\ref{eq:ham}) we have let $x_0= x_{N+1} \equiv 0$ representing grounded (or fixed) ends.

We can write the Hamilton's equations of motion for such finite chain,
\begin{subequations}
\begin{align}
\dot{p}_j &= -\frac{\partial H}{\partial x_j},\\
\dot{x}_j &= \frac{\partial H}{\partial p_j}.
\end{align}
\end{subequations}
Let $\mathbf{X} \equiv (\mathbf{p};\mathbf{x})$, and the above two sets of equations can be written in a more compact form,
\begin{equation}\label{eq:hams}
    \dot{\mathbf{X}} = \mathbf{E}\frac{\partial H}{\partial \mathbf{X}}, \; \mathbf{E}=
    \begin{pmatrix}
    \mathbf{0} & -\mathbf{1}\\
    \mathbf{1} & \mathbf{0}
    \end{pmatrix},
\end{equation}
where the bold $\mathbf{1}$ and $\mathbf{0}$ in matrix $\mathbf{E}$ are $N\times N$ identity and zero matrices, respectively.
$\tfrac{\partial H}{\partial \mathbf{X}}$ is a $2N\times 1$ column vector with each of the components evaluated as $\tfrac{\partial H}{\partial X_j}$. It prescribes a linear operation on $\mathbf{X}$ given that $H$ is a homogeneous quadratic polynomial in $X_j$.
Eq.~(\ref{eq:hams}) can be written as a set of linear differential equations as
$\dot{\mathbf{X}} = \mathsf{H} \mathbf{X}$
by letting
$\mathbf{E}\frac{\partial H}{\partial \mathbf{X}}
    \equiv
    \mathsf{H}\mathbf{X}$,
where $\mathsf{H}$ is the coefficient matrix of the Hamilton's equations.
Substituting the time-harmonic ansatz $\mathbf{X}\rightarrow\mathbf{X}\ee^{-\ii\omega t}$ gives the eigenvalue problem
\begin{equation}\label{eq:hameig}
    \mathsf{H}\mathbf{X}
    =-\ii \omega \mathbf{X}.
\end{equation}
With the system being time-reversal invariant, i.e., $H(\mathbf{p},\mathbf{x},t)=H(-\mathbf{p},\mathbf{x},-t)$,
the matrix $\mathsf{H}$ always has symmetric spectra $\pm \omega$ corresponding to time-reversal pairs of eigenmodes $\mathbf{X}$ and $\mathbf{X}^\ast$ \cite{dulock1965degeneracy}, reminiscent of particle-antiparticle symmetry in the Dirac equation.

However, it is generally impossible to open up a band gap at zero frequency for a classical linearly elastic mechanical chain
and hence impossible for the zero-frequency bound states to exist.
Even if possible, a zero-frequency (static) mode would not carry a phase information other than $0$ or $\pi$ thus have less significance in signal processing, information, as well as vibration and noise control applications.

In the following, we will show that the 1D dimerized lattice has symmetric $\omega^2$-spectrum with respect to a non-zero reference level. This situation is rather similar to particle-hole symmetry in solid state systems, where the symmetry of the energy spectrum is with respect to a reference level (Fermi level) instead of zero energy.
Such analogue particle-hole symmetry results from the combination of time-reversal, space-inversion, and odd sublattice symmetry of the chain
\ed{(namely, a negative sign in the perturbation terms showing up upon exchanging the positions of the two internal degrees of freedom, i.e. the sublattices)}. 
Under these conditions, both topological transitions and bound states can be created at either internal interfaces or terminals of the chain, \ed{with the bound state Hamiltonian being} similar \ed{at least, in mathematical form} to the solid state Majorana zero modes.

\subsection{Dimerized mechanical chain}
In the following, we will consider the particle-spring chain with staggered mass and spring constants. 
\label{ParaB1}
{
Such alternating pattern is also reminiscent of the SSH model \cite{su1979solitons}
\ed{when only the spring constants are varied},
and of the Rice-Mele \cite{rice1982elementary} model
\ed{when both variations in the spring and the mass constants are considered. The SSH and Rice-Mele models have been used in the literature to describe dimerized polymer chains}%
, hence we will also use the word \emph{dimerized} to refer to the chain with alternating masses and spring constants.
\ed{
Previous studies have suggested that the classical analog SSH chain with alternating springs exhibits a nontrivial topological phase \cite{xiao2015geometric,xiao2017topological,chaunsali2017demonstrating,yin2018band,chen2018study,vila2019role,shi2021disorder}, while chains with alternating particle masses are associated with the Rice-Mele model and lack well-defined topological phases \cite{chen2018study},
which include the diatomic chain with only alternating particle masses but a constant spring constant.
However, an important difference between the SSH and the Rice-Mele models lies in the fact that the former respects inversion symmetry whereas the latter does not. We will show that the diatomic chain with nonzero particle mass variation but no variation in the spring constant can still produce similar topological phase, as this configuration restores inversion symmetry in the chain. However, to ensure global inversion symmetry, the chain must have an odd number of particles; this latter case does not have a real-world quantum mechanical counterpart because polymer chains comprising diatomic unit cells always have an even number of atoms.
In later analyses, we will discuss how the classical chain under investigation exhibits a duality between the spring and mass constants variations, and how nontrivial topological phases and bound states can exist in a system possessing either one of the two variations.
}
}

The matrix $\mathsf{H}$ of a nonuniform chain with varying $m_j$ and $\kappa_j$ can be expressed in terms of the dimerization parameters.
Note that $m_j$ appears in the denominator of the kinetic energy term, while $\kappa_j$ appears in the numerator of the potential energy term in the Hamiltonian function.
Then, we can set the \emph{reference} mass $m_0$ and the spring constant $\kappa_0$ and rewrite $m_j$ and $\kappa_j$ with two dimensionless dimerization parameters
$r_j$ and $\epsilon_j$ which control the strength of the staggering process in the following way ($\myabs{r_j}<1$ and $\myabs{\epsilon_j}<1$):
\begin{subequations}\label{eq:mjkj}
\begin{align}
    m_j &=  m_0 \left(1- \frac{r_j}{1+r_j}\right),\label{eq:mj}\\
    \kappa_j &= \kappa_0 (1+\epsilon_j),\label{eq:kj}
\end{align}
\end{subequations}
\begin{figure}
\centering
\includegraphics[width=\columnwidth]{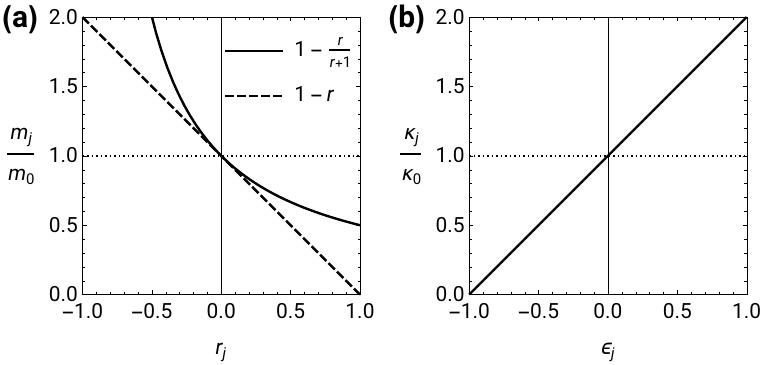}
\caption{\label{fig:mjkj}
Plots of (a) the particle mass $m_j$ and (b) the spring constant $\kappa_j$ versus the dimerization parameters $r_j$ and $\epsilon_j$, respectively.
The dimerization in $m$ follows a constant harmonic mean as $m$ appears in the denominator of the kinetic energy terms of the Hamiltonian function. The dimerization in $\kappa$ satisfies a constant arithmetic mean as it shows in the numerators of the potential energy terms in the Hamiltonian function.}
\end{figure}
These relations are plotted in Fig.~\ref{fig:mjkj}. Such setup (particularly the counter intuitive Eq.~(\ref{eq:mj})) is devised so that {each term in the Hamiltonian function}
(see
Eq.~\ref{eq:ham}) is linear in \emph{finite} $r_j$ and $\epsilon_j$.
Note that if we simply let $m_j=m_0 (1-r_j)$, the parameter $r$ will appear in the denominator and the Hamiltonian can only be linearized in the case of \emph{infinitesimal} perturbation $r$.

\begin{figure}[!htb]
\centering
\includegraphics[width=.85\columnwidth]{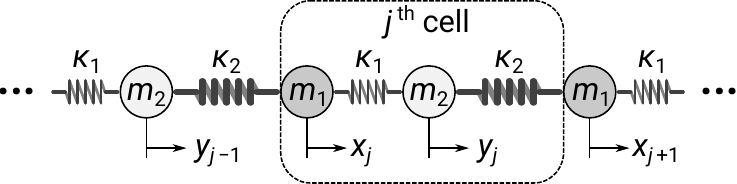}
\caption{\label{fig:di}
A sketch of the 1D dimerized (%
staggered, or diatomic) mechanical chain. Two particles with mass $m_1$, $m_2$, and two springs with constants $\kappa_1$, $\kappa_2$ compose a unit cell. For the $j^\mathrm{th}$ cell, the conjugate momenta and coordinates of the two particles are labeled $(p_j,q_j;x_j,y_j)$.
}
\end{figure}

For a diatomic lattice, the mass of the particles and of the spring constants repeat identically every other element, as graphically shown in Fig.~\ref{fig:di}.
These staggered mass and spring constants are described by Eqs.~\ref{eq:mjkj} with $\pm r$ and $\pm \epsilon$, respectively
(``$+$'' for index 1 and ``$-$'' for index 2), 
 \begin{subequations}\label{eq:mk12}
 \begin{align}
     m_{1,2}&=m_0 \left(1- \frac{\pm r}{1 \pm r}\right),\\ 
     \kappa_{1,2}&=\kappa_0 (1\pm\epsilon).
 \end{align}
 \end{subequations}
The reference spring constant $\kappa_0$ is the arithmetic mean of $\kappa_1$ and $\kappa_2$, while the reference mass $m_0$ is the \emph{harmonic mean} of $m_1$ and $m_2$,
 \begin{subequations}\label{eq:mk0}
 \begin{align}
     \kappa_0&=\frac{1}{2}\left( \kappa_1 + \kappa_2 \right),\\
     \frac{1}{m_0}&=\frac{1}{2}\left(\frac{1}{m_1}+\frac{1}{m_2}\right), \mathrm{or}\;
     m_0 = \frac{2m_1m_2}{m_1+m_2}.
 \end{align}
 \end{subequations}
The reference frequency can then be defined as
\begin{equation}\label{eq:o0}
    \omega_0=\sqrt{\frac{2\kappa_0}{m_0}},
\end{equation}
which stands for the resonance frequency of a reference oscillator composed of a particle with mass $m_0$ and attached to the ground on both sides via springs of constant $\kappa_0$.

Since the lattice is dimerized, it is convenient to separate the two inner degrees of freedom (sublattices) within a unit cell. Hence, we will employ $(p_j, x_j)$ for the first particle (sublattice $A$) and $(q_j, y_j)$ for the second one (sublattice $B$), obtained via the following substitutions:
\begin{equation}
\begin{cases}
    p_{2j-1} \rightarrow p_j,
    & p_{2j} \rightarrow q_j\\
    x_{2j-1} \rightarrow x_j,
    & x_{2j} \rightarrow y_j
\end{cases}
     ,\; j=1,\dots,N,
\end{equation}
where now $j$ is the unit cell index and $N$ is the number of unit cells in a chain. It follows that there are $2N$ particles in the chain, the state vector $\mathbf{X}=(\dots,p_j,q_j,\dots;\dots,x_j,y_j,\dots)$ becomes a $4N$-tuple, and the Hamiltonian function reads
\begin{align}\label{eq:hamdi}\begin{split}
    H
    &=
    \sum_{j=1}^N{\left[
    (1+r)\frac{p_j^2}{2m_0}+
    (1-r)\frac{q_j^2}{2m_0}
    \right]}\\
    &+\sum_{j=0}^N{\left[
    (1+\epsilon)\frac{\kappa_0}{2}(y_{j}-x_j)^2+
    (1-\epsilon)\frac{\kappa_0}{2}(x_{j+1}-y_j)^2
    \right]}.
\end{split}\end{align}
%
 
\subsection{Second quantization formalism for classical systems}
In this section we develop the ``second quantization'' formalism for the classical mechanical lattice. Second quantization is the standard language in quantum many-body physics. While clearly there is no real notion of quantization in classical systems, this formalism can help drawing closer comparisons between the classical and quantum systems, based on their Hamiltonian's representations. 

We consider the coordinate transformation from $(p_j,q_j;x_j,y_j)$ to $(a_j^+,b_j^+;a_j^-,b_j^-)$ following
\begin{subequations}\label{eq:ajbj}
\begin{align}
    a_j^\pm &\equiv
    \frac{1}{\sqrt {\ii\omega_0}}
    \frac{1}{\sqrt {2}}
    \left(
    \frac{p_j}{\sqrt {m_0}}
    \mp i\sqrt{2\kappa_0}x_j
    \right),\\
    b_j^\pm &\equiv
    \frac{1}{\sqrt {\ii\omega_0}}
    \frac{1}{\sqrt {2}}
    \left(
    \frac{q_j}{\sqrt {m_0}}
    \mp i\sqrt{2\kappa_0}y_j
    \right).
\end{align}
\end{subequations}
The coefficients multiplying the terms parenthesis ensures that the new coordinate system remains canonical (and therefore the Hamilton's equations still hold, see Appendix \ref{sec:pf}). The terms in parentheses are the two eigenmodes (time-reversed counterparts) considering each particle as a stand-alone oscillator. \textcite{dulock1965degeneracy} presented a ladder-operator approach applied to a single classical oscillator, but a similar approach has not been applied to classical many-body systems.
\label{ParaC4}
{
\ed{Such classical second quantized variables ($a^\pm$) share some properties of the quantum fermionic creation/annihilation operators ($\hat{c}^\dagger$, $\hat{c}$) but also exhibit some differences. For example, the classical variables are scalars, thus they always commute under the multiplication operation. The anticommutation relations of fermionic operators are replaced by Poisson brackets (please see Appendix \ref{sec:pr} for details).
It is worth noting that in the second quantization formalism of quantum mechanics, the creation and annihilation operators operate on the Fock space by adding or removing particles from a specific mode (i.e., at a specific site). However, in the proposed classical description, scalar quantities are used instead of operators, and the idea of energy quantization does not apply. Also in this classical framework, the Hamiltonian is represented as a scalar quantity.
Despite these differences, we will show how the classical and quantum systems share similar mathematical structures, which allow applying the proposed representation to classical many-body systems in a way that resembles the quantum mechanical second quantization formalism.
}}

\subsubsection{Hamiltonian function and equations of motion}

In the dimerized mechanical lattice, consider an admissible state vector whose motion only localizes at a single particle, while all others are fixed, i.e., $\mathbf{X}=(0,\dots,p_j,\dots,0;0,\dots,x_j,\dots,0)$ or $\mathbf{X}=(0,\dots,q_j,\dots,0;0,\dots,y_j,\dots,0)$. Substituting into Eq.~(\ref{eq:hamdi}) gives the ``on-site energy'' at the specific particle site,
\begin{subequations}
\begin{align}
    H_{A,j} &=
    (1+r)\frac{p_j^2}{2m_0}
    +\frac{2\kappa_0 x_j^2}{2},\\
    H_{B,j} &=
    (1-r)\frac{q_j^2}{2m_0}
    +\frac{2\kappa_0 y_j^2}{2},
\end{align}
\end{subequations}
where the effect of staggered spring constants $\epsilon$ is canceled by two neighboring springs, resulting in a equivalent spring constant $2\kappa_0$, and only $r$ appears in the kinetic energy term. The same terms can be rewritten via $a_j^\pm$, $b_j^\pm$, and $r$ by using the transformation rules in Eqs.~(\ref{eq:ajbj}), obtaining
\begin{subequations}
\begin{align}
    H_{A,j} (r) &=\ii \omega_0 \frac{1}{4}
    \left(
        (1-r)(a^+_j+a^-_j)^2 - (a^+_j-a^-_j)^2
    \right),
    \\
    H_{B,j} (r) &=\ii \omega_0 \frac{1}{4}
    \left(
        (1+r)(b^+_j+b^-_j)^2 - (b^+_j-b^-_j)^2
    \right).
\end{align}
\end{subequations}
Note that when $r=0$, they reduce to
\begin{subequations}
\begin{align}
    H_{A,j} (r=0) &=\ii\omega_0\, a_j^+ a_j^-,
    \\
    H_{B,j} (r=0) &=\ii\omega_0\, b_j^+ b_j^-,
\end{align}
\end{subequations}
which is reminiscent of the on-site energy of a quantum mechanical system.

These ``on-site energy'' terms represent part of the total Hamiltonian shown in Eq.~(\ref{eq:hamdi}). The remaining terms are the inter-particle terms
$-(1+\epsilon)\kappa_0\, x_j y_j$ and
$-(1-\epsilon)\kappa_0\, x_{j+1} y_j$ 
found in the second line of Eq.~(\ref{eq:hamdi}).
They can be expressed in terms of the analogue ``hopping'' and ``superconducting'' terms {defined} as
\begin{subequations}
\begin{align}\label{eq:habh}
\begin{split}
    H_{hop,A\leftrightarrow B,j} &\equiv \ii \omega_0
    \left(
        a^-_j b^+_j + \mathrm{c.c.}
    \right)\\
    &=p_j q_j/m_0 + 2 \kappa_0 x_j y_j,
\end{split}
\\
\begin{split}
    H_{sc,A\leftrightarrow B,j} &\equiv \ii \omega_0
    \left(
        a^-_j b^-_j + \mathrm{c.c.}
    \right)\\
    &=p_j q_j/m_0 - 2 \kappa_0 x_j y_j,
\end{split}
\\
\begin{split}
    H_{hop,B\leftrightarrow A,j} &\equiv \ii \omega_0
    \left(
        b^-_j a^+_{j+1} + \mathrm{c.c.}
    \right)\\
    &=p_{j+1} q_j/m_0 + 2 \kappa_0 x_{j+1} y_j,
\end{split}
\\
\begin{split}
    H_{sc,B\leftrightarrow A,j} &\equiv \ii \omega_0
    \left(
        b^-_j a^-_{j+1} + \mathrm{c.c.}
    \right)\\
    &=p_{j+1} q_j/m_0 - 2 \kappa_0 x_{j+1} y_j.
\end{split}
\end{align}
\end{subequations}
In the mechanical lattice, the hopping and superconducting terms always appear with opposite coefficients as there is no $pq$ coupled terms in the Hamiltonian function. Finally, the Hamiltonian in the second quantized form is found as
\begin{multline}
H
\begin{aligned}[t]
    =&
    \sum_{j=1}^N{
        H_{A,j}(r)+H_{B,j}(r) }\\
    +\frac{1}{4}&\sum_{j=0}^N\big[
    \begin{aligned}[t]
    &(1+\epsilon)\left(
        H_{sc,A\leftrightarrow B}-H_{hop,A\leftrightarrow B}
    \right)\\
    +&(1-\epsilon)\left(
        H_{sc,B\leftrightarrow A}-H_{hop,B\leftrightarrow A}
    \right)\;
    \big]_j
    .
    \end{aligned}
\end{aligned}
\end{multline}
Or explicitly in terms of $a_j^\pm$ and $b_j^\pm$, it reads
\begin{widetext}
\begin{equation}\label{eq:habexp}
\begin{alignedat}{5}
  H  =\frac{\ii \omega_0}{4} &\Bigg\{%
      &&\sum_{j=1}^N
      \bigg[
      (1+r)
      \left(
      a_j^+ + a_j^-
      \right)^2
      -
      \left(
      a_j^+ - a_j^-
      \right)^2
      && {}+{} &&
      (1-r)
      \left(
      b_j^+ + b_j^-
      \right)^2
      -
      \left(
      b_j^+ - b_j^-
      \right)^2
      && \bigg]
      \\
    & +
    &&\sum_{j=0}^N
    \bigg[
      (1+\epsilon)
      \left(
      a_j^- b_j^-
      -
      a_j^- b_j^+
      +\mathrm{c.c.}
      \right)
      && {}+{} &&
      (1-\epsilon)
      \left(
      b_j^- a_{i+1}^-
      -
      b_j^- a_{i+1}^+
      +\mathrm{c.c.}
      \right)
      && \bigg]\Bigg\}.
\end{alignedat}
\end{equation}
\end{widetext}

The Hamiltonian expressions identified above (either in $(p,q,x,y)$ or $(a^\pm,b^\pm)$, see Eqs.~\ref{eq:hamdi}, \ref{eq:habexp}) contains $j=0$ and $j=N+1$ variables, where we let,
for $j=0$ and $j=N+1$,
$x_j=y_j=0$, $p_j=q_j=0$, and $a^{\pm}_j=b^{\pm}_j=0$
representing the ground. The current Hamiltonian describes a chain with both ends connected to the ground by springs with constant $(1-\epsilon)\kappa_0$ at both the left and the right ends. To study a chain with different boundary conditions, one must simply modify the corresponding terms in the Hamiltonian. For example, the Hamiltonian of the chain with free-free ends is given by the current Hamiltonian to which we subtract the potential energy contributed by the two terminal springs,
\begin{align}
    H_\mathrm{free} 
    &=
    H - (1-\epsilon)\frac{\kappa_0}{2} \left(x_1^2 + y_N^2\right)\\
    &=
    H - (1-\epsilon)\frac{-\ii \omega_0}{8}
    \left[
        \left(a_1^+ - a_1^-\right)^2 + \left(b_N^+ - b_N^-\right)^2
    \right].
\end{align}
Similarly, we could remove the last particle from the chain, hence resulting in an odd number of particles (spring-spring boundary conditions), and the Hamiltonian would be given by,
\begin{multline}
    H_\mathrm{odd} 
    =
    H - H_{B,N} -\frac{1}{4}\Big[
    (1+\epsilon)\left(
        H_{sc,A\rightarrow B}-H_{hop,A\rightarrow B}
    \right)\\
    +(1-\epsilon)\big(
        H_{sc,B\rightarrow A}-H_{hop,B\rightarrow A}
    \big)
    \Big]_N
\end{multline}

\begin{figure}[!htb]
\centering
\includegraphics[width=\columnwidth]{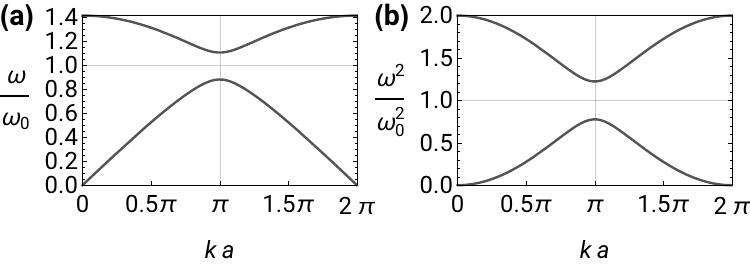}
\caption{\label{fig:2omgk}
(a) $\omega$-$k$ and (b) $\omega^2$-$k$ band structures of a classical dimerized mechanical chain \ed{with $r=0.2$ and $\epsilon=0.1$}. The $\omega^2$-spectrum is symmetric about the level $\omega_0^2=2\kappa_0/m_0$.}
\end{figure}

From the knowledge of the Hamiltonian, the system matrix $\mathsf{H}$ can then be obtained from the Hamilton's equations with the new second quantized basis $\mathbf{X}=(a^+_1,b^+_1,\dots,a^+_N,b^+_N;a^-_1,b^-_1,\dots,a^-_N,b^-_N)^\intercal$, in the following form,
\begin{align}
\begin{split}
    \mathsf{H}
    &=\sigma_3\otimes \mathbf{H}_0 + \ii \sigma_2 \otimes\bm{\Delta}_0\\
    &=
    \left(
    \begin{array}{c|c}
         \mathbf{H}_0 & \bm{\Delta}_0\\
         \hline
         -\bm{\Delta}_0 & -\mathbf{H}_0
    \end{array}
    \right),
\end{split}
\end{align}
where
$\otimes$ indicates the Kronecker product, $\sigma_j$ are the Pauli matrices, while
$\mathbf{H}_0$ and $\bm{\Delta}_0$ are $2N\times 2N$ matrices, with $N$ the number of unit cells (such that each unit cell contains two particles).
In explicit form, they read
\begin{align}
\begin{split}
    \mathbf{H}_0%
    =-\ii \omega_0 \Bigg[ \; \sum_{j=1}^N
    \;&
    \left(1+\frac{r}{2}\right)
    \ket{2j-1}\bra{2j-1}\\
    +&
    \left(1-\frac{r}{2}\right) \ket{2j}\bra{2j}\\
    -&
    \left(\frac{1+\epsilon}{4}\right) (\ket{2j-1}\bra{2j}+\mathrm{H.c.})\\
    -&
    \left(\frac{1-\epsilon}{4}\right) (\ket{2j}\bra{2j+1}+\mathrm{H.c.})
    \Bigg]+\mathbf{H}_\mathrm{B.C.},
\end{split}\\
\begin{split}
    \bm{\Delta}_0%
    =-\ii \omega_0 \Bigg[ \; \sum_{j=1}^N
    \;&
    \frac{r}{2}
    \ket{2j-1}\bra{2j-1}\\
    -&
    \frac{r}{2} \ket{2j}\bra{2j}\\
    +&
    \left(\frac{1+\epsilon}{4}\right) (\ket{2j-1}\bra{2j}+\mathrm{H.c.})\\
    +&
    \left(\frac{1-\epsilon}{4}\right) (\ket{2j}\bra{2j+1}+\mathrm{H.c.})
    \Bigg]+\bm{\Delta}_\mathrm{B.C.},
\end{split}
\end{align}
where we use the notation $\ket{j}$ to represent a $2N\times 1$ column vector with its $l$-th component equal to $\delta_{jl}$,
and $\mathbf{H}_\mathrm{B.C.}$ and $\bm{\Delta}_\mathrm{B.C.}$ are the contributions from boundary conditions that are different from the default spring-spring condition. The effects of dimerization $(r,\epsilon)$ manifests itself in the system matrix. As an example and in order to illustrate the specific pattern of the Hamiltonian matrix, we report here below the system matrix of a 4-particle (2-cell) chain under spring-spring boundary conditions
\begin{widetext}
\begin{equation}
\mathsf{H}=-\ii \omega_0
\left(
\begin{array}{cccc|cccc}
 1+\frac{r}{2} & -\frac{1+\epsilon}{4} & 0 & 0 & \frac{r}{2} & \frac{1+\epsilon }{4} & 0 & 0\\
 -\frac{1+\epsilon}{4} & 1-\frac{r}{2} & -\frac{1-\epsilon}{4} & 0 &\frac{1+\epsilon }{4} & -\frac{r}{2} & \frac{1-\epsilon }{4} & 0\\
 0 & -\frac{1-\epsilon}{4} & 1+\frac{r}{2} & -\frac{1+\epsilon}{4} & 0 & \frac{1-\epsilon }{4} & \frac{r}{2} & \frac{1+\epsilon }{4}\\
 0 & 0 & -\frac{1+\epsilon}{4} & 1-\frac{r}{2} & 0 & 0 & \frac{1+\epsilon }{4} & -\frac{r}{2}\\
   \hline
 -\frac{r}{2} & -\frac{1+\epsilon}{4} & 0 & 0 & -1-\frac{r}{2} & \frac{1+\epsilon }{4} & 0 & 0\\
 -\frac{1+\epsilon}{4} & \frac{r}{2} & -\frac{1-\epsilon}{4} & 0 & \frac{1+\epsilon }{4} & -1+\frac{r}{2} & \frac{1-\epsilon }{4} & 0\\
 0 & -\frac{1-\epsilon}{4} & -\frac{r}{2} & -\frac{1+\epsilon}{4} & 0 & \frac{1-\epsilon }{4} & -1-\frac{r}{2} & \frac{1+\epsilon }{4}\\
 0 & 0 & -\frac{1+\epsilon}{4} & \frac{r}{2} & 0 & 0 & \frac{1+\epsilon }{4} & -1+\frac{r}{2}\\
\end{array}
\right)
\end{equation}
\end{widetext}
This matrix possesses a block structure reminiscent of the Bogoliubov–de Gennes (BdG) formalism \cite{de1966superconductivity} of the Hamiltonian of the 1D superconductor, which has the form (when $\Delta \in \mathbb{R}$),
\begin{equation}
    \mathsf{H}_{\textrm{BdG}}
    =
    \left(
    \begin{array}{c|c}
         \mathbf{H} & \bm{\Delta}\\
         \hline
         -\bm{\Delta} & -\mathbf{H}
    \end{array}
    \right),
\end{equation}
Such system has a particle-hole symmetry,
\begin{equation} \mathcal{P}\mathsf{H}_{\textrm{BdG}}\mathcal{P}
=\mathcal{\bm{\sigma}}_1\mathsf{H}_{\textrm{BdG}}^\ast \bm{\sigma}_1
=-\mathsf{H}_{\textrm{BdG}},
\end{equation}
where the particle-hole symmetry operator $\mathcal{P}=\bm{\sigma}_1\mathcal{K}$, $\mathcal{K}$ is the complex conjugate operator, and $\bm{\sigma}_1=\mathbf{1}\otimes\sigma_1$. It is this symmetry that leads the system to acquire a symmetric $\pm E$ energy spectrum.
Having the same block matrix structure,
the classical diatomic chain certainly possesses the same $\pm \omega$ symmetry in the spectrum, interpreted as time-reversal symmetry, but does it show any additional hidden symmetry?
The $\omega$-$k$ dispersion of the dimerized mechanical chain is found as (see Appendix~\ref{sec:disperion})
\begin{subequations}
\begin{align}
\label{eq:dispersionF}
\omega(k) &= \pm\omega_0\sqrt{1\pm F(k)},\\
F(k) &= \sqrt{1+(1-r^2)(1-\epsilon^2)(\cos{k-1})/2}.
\end{align}
\end{subequations}
which indicates that the band structure $\omega^2$-$k$ is symmetric about $\omega_0^2$, (and also about $ka=n \pi$). Fig.~\ref{fig:2omgk} \ed{shows the typical $\omega$-$k$ and $\omega^2$-$k$ band structures of a dimerized mechanical chain. In this example, $r=0.2$ and $\epsilon=0.1$}.

Therefore we can reformulate the original eigenvalue problem $\mathsf{H}\mathbf{X} = -\ii \omega \mathbf{X}$ as $\mathsf{H}^2\mathbf{X} = -\omega^2 \mathbf{X}$ and study the new system matrix $\mathsf{H}^2$.

The matrix $\mathsf{H}^2$ has a constant term
$-\omega_0^2$ along its main diagonal, which shifts the entire spectrum to the reference level $-\omega_0^2$ without affecting the eigenvectors. It is convenient to drop the constant term and focus on the rest of the matrix $\tilde{\mathsf{H}}^2$,
\begin{equation}
    \mathsf{H}^2 = \omega_0^2 \left(-
        \mathbf{1} +\tilde{\mathsf{H}}^2
    \right).
\end{equation}
Note that $\tilde{\mathsf{H}}^2$ is nondimensionalized, and has the form,
\begin{align}
\begin{split}
    \tilde{\mathsf{H}}^2
    &=\mathbf{1}\otimes \mathbf{H}_0' + \sigma_1 \otimes\bm{\Delta}_0'\\
    &=
    \left(
    \begin{array}{c|c}
         \mathbf{H}_0' & \bm{\Delta}_0'\\
         \hline
         \bm{\Delta}_0' & \mathbf{H}_0'
    \end{array}
    \right),
\end{split}
\end{align}
where $\mathbf{1}$ represents the $2\times2$ identity matrix. $\tilde{\mathsf{H}}^2$ is
composed of the $2N\times 2N$ blocks
$\mathbf{H}_0'$ and $\bm{\Delta}_0'$ that, in explicit form, are expressed as
\begin{align}
\begin{split}
    \mathbf{H}_0'%
    =\sum_{j=1}^N
    \Bigg[
    &
    r \ket{2j-1}\bra{2j-1}\\
    -&
    r \ket{2j}\bra{2j}\\
    -&
    \left(\frac{1+\epsilon}{2}\right) (\ket{2j-1}\bra{2j}+\mathrm{H.c.})\\
    -&
    \left(\frac{1-\epsilon}{2}\right) (\ket{2j}\bra{2j+1}+\mathrm{H.c.})\Bigg]
    +\mathbf{H}'_\mathrm{B.C.},
\end{split}
\end{align}
\begin{align}
\begin{split}
    \bm{\Delta}'_0%
    =r\sum_{j=1}^N
    \Bigg[
    &
    \left(\frac{1+\epsilon}{2}\right) (\ket{2j-1}\bra{2j}-\mathrm{H.c.})\\
    -&
    \left(\frac{1-\epsilon}{2}\right) (\ket{2j}\bra{2j+1}-\mathrm{H.c.})
    \Bigg]+\bm{\Delta}'_\mathrm{B.C.},
\end{split}
\end{align}
The matrix $\tilde{\mathsf{H}}^2$ for the same 4-particle chain is shown here below,
\begin{widetext}
\begin{equation}\label{eq:h2tilde}
    \tilde{\mathsf{H}}^2=
    \left(
    \begin{array}{cccc|cccc}
     r & -\frac{1+\epsilon}{2} & 0 & 0 & 0 & \frac{r (1+\epsilon)}{2} & 0 & 0 \\
     -\frac{1+\epsilon}{2} & -r & -\frac{1-\epsilon}{2} & 0 & -\frac{r (1+\epsilon)}{2}  & 0 & -\frac{r (1-\epsilon)}{2} & 0 \\
     0 & -\frac{1-\epsilon}{2} & r & -\frac{1+\epsilon}{2} & 0 & \frac{r(1-\epsilon )}{2} & 0 & \frac{r (1+\epsilon)}{2} \\
     0 & 0 & -\frac{1+\epsilon}{2} & -r & 0 & 0 & -\frac{r(1+\epsilon)}{2} & 0 \\
     \hline
     0 & \frac{r (1+\epsilon)}{2} & 0 & 0 & r & -\frac{1+\epsilon}{2} & 0 & 0 \\
     -\frac{r(1+\epsilon)}{2} & 0 & -\frac{r (1-\epsilon)}{2} & 0 & -\frac{1+\epsilon}{2}
       & -r & -\frac{1-\epsilon}{2} & 0 \\
     0 & \frac{r(1-\epsilon)}{2} & 0 & \frac{r (1+\epsilon)}{2} & 0 & \frac{\epsilon
       -1}{2} & r & -\frac{1+\epsilon}{2} \\
     0 & 0 & -\frac{r(1+\epsilon)}{2} & 0 & 0 & 0 & -\frac{1+\epsilon}{2} & -r \\
    \end{array}
    \right).
\end{equation}
\end{widetext}

At first glance, $\tilde{\mathsf{H}}^2$ lost the BdG-like block-antisymmetry pattern seen in $\mathsf{H}$. However, it gains additional antisymmetry in the $A$-$B$ sublattices; the dimerization parameters $r$ and $\epsilon$ appear with opposite signs in $2j-1$ and $2j$ components in $\tilde{\mathsf{H}}^2$. In other words, the $A$-$B$ sublattice degree of freedom contributes to the determination of the analogue particle-hole degree of freedom.
Note that both the SSH \cite{su1979solitons} and the Rice-Mele \cite{rice1982elementary} models  also consider dimerized 1D lattices. However, they are not superconducting models and do not have the pairing terms $\Delta \hat{c}_j^\dagger\hat{c}_{j+1}^\dagger + \mathrm{H.c.}$ in their Hamiltonians.
\label{ParaB3}
{
In the case of dimerized mechanical lattices,
\ed{the use of the dynamical matrix approach allowed identifying these systems as classical analogue to the SSH systems \cite{xiao2015geometric,xiao2017topological,chaunsali2017demonstrating,yin2018band,chen2018study,vila2019role,shi2021disorder}, however when employing the second quantized form of the Hamiltonian
}
the analog ``superconducting'' and ``hopping'' terms naturally appear simultaneously. These terms also make the system very similar (at a mathematical level) to a Kitaev chain model. Also, 
\ed{
under the new coordinates of $a_j^\pm, b_j^\pm$,
}
the off-diagonal block $\bm\Delta_0'$ in $\tilde{\mathsf{H}}^2$ becomes antisymmetric, which is aligned with the BdG form of the Hamiltonian.
}

Note that $\tilde{\mathsf{H}}^2$ is dimensionless and so are its eigenvalues. Let $\tilde{\omega}^2 = \frac{\omega^2-\omega_0^2}{\omega_0^2}$ be the normalized eigenvalue, then $\tilde{\mathsf{H}}^2\mathbf{X} = \tilde{\omega}^2\mathbf{X}$, where $\mathbf{X}$ is the same eigenvector satisfying $\mathsf{H}\mathbf{X}=-\ii\omega\mathbf{X}$. 
Given that $\omega$ always shows in positive-negative pairs, the $\omega^2$ and $\tilde{\omega}^2$ spectra are always doubly degenerate, hence corresponding to time-reversal pairs of eigenvectors. In addition, the number of distinct $\tilde{\omega}^2$ values matches the number of particles in a chain.

\begin{figure*}[!htb]
\centering
\includegraphics[width=.9\textwidth]{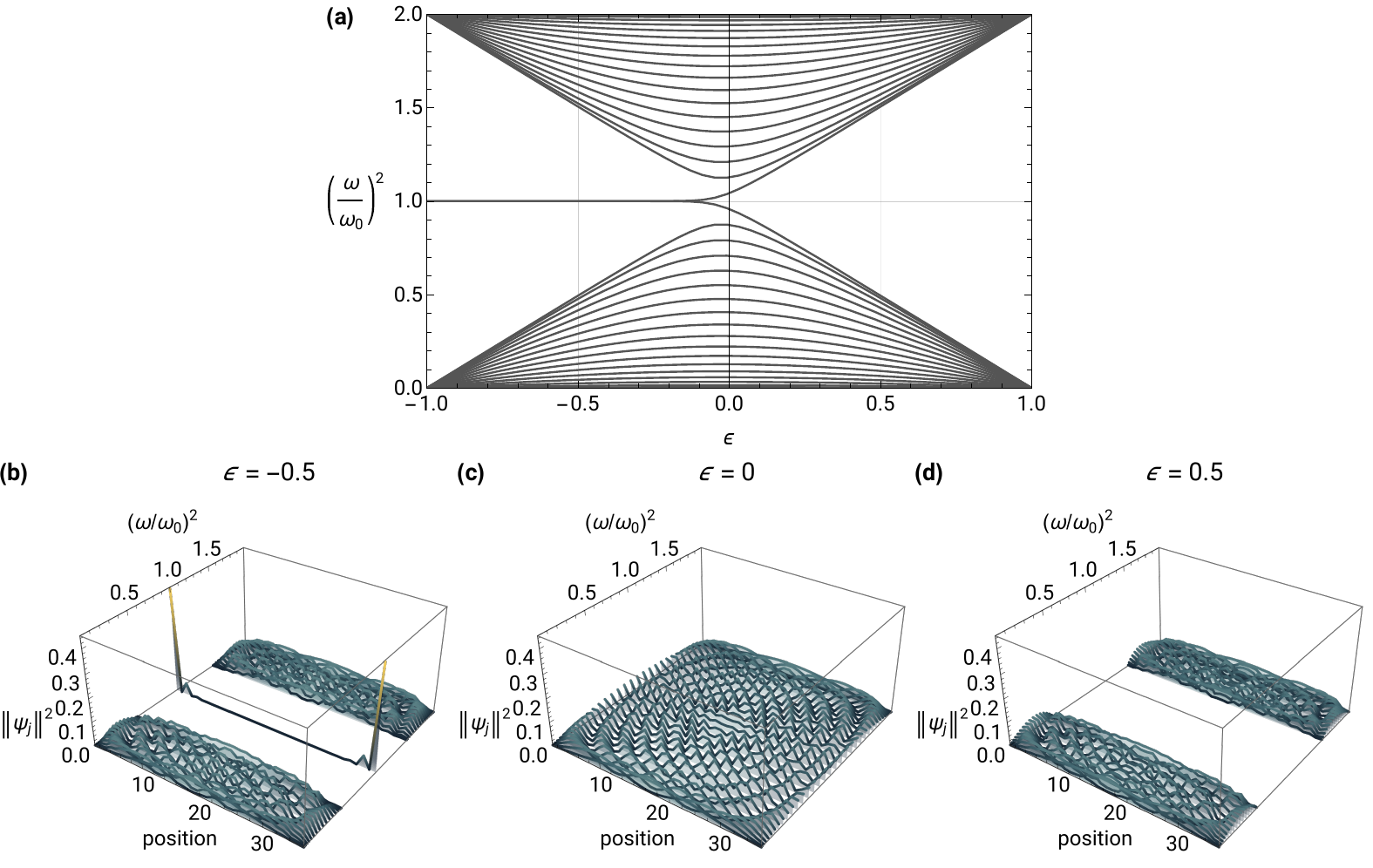}
\caption{\label{fig:di_ess_e}
(a) Spectrum and (b-d) mode shapes of the diatomic chain with 36 (even) particles (18 cells), both ends terminated by springs connected to ground, $r=0$, and varying $\epsilon$ values.
When $\epsilon<0$, Majorana-like bound states with $\omega=\omega_0$ appear at the two ends of the mechanical chain. Symbolic representations of the chains for $\epsilon<0$:
$\lvert=\bullet-\bullet\cdots\bullet-\bullet=\rvert$,
and $\epsilon>0$:
$\lvert-\bullet=\bullet\cdots\bullet=\bullet-\rvert$.}
\end{figure*}

Before analyzing the symmetry of the system, let us first take a look at some selected numerical results within a specific scenario.
Fig.~\ref{fig:di_ess_e} shows the spectrum and the mode shapes of a classical diatomic chain with 36 particles (18 cells) and spring-terminated ends. $r$ is fixed at 0, with $\epsilon$ varying from $-1$ to $1$. When $\epsilon<0$, bound states with $\omega=\omega_0$ appear at both ends of the mechanical chain. The entire spectrum is symmetric with respect to $\omega=\omega_0$, which can be ascribed to the synthetic particle-hole symmetry in the dimerized mechanical chain. Further examples under different parameters are shown in Appendix~\ref{sec:examples} for reference.

\subsubsection{Synthetic particle-hole symmetry}
The synthetic particle-hole symmetry operator $\mathbf{P}$ under the second quantized basis $(a^+_j,b^+_j;a^-_j,b^-_j)^\intercal$ can be expressed as
\begin{equation}
    \mathbf{P} = \sigma_3 \otimes \mathbf{D} \otimes \left(\ii\sigma_2\right),\;
    \mathbf{D} = 
    \begin{pmatrix}
    & & & &\diagup\\
    & &1& &\\
    \diagup& & & &
    \end{pmatrix}_{N\times N},
\end{equation}
with $\mathbf{D}$ the $N\times N$ skew-diagonal identity matrix. Explicitly, $\mathbf{P}$ in matrix form looks like,
\begin{equation}
    \mathbf{P}=\left(
    \begin{array}{ccc|ccc}
         & & +1 & & &\\
         & -1 & & & \mathbf{0} &\\
         \iddots& & & & &\\
         \hline
          & & & & & -1\\
          & \mathbf{0} & & & +1 &\\
          & & & \iddots& &\\
    \end{array}
    \right)_{4N\times 4N}.
\end{equation}
The action of $\mathbf{P}$ on a state vector can be interpreted as follows: 1) the $\sigma_3$ term adds minus signs to the ``hole'' components i.e., $a_j^-$ and $b_j^-$ components. This is effectively a complex conjugate (time reversal) operator if one transform the vector back to the $(p,x)$ phase space, as it will reverse the phase difference between $p$ and $x$, 2) $\mathbf{D}$ reverses the cell order in the chain, and 3) $\ii\sigma_2$ swaps and adds alternating signs to the $a$ and $b$ components. The intra-cell swapping and cell-order-reversing $\mathbf{D}$ together make a space inversion operator for a state vector. The alternating-sign operation represents the sublattice (or chiral) symmetry. Its matrix representation read $\sigma_3$ acting on $(A,B)$ sublattice space. Essentially,
such operation multiplies the state vector $\bm{\psi}$ by the vector $\mathbf{v}=(+1, -1, +1, -1, \dots)$ in real space. Note that $v_j=e^{i\pi j}=e^{i (2\pi/a) j}$, where $a=2$ is the lattice constant. In $k$-space, $v(k)$ is a unit impulse located at $k=2\pi/a$. The multiplication in real space is performed as convolution in $k$-space, and the operation manifests in $\psi(k)\rightarrow\psi(k+2\pi/a)$, i.e., the wavenumber increases by $2\pi/a$. In the first Brillouin zone, where $k$ is limited to the interval $[-\pi/a,\pi/a]$,
the shift in the wavenumber does not change $k$ but it does produce a switch from the low-frequency (acoustic) branch to the high-frequency (optical) branch, and vice versa.
It follows that, if vectors before and after the sublattice symmetry operation are both eigenvectors of a chain, they are of identical wavenumber, and such operation will swap them vertically in the $\omega^2$-$k$ spectrum.

In summary, the synthetic particle-hole symmetry operator $\mathbf{P}$ is a composite operator comprising time-reversal, space-inversion, and sublattice symmetry operations.
\ed{As will be seen in
Sec.~\ref{sec:PHSk}, the k-space analysis will lead to the same conclusion.
}

Under current $(a^+_j,b^+_j;a^-_j,b^-_j)^\intercal$-basis representation, $\mathbf{P}$ is an orthogonal (real unitary%
) matrix, $\mathbf{P}^\intercal=\mathbf{P}^{-1}$.
For a chain with an even number of particles and both ends spring-terminated (composed of complete unit cells), its system matrix $\tilde{\mathsf{H}}^2$ (e.g., Eq.~\ref{eq:h2tilde}) satisfies the synthetic particle-hole symmetry,
\begin{equation}
    \mathbf{P}\tilde{\mathsf{H}}^2\mathbf{P}^{-1}=-\tilde{\mathsf{H}}^2.
\end{equation}
For an eigenmode $\mathbf{X}$ of the chain, which satisfies $\tilde{\mathsf{H}}^2 \mathbf{X}= \tilde{\omega}^2\mathbf{X}$, we will have $\mathbf{P}\tilde{\mathsf{H}}^2 \mathbf{X} = \tilde{\omega}^2 \mathbf{P}\mathbf{X}$. With the manipulation $\mathbf{P}\tilde{\mathsf{H}}^2 \mathbf{P}^{-1}\mathbf{P} \mathbf{X} = \tilde{\omega}^2 \mathbf{P}\mathbf{X}$, we obtain $-\tilde{\mathsf{H}}^2 \mathbf{P} \mathbf{X} = \tilde{\omega}^2 \mathbf{P}\mathbf{X}$, or $\tilde{\mathsf{H}}^2 \mathbf{P} \mathbf{X} = -\tilde{\omega}^2 \mathbf{P}\mathbf{X}$. That is, for any eigenvector $\mathbf{X}$ of $\tilde{\mathsf{H}}^2$ with eigenvalue $\tilde{\omega}^2$, its synthetic-particle-hole-exchanged vector $\mathbf{P}\mathbf{X}$ is still an eigenvector of $\tilde{\mathsf{H}}^2$, with the new eigenvalue $-\tilde{\omega}^2$.

\begin{sloppypar}
This synthetic PHS holds for chains with even numbers of particles with spring terminations (complete cells),
having the patterns like
${\lvert-\bullet=\circ-\bullet=\circ-\rvert}$.
Here we introduce the notations
``$\bullet$,''
``$\circ$,''
``$-$,''
``$=$,'' and
``$|$,''
which stands for heavier and lighter particles, softer and stiffer springs, and the ground, respectively. By printing only 2 (or 1.5) cells (which is enough to observe chains' patterns and their symmetry), we can symbolize the same kind of chains with minimal notation.
For a chain with an odd number of particles (see Fig.~\ref{fig:di_off_r} in Appendix~\ref{sec:examples}), there is an odd number of distinct eigenvalues and the spectrum cannot be perfectly symmetric about the reference level, unless there is one uniformly lying on the reference level (see Fig.~\ref{fig:di_oss_e} in Appendix~\ref{sec:examples}).
Also, for a chain with an odd number of particles, its system matrix $\tilde{\mathsf{H}}^2_\mathrm{odd}$ is of dimensions $(4N-2)\times(4N-2)$, and the synthetic PHS operator $\mathbf{P}_\mathrm{odd}$ with the same dimensions can be built by dropping the last columns and rows in each of the four blocks. It turns out that the synthetic PHS operator for odd-particle chains does not map between eigenvectors of the same chain. Instead, it maps between two chains with opposite dimerization parameters, e.g.,
${\lvert-\bullet=\circ-\bullet=\rvert}$ $\leftrightarrow$
${\lvert=\circ-\bullet=\circ-\rvert}$,
namely, two chains with patterns displaced by a half lattice,
\begin{equation}\label{eq:phsodd}
    \mathbf{P}_\mathrm{odd}
    \tilde{\mathsf{H}}^2_\mathrm{odd} (r,\epsilon)
    \mathbf{P}_\mathrm{odd}^{-1}
    = -
    \tilde{\mathsf{H}}^2_\mathrm{odd} (-r,-\epsilon).
\end{equation}
Note that shifting the pattern by half of a lattice in a periodic (infinite) chain does note affect the response of the chain. For a finite chain with a large number $(N\gg 1)$ of cells (i.e., in the thermodynamic limit), the contribution to the bulk modes due to the boundary is minimal, so the bulk spectra shown in Fig.~\ref{fig:di_off_r} in Appendix~\ref{sec:examples} and Fig.~\ref{fig:di_all3D} still all look symmetric.
\end{sloppypar}

In addition, in the off-diagonal blocks $\bm\Delta'_0$, there are quadratic $r\epsilon$ coupled terms. When $r\epsilon=0$, that is, either $r$ or $\epsilon$ vanishes, the coupled terms disappear. In terms of symmetry in the chain, the bulk pattern will then acquire inversion symmetry
(e.g., ${\cdots-\bullet=\bullet-\cdots}$, or ${\cdots\circ-\bullet-\circ\cdots}$).
Concerning the inversion symmetry of a finite chain, it needs accounting also for the boundary conditions. Particularly, the example just shown (that is (1) even, spring-spring, $r=0$, $\lvert=\bullet-\bullet=\bullet-\bullet=\rvert$, see Fig.~\ref{fig:di_ess_e} and (2) odd, free-free, $\epsilon=0$, $\bullet-\circ-\bullet$, see Fig.~\ref{fig:di_off_r} in Appendix~\ref{sec:examples}) are chains with inversion symmetry.
For chains with inversion symmetry, the bound states always show in pairs at the two ends because the two ends appear to be identical when viewed from each side.
This is in line with the Kitaev model \cite{kitaev2001unpaired}, where Majorana zero modes must show in pairs, since they are obtained by splitting electrons into halves.

Fig.~\ref{fig:di_all3D} shows the spectra of the diatomic chains under $(r,\epsilon)$-parametric space with various combinations of boundary conditions.
\begin{figure*}[!htb]
\centering
\includegraphics[width=\textwidth]{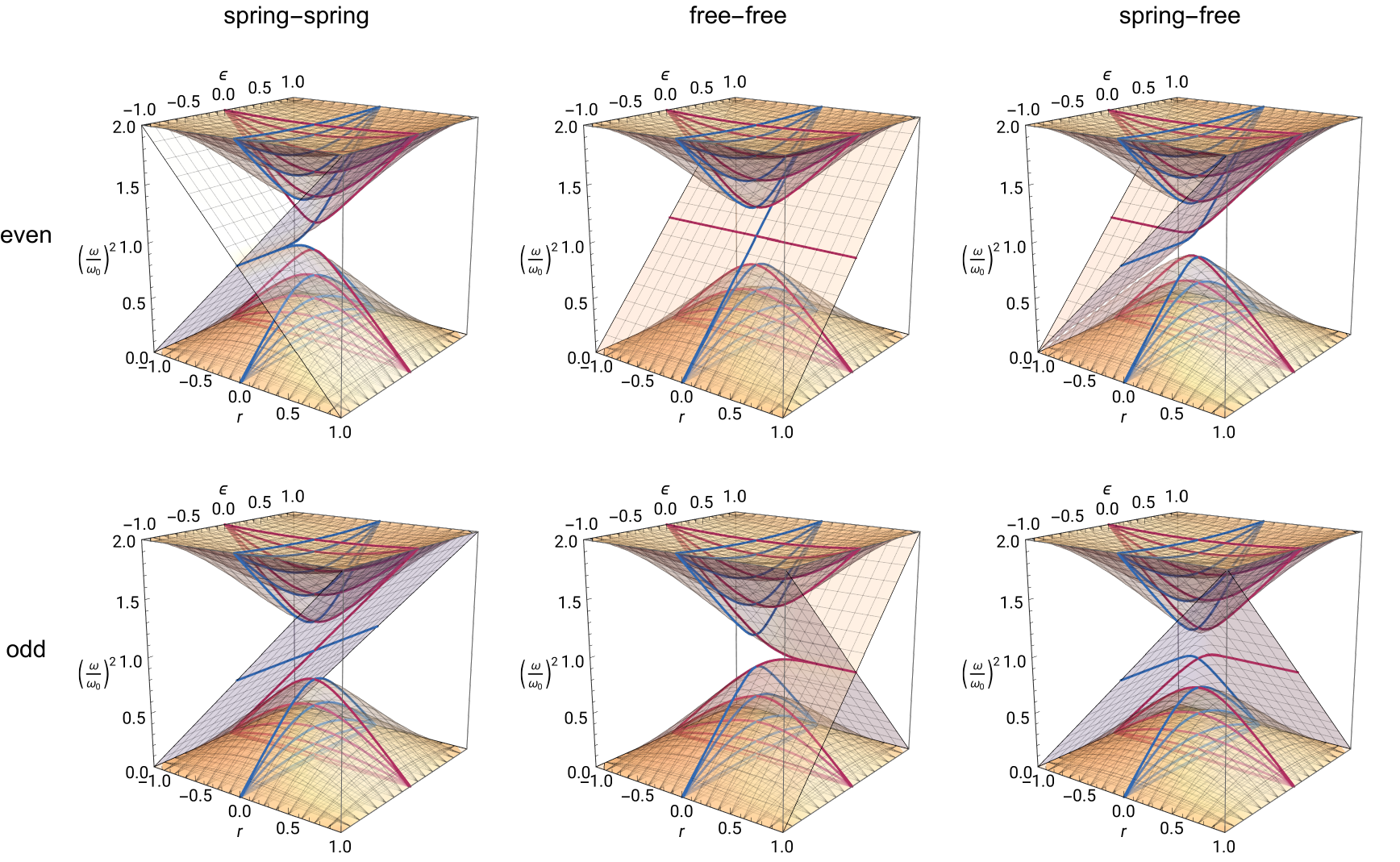}
\caption{\label{fig:di_all3D}
Spectra of the diatomic chains in the $(r,\epsilon)$-parametric space with various combinations of boundary conditions.
Section curves on the high-symmetry planes $r=0$ and $\epsilon=0$ are emphasized with blue and magenta curves, respectively.
}
\end{figure*}
Spectra on the cross-sections with inversion symmetry bulk patterns ($r=0$ or $\epsilon=0$) are emphasized with blue or magenta curves, respectively.
For chains with inversion symmetry, there can be either 1) two Majorana-like zero modes (one at each end), or 2) no Majorana-like modes at all; these two cases correspond to topological and trivial phases. For a chain with the bulk pattern respecting inversion symmetry (although the chain itself does not, due to terminal conditions), there can be a single Majorana zero mode at one of the ends, or at none of them. 

\subsubsection{Single bound state in the mechanical chain}
1D topological quantum systems such as the Kitaev model or the SSH model always have the bound states appearing in pairs at the two ends of the chain. The bound states can be interpreted as quasiparticles with half of the degrees of freedom of a unit cell of the original chains.
\ed{The Majorana bound states and the topological bound states of the SSH chain must show in pairs due to the fact that}
in the Kitaev chain, the total number of electrons must be an integer, while in the SSH chain (polyacetylene, $[\text{C}_2\text{H}_2]_n$) there is always an even number of carbon atoms.

The classical mechanical lattices presented in this work possesses even greater flexibility, as the chain can be truncated at any point (including locations in between sublattices), hence also leading to individual bound states, as shown in Fig.~\ref{fig:di_oss_e}  in Appendix~\ref{sec:examples}. Nonetheless, whenever $r\epsilon=0$, that is, bulk pattern has inversion symmetry \emph{locally} (while the global inversion symmetry could be broken due to the number of particles or different boundary conditions), the bound states always have zero frequency $\tilde{\omega}=0$ with respect to the reference level, or $\omega=\omega_0$, and hence they are zero modes. In the following paragraph we show with asymptotic analysis the presence of the zero modes, and its Hamiltonian akin to the Majorana zero modes of the Kitaev model.

\subsubsection{Majorana-like Hamiltonian of the zero modes}
The topological analysis of the band structure necessary to show the nontrivial (topological) nature of the zero modes will be addressed in the next section after presenting the $k$-space representation of the system. In this section, we offer a phenomenological point of view which explains the existence of localized modes at zero frequency (about the reference level) under certain parameter ranges.
\textcite{kitaev2001unpaired} considered two extreme conditions in his model (see Eq.~(\ref{eq:HK})): the on-site energy ($\mu$) being much greater than the hopping and superconducting amplitude $\mu\gg t=\lvert\Delta\rvert$, and vice versa, $\mu\ll t=\lvert\Delta\rvert$. They correspond to different ways of pairing the Majorana fermions. In the latter case, unpaired Majorana fermions can be found at the ends.
Here, a similar idea can be utilized to understand the existence of the zero bound states.

\begin{figure*}[!htb]
\centering
\includegraphics[width=.8\textwidth]{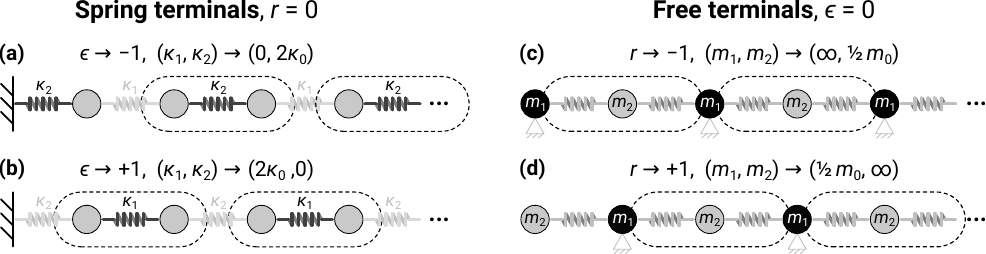}
\caption{\label{fig:interpretation}
Phenomenological interpretation of the bound states. (a,b) Chains with spring terminals and $r=0$ under two extreme conditions: $\epsilon$ approaching $-1$ and $1$, respectively. An unpaired simple resonator with $m_0$ and $2\kappa_0$ ($\omega=\omega_0$) appears at the end of the chain in the former case.
(c,d) Chains with free terminals and $\epsilon=0$ under two extreme conditions: $r$ approaching $-1$ and $1$, respectively. An unpaired simple resonator with $0.5m_0$ and $\kappa_0$ ($\omega=\omega_0$) appears at the end of the chain in the latter case.
}
\end{figure*}

Let us first consider a spring-terminated chain with $r=0$ (all particle with identical mass $m_0$) under two extreme conditions: 1) $\epsilon\rightarrow-1$ and 2) $\epsilon\rightarrow+1$, as shown in the left of Fig.~\ref{fig:interpretation}.

In both cases, the dimerization leads to a sequence of \emph{decoupled} diatomic oscillators given that one of the two spring constant vanishes.
These oscillators have two eigenmodes: the internal contraction/extension mode $(1,-1)$ with eigenfrequency $\omega=\sqrt{4\kappa_0/m_0}=\sqrt{2}\omega_0$ and the rigid body mode $(1,1)$ with $\omega=0$. They compose the $N$-fold degeneracy (assuming $N$ pairs) at $\omega^2=2\omega^2_0$ and $\omega^2=0$, on the left and right sides of the spectrum shown in Fig.~\ref{fig:di_ess_e}.
Nevertheless, in the first case ($\epsilon-\rightarrow-1$), 
the first particle is not paired with its neighboring particle, but attached to the ground with a spring having $\kappa_2=2\kappa_0$.
This condition leads to a local resonant mode with $\omega=\sqrt{2\kappa_0/m_0}=\omega_0$. Further, if the chain possesses inversion symmetry, the same thing will happen at the other terminal, hence resulting in a two-fold degeneracy at $\omega=\omega_0$ as $\epsilon\rightarrow-1$, as shown in Fig.~\ref{fig:di_ess_e}.

The Hamiltonian under the first condition $(r=0,\epsilon=-1)$ reads (c.f. Eq.~\ref{eq:habexp})
\begin{multline}
    H_1=\ii\omega_0\sum_j \Big[ a_j^+a_j^- + b_j^+b_j^- \\
    +
    \frac{1}{2}\left(
        b_j^-a_{j+1}^-+b_j^+a_{j+1}^+ - b_j^-a_{j+1}^+-b_j^+a_{j+1}^-
    \right)
    \Big].
\end{multline}
The first two terms in the summation only provide the reference level $\omega_0$.
Recall that, in condensed matter systems, the Majorana operator $\hat{\gamma}$ is obtained by taking either the real or the imaginary part of the fermion creation and annihilation operator, $\hat{c}^\dagger =\frac{1}{2}\left(\hat{\gamma}_1+\ii \hat{\gamma}_2\right)$,
$\hat{c} =\frac{1}{2}\left(\hat{\gamma}_1-\ii \hat{\gamma}_2\right)$.
Following the same idea,
if we define ${\gamma_j^A}=\frac{\ii}{\sqrt{2}}(a_j^- - a_j^+)$, and ${\gamma_j^B}=\frac{\ii}{\sqrt{2}}(b_j^- - b_j^+)$ (that are self-conjugate
$
\left(\gamma_j^{A/B}\right)^\ast=\gamma_j^{A/B}
$),
the remaining terms in the Hamiltonian are then
\begin{equation}
    \tilde{H}_1=-\ii\omega_0\sum_j
        \gamma_j^B 
        \gamma_{j+1}^A.
\end{equation}
The Hamiltonian says that the $B$ sublattice of the $j^\mathrm{th}$ cell is paired with the $A$ sublattice of the $(j+1)^\mathrm{th}$ cell. Such pairing leaves the $A$ sublattice of the first cell and the $B$ sublattice of the last cell behind, and they become unpaired bound states, similar to the nontrivial Kitaev chain Hamiltonian (Eq.~(\ref{eq:HK2})).

On the other hand, the Hamiltonian under the second condition $(r=0,\epsilon=+1)$ reads
\begin{multline}
    H_2 = \ii\omega_0\sum_j \Big[ a_j^+a_j^- + b_j^+b_j^-\\
    +
    \frac{1}{2}\left(
        a_j^-b_{j}^-+a_j^+b_{j}^+ - a_j^-b_{j}^+-a_j^+b_{j}^-
    \right)
    \Big],
\end{multline}
or,
\begin{equation}
    \tilde{H}_2 = -\ii\omega_0\sum_j
        \gamma_j^A 
        \gamma_j^B,
\end{equation}
which pairs the $A/B$ sublattices within each unit cell, without leaving any unpaired states; this latter case is reminiscent of the trivial Kitaev chain (Eq.~(\ref{eq:HK1})).

On the right-hand side of Fig.~\ref{fig:interpretation}, another situation is considered: a diatomic chain with free ends and with $\epsilon=0$. Now, let us consider two extreme conditions: 1) $r\rightarrow-1$ and 2) $r\rightarrow+1$.
In both cases, $r\rightarrow\pm1$ corresponds to one mass approaching infinity and the other approaching a value of half of the reference mass (according to Eq.~\ref{eq:mj}).
A particle with infinite mass could be interpreted as the ground, or considered as almost fixed.
So, again, the dimerized chain becomes a series of decoupled oscillators, separated by the infinite-mass particles which hardly transfer any force or displacement. Inside each oscillator there is a particle with mass $\frac{1}{2}m_0$ connected by two springs of constant $\kappa_0$, and the resonant frequency is $\sqrt{4\kappa_0/m_0}=\sqrt{2}\omega_0$. All the local resonators contribute a $N$-fold degeneracy at $\omega^2=2\omega_0^2$ in the spectrum as $r\rightarrow\pm1$. The $N$-fold degenerate eigenfrequencies at $\omega^2\rightarrow0$ are ascribed to the heavy particles ---they are not truly fixed--- but all the eigenmodes involving their motion have eigenfrequencies approaching zero since the modal mass approaches infinity. Particularly, in case 2, there is a particle not bounded by two heavy ones but suspended at the end with only one connected spring. It has natural frequency $\omega=\sqrt{\kappa/(m_0/2)}=\omega_0$ and results in the zero bound state.

At first glance, there is no direct analog to the pairing of Majorana modes in a unit cell, given in every oscillator there is only one moving particle. In fact, the analog lies in pairing of springs rather than particles, as in each oscillator there are two springs with deflections of equal amount but opposite signs. Nevertheless, given that we are using ``particle-standard'' notations (including $p_j,q_j,x_j,y_j$ and their derivative quantities $a_j^\pm,b_j^\pm$), the Majorana states cannot be efficiently formulated for such chains. If instead, ``spring-standard'' notations were adopted (using spring deflections as variables such as $\xi_j=y_j-x_j, \eta_j=x_{j+1}-y_j$ and so on), a dual formulation reminiscent of the previous examples could be obtained. Nevertheless, the duality between both cases have been shown in various examples (e.g., Fig.~\ref{fig:di_ess_e} and Fig.~\ref{fig:di_off_r}).



\subsection[Momentum space representation and the topological invariant]{\textit{k}-space representation and the topological invariant}\label{sec:kspace}
To express the Hamiltonian in $k$-space and second quantization formalism, consider the discrete Fourier transform pairs,
\begin{equation}\label{eq:ftaj2ak}
    \begin{cases}
			\;a_k^\pm &= \displaystyle\frac{1}{\sqrt{N}} \sum_{j=1}^N \ee^{-\ii k j} a_j^\pm,\\
            \;a_j^\pm &= \displaystyle\frac{1}{\sqrt{N}} \sum_{q=1}^N \ee^{+\ii k j} a_k^\pm,
	\end{cases}\;\;
	\begin{cases}
			\;b_k^\pm &= \displaystyle\frac{1}{\sqrt{N}} \sum_{j=1}^N \ee^{-\ii k j} b_j^\pm,\\
            \;b_j^\pm &= \displaystyle\frac{1}{\sqrt{N}} \sum_{q=1}^N \ee^{+\ii k j} b_k^\pm.
	\end{cases}
\end{equation}
For all $k\in(0,2\pi]$, the transformed equations become
\begin{equation}\label{eq:ft}
    \tilde{\mathsf{H}}_k^2(k)
    \left(
    \begin{array}{c}
         a_{k}^+ \\
         b_{k}^+ \\
         a_{k}^- \\
         b_{k}^- \\
    \end{array}
    \right)
    =
    \tilde{\omega}^2_k
    \left(
    \begin{array}{c}
         a_{k}^+ \\
         b_{k}^+ \\
         a_{k}^- \\
         b_{k}^- \\
    \end{array}
    \right),
\end{equation}
in which the system matrix reads
\begin{widetext}
\begin{equation}\label{eq:hk2t}
    \tilde{\mathsf{H}}_k^2(k)=\left(
    \begin{array}{cc|cc}
         r & -\frac{1}{2} \left(\ee^{-\ii k} (1-\epsilon )+\epsilon +1\right) & 0 & \frac{1}{2} r \left(\ee^{-\ii k} (1-\epsilon )+\epsilon +1\right) \\
         -\frac{1}{2} \left(\ee^{\ii k} (1-\epsilon )+\epsilon +1\right) & -r & -\frac{1}{2} r \left(\ee^{\ii k} (1-\epsilon )+\epsilon +1\right) & 0 \\
         \hline
         0 & \frac{1}{2} r \left(\ee^{-\ii k} (1-\epsilon )+\epsilon +1\right) & r & -\frac{1}{2} \left(\ee^{-\ii k} (1-\epsilon )+\epsilon +1\right) \\
         -\frac{1}{2} r \left(\ee^{\ii k} (1-\epsilon )+\epsilon +1\right) & 0 & -\frac{1}{2} \left(\ee^{\ii k} (1-\epsilon )+\epsilon +1\right) & -r \\
    \end{array}
    \right).
\end{equation}
\end{widetext}
Note that the matrix is not Hermitian particularly due to the fact that the selected basis $a_k^\pm$ ($b_k^\pm$) are not complex conjugate pairs. Based on Eq.~(\ref{eq:ft}), ${a_k^\pm}^\ast=a_{-k}^\mp$ (note the sign change in $k$), which is different from the quantum mechanics convention, where $\hat{c}_k^\dagger$ is the Hermitian conjugate of $\hat{c}_k$. The detail of the Fourier transform is provided in Appendix~\ref{sec:FT} for reference.

$\tilde{\mathsf{H}}_k^2(k)$ can be decomposed into two pairs of identical $2\times2$ blocks and can be expressed as
\begin{equation}
    \tilde{\mathsf{H}}_k^2(k) =
    \mathbf{1} \otimes \tilde{\mathsf{H}}_k^2(k)_{1,1} + \sigma_1 \otimes \tilde{\mathsf{H}}_k^2(k)_{1,2},
\end{equation}
where
\begin{subequations}
\begin{align}
    \begin{split}
    \tilde{\mathsf{H}}_k^2(k)_{1,1} =
    -\frac{1}{2}\left[(1-\epsilon)\cos{k}+(1+\epsilon)\right] & \sigma_1\\
    -\frac{1-\epsilon}{2}\sin{k}\; \sigma_2
    + r & \sigma_3,
    \end{split}
    \\
    \begin{split}
    \tilde{\mathsf{H}}_k^2(k)_{1,2} =
    \frac{\ii r}{2} \big\{ \left[(1-\epsilon ) \cos{k} +(1+\epsilon)\right] &\sigma_2\\
    -  \left[(1-\epsilon ) \sin{k}\right] &\sigma_1\big\}.
    \end{split}
\end{align}
\end{subequations}
As a $4\times4$ matrix, $\tilde{\mathsf{H}}_k^2(k)$ has four eigenvalues that form two doubly degenerate pairs. The two distinct eigenvalues are found to be
\begin{equation}
    \tilde{\omega}^2_\pm(k) = \pm \sqrt{1+(1-r^2)(1-\epsilon^2)(\cos{k-1})/2},
\end{equation}
which is identical to $F(k)/\omega_0$ (see Eq.~\ref{eq:dispersionF} in p.~\pageref{eq:dispersionF}) obtained using conventional dynamic matrix approach. Again, double degeneracy is due to time reversal symmetry, and each eigenvalue corresponds to a pair of time reversal eigenmodes.

Also, remember that $\tilde{\omega}^2_\pm$ are defined with respect to the reference squared frequency $\omega_0^2=2\kappa_0/m_0$, around which the spectrum is symmetric,
\begin{multline}
    \omega^2=\omega_0^2 \left(1\pm \tilde{\omega}^2\right)\\
    =\frac{2\kappa_0}{m}
    \left[1\pm \sqrt{1+\left(1-r^2\right)\left(1-\epsilon^2\right)(\cos{k-1})/2}\right].
\end{multline}

\subsubsection[Synthetic particle-hole symmetry in momentum space]{Synthetic particle-hole symmetry in \textit{k}-space}\label{sec:PHSk}
The $k$-space representation of the synthetic particle-hole symmetry operator, which is an antiunitary operator, can be expressed as $\mathbf{P}=\mathbf{U}_P\mathcal{K}$, where $\mathbf{U}_P$ is a unitary operator and $\mathcal{K}$ the complex conjugation.
The matrix form of $\mathbf{U}_P$ can be deduced from the real-space version of $\mathbf{P}$. Under current $(a_k^+,b_k^+,a_k^-,b_k^-)^\intercal$ representation,
\begin{equation}
    \mathbf{U}_P=\begin{pmatrix}
    0 & -1 & 0 & 0\\
    1 & 0 & 0 & 0\\
    0 & 0 & 0 & 1\\
    0 & 0 & -1 & 0
    \end{pmatrix}
    = - \sigma_3 \otimes \left(\ii\sigma_2\right).
\end{equation}
As an example, $\mathbf{P}=\mathbf{U}_P\mathcal{K}$ acting on a vector $\mathbf{X}=(u, v, \mu, \nu)^\intercal$ yields
\begin{equation}
    \mathbf{P}\begin{pmatrix}
    u\\ v\\ \mu \\ \nu
    \end{pmatrix}
    =
    \mathbf{U}_P\mathcal{K}
    \begin{pmatrix}
    u\\ v\\ \mu \\ \nu
    \end{pmatrix}
    =
    \mathbf{U}_P\begin{pmatrix}
    u^\ast\\ v^\ast\\ \mu^\ast \\ \nu^\ast
    \end{pmatrix}
    =
    \begin{pmatrix}
    -v^\ast\\ u^\ast\\ \nu^\ast \\ -\mu^\ast
    \end{pmatrix}.
\end{equation}

\begin{figure}[!htb]
\centering
\includegraphics[width=.8\columnwidth]{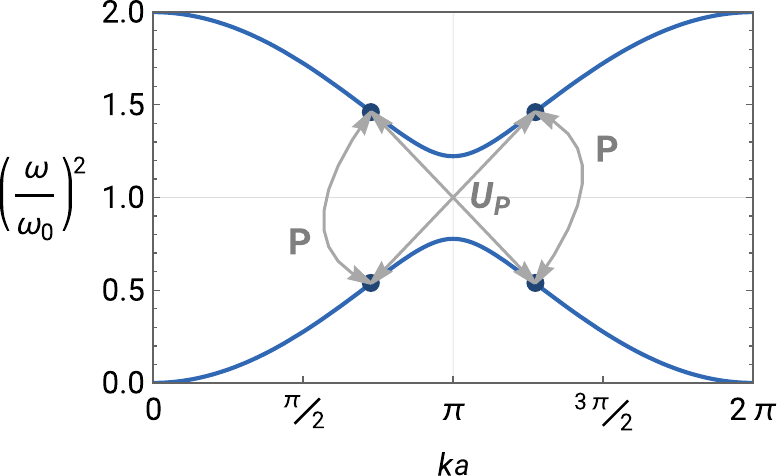}
\caption{\label{fig:spctphs}
Illustration of mapping produced by the applications of the synthetic particle-hole symmetry operator in $k$-space.}
\end{figure}

It can be easily checked that the synthetic particle-hole symmetry holds for the system; applying $\mathbf{P}$ on the system matrix $\tilde{\mathsf{H}}^2_k(k)$ gives
\begin{align}
    \mathbf{P} \tilde{\mathsf{H}}^2_k(k) \mathbf{P}^{-1}
    &=
    \mathbf{U}_P \mathcal{K}\, \tilde{\mathsf{H}}^2_k(k)\, \mathcal{K}^{-1} \mathbf{U}_P^{-1}\\
    &=
    \mathbf{U}_P \left[\tilde{\mathsf{H}}^2_k(k)\right]^\ast \mathbf{U}_P^{-1}\\
    &=
    -\tilde{\mathsf{H}}^2_k(k).
\end{align}
It leads to the fact that for a given eigenvector $\mathbf{X}_j(k)$ of $\tilde{\mathsf{H}}^2_k(k)$ with eigenvalue $\tilde{\omega}_j^2(k)$, there is another eigenvector $\mathbf{X}_l(k)=\mathbf{P}\mathbf{X}_j(k)$ with opposite eigenvalue $\tilde{\omega}_l^2(k)=-\tilde{\omega}_j^2(k)$. Furthermore, the system matrix has the attribute,
$\tilde{\mathsf{H}}^2_k(-k) = \left[\tilde{\mathsf{H}}^2_k(k)\right]^\ast$ (see Eq.~\ref{eq:hk2t}), thus
\begin{align}
    \mathbf{U}_P \tilde{\mathsf{H}}^2_k(-k) \mathbf{U}_P^{-1} &=
    -\tilde{\mathsf{H}}^2_k(k),\\
    \mathrm{or}\;\;\;
    \mathbf{U}_P^{-1} \tilde{\mathsf{H}}^2_k(k) \mathbf{U}_P &=
    -\tilde{\mathsf{H}}^2_k(-k).
\end{align}
Therefore, given an eigenvector $\mathbf{X}_j(k)$ of $\tilde{\mathsf{H}}^2_k(k)$ with eigenvalue $\tilde{\omega}_j^2(k)$, there is an eigenvector $\mathbf{X}_l(-k)=\mathbf{U}_P^{-1}\mathbf{X}_j(k)$ of $\tilde{\mathsf{H}}^2_k(-k)$ with eigenvalue $\tilde{\omega}_l^2(-k)=-\tilde{\omega}_j^2(k)$.
Fig.~\ref{fig:spctphs} shows the discussed correspondences in the $\omega^2$-band structure (of a chain with $r=0.1$, $\epsilon=0.2$). Note that $-k$ is equivalent to $2\pi-k$, hence symmetry about $k=0$ also is identical to symmetry about $k=\pi$.

\phantomsection \subsubsection{Topological invariant, Pfaffian approach}
The Pfaffian is an invariant quantity of an even-dimensional antisymmetric matrix, and can be used in the calculation of the topological invariant of the present system. Another approach to derive a topological invariant can be based on first-principle Berry connection integral,
which is known as the Zak phase for 1D systems, and can be interpreted as the winding number in the parametric space,
as shown in Appendix~\ref{sec:topoint}. Both approaches return equivalent results.

\textcite{kitaev2001unpaired} showed that the $\mathbb{Z}_2$ topological invariant $\nu\; (=0\mathrm{\;or\;}1)$ of a 1D topological superconductor system can be defined as the product of the sign of the Pfaffian of the Hamiltonian matrix at the two high-symmetry points $k=0,\pi$,
\begin{equation}
    (-1)^\nu = \mathrm{sign} \left[\mathrm{Pf}(\mathbf{A}\rvert_{k=0}) \right] \; \mathrm{sign} \left[\mathrm{Pf}(\mathbf{A}\rvert_{k=\pi}) \right],
\end{equation}
As in the Kitaev chain, the band gap can only close and reopen (when the band topology may change) at these two points as parameters evolve.

\begin{figure*}[!htb]
\centering
\includegraphics[width=.75\textwidth]{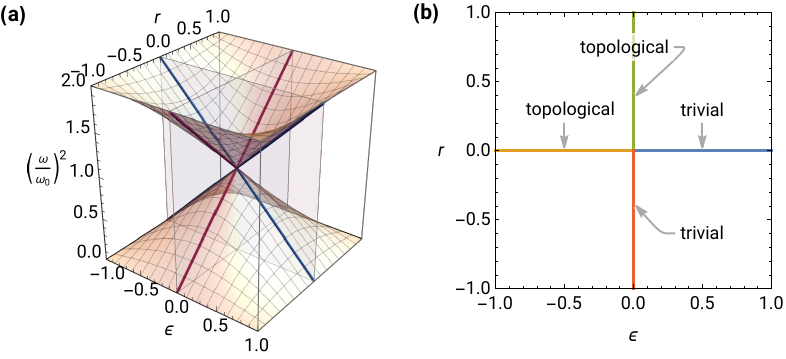}
\caption{\label{fig:topoclass}
(a) $\omega^2$-spectrum at the high-symmetry point $ka=\pi$ in the $(r,\epsilon)$ parametric space.
Section curves on the high-symmetry planes $r=0$ and $\epsilon=0$ are emphasized with blue and magenta curves, respectively.
(b) Topological phases defined by the Pfaffian of system matrices under
enforced symmetry condition $r\epsilon=0$.
}
\end{figure*}

In reference to the diatomic chain considered in this study, the band gap only closes at $k=\pi$ when both $r=\epsilon=0$ (condition at which the diatomic chain degenerates into a monoatomic chain); therefore we will only focus on $k=\pi$.
Substituting $k=\pi$ into Eq.~\ref{eq:hk2t}, the system matrix can be greatly simplified as
\begin{equation}\label{eq:hkpi}
    \tilde{\mathsf{H}}_k^2(\pi)\equiv\tilde{\mathsf{H}}_\pi^2=\left(
    \begin{array}{cc|cc}
         r & -\epsilon & 0 & r\epsilon \\
        -\epsilon & -r & -r\epsilon & 0\\
        \hline
         0 & r\epsilon & r & -\epsilon \\
        -r\epsilon & 0 & -\epsilon & -r
    \end{array}
    \right).
\end{equation}
Similarly to what shown for the real-space system matrix, the off-diagonal blocks contain nonlinear terms of $r\epsilon$. These terms can be eliminated only if $r=0$ or $\epsilon=0$, which is when the diatomic chain acquires inversion symmetry. We note that such inversion symmetry constraint enriches the topology of the diatomic chain system, as will be elaborated later. 
Fig.~\ref{fig:topoclass} (a) plots $\omega^2$ at $k=\pi$ in the $(r,\epsilon)$-parametric space.
As shown, the band gap closes only at an isolated point $(r,\epsilon)=(0,0)$. Given that topological transition only takes place when the band gap closes and reopens, and the entire gapped domain is connected,
it seems that the system with all possible configurations should stay in the same topological phase.
However, if we exclude $r$ (or $\epsilon$) in the parametric space and force it to vanish, so that the only degree of freedom is $\epsilon$ ($r$), then as $\epsilon$ ($r$) continuously evolves from $\epsilon_0$ ($r_0$) to $-\epsilon_0$ ($-r_0$), the band gap must close and reopen once, indicating a possible topological transition at $\epsilon=0$ ($r=0$).
In Fig.~\ref{fig:topoclass} (a), two sectional planes of $r=0$ and $\epsilon=0$ are shown. The eigenvalues under such constraints are simple linear relations, $\tilde{\omega}^2=\pm\epsilon$ and $\tilde{\omega}^2=\pm r$, as depicted in magenta and blue lines, respectively.

\begin{figure}[!htb]
    \centering
\includegraphics[width=\columnwidth]{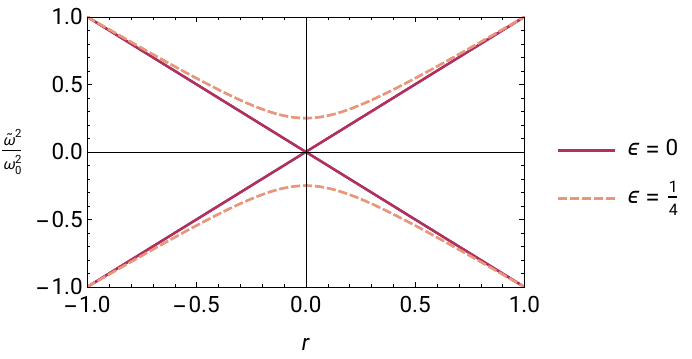}
    \caption{{
    The $\tilde\omega$ spectra at $ka=\pi$ on sectional planes $\epsilon=0$ (magenta) and $\epsilon=1/4$ (light red), with varying $r$. The former exhibits a crossing at $r=0$ as a result of synthetic PHS. The latter shows a repulsion between the two bands, indicating the absence of synthetic PHS, due to broken inversion symmetry, and no topological phases can be defined thereon.
    }}
    \label{fig:repul}
\end{figure}

\label{ParaB4}
{
\ed{
Fig.~\ref{fig:repul} provides a comparison of the spectra at $ka=\pi$ on two different sectional planes, namely $\epsilon=0$ (shown in magenta) and $\epsilon=1/4$ (shown in dashed light red), with varying $r$. The former curve exhibits a crossing at $r=0$, which is a typical signature of a system possessing particle-hole symmetry. Conversely, the latter curve demonstrates a repulsion between the two bands, despite being symmetric with respect to $\tilde{\omega}^2=0$. This repulsion indicates the absence of synthetic PHS, resulting from the lack of inversion symmetry.
}}

To identify the topological phases \ed{of chains possessing synthetic PHS}, we can evaluate the Pfaffian of the system matrix. When $r=0$ or $\epsilon=0$, the system matrix becomes
\begin{subequations}\label{eq:Phaf1}
\begin{align}
    \tilde{\mathsf{H}}_\pi^2\rvert_{\epsilon=0} &=
    \left(
    \begin{array}{cc|cc}
         r & 0 & 0 & 0 \\
        0 & -r & 0 & 0\\
        \hline
         0 & 0 & r & 0 \\
        0 & 0 & 0 & -r
    \end{array}
    \right)
    = \mathbf{1} \otimes \left(r \sigma_3\right),\\
    \tilde{\mathsf{H}}_\pi^2\rvert_{r=0} &=
    \left(
    \begin{array}{cc|cc}
        0 & -\epsilon & 0 & 0 \\
        -\epsilon & 0 & 0 & 0\\
        \hline
        0 & 0 & 0 & -\epsilon \\
        0 & 0 & -\epsilon & 0
    \end{array}
    \right)
    = \mathbf{1} \otimes \left(-\epsilon \sigma_1\right).
\end{align}
\end{subequations}
Each of the two matrices contains two identical $2\times2$ blocks representing two sets of decoupled equations. This corresponds to the two-fold degeneracy in the $\omega^2$-spectrum. To evaluate the Pfaffian, we need to first make the matrices antisymmetric via a change of basis. Given the matrix
\begin{equation}
    \mathbf{Q}=\frac{1}{\sqrt{2}}
    \begin{pmatrix}
    1 & 1\\
    -\ii & \ii
    \end{pmatrix},
\end{equation}
$\mathbf{Q}$ rotates the three Pauli matrices, $\mathbf{Q}\sigma_{1,2,3}\mathbf{Q}^\dagger=\sigma_{3,1,2}$ or $\mathbf{Q}^\dagger\sigma_{1,2,3}\mathbf{Q}=\sigma_{2,3,1}$. Given $\sigma_2$ is antisymmetric, we transform the matrices so that they align with $\sigma_2$, then the Pfaffian can be evaluated,
\begin{subequations}\label{eq:Phaf2}
\begin{align}
    \mathbf{Q}^\dagger\left(\epsilon\sigma_1\right)\mathbf{Q}=\epsilon\sigma_2=\ii
    \begin{pmatrix}
    0 & -\epsilon\\
    \epsilon & 0
    \end{pmatrix},\;
    & \mathrm{Pf} \left(\ii \epsilon\sigma_2\right) = \epsilon,\\
    \mathbf{Q}\left(r\sigma_3\right)\mathbf{Q}^\dagger=-r\sigma_2=\ii
    \begin{pmatrix}
    0 & r\\
    -r & 0
    \end{pmatrix},\;
    & \mathrm{Pf}\left(\ii (-r)\sigma_2\right) = -r.
\end{align}
\end{subequations}
\label{ParaB2}
{
It turns out, under the constraint $r\equiv0$, that the chain is topological when $\epsilon<0$ as $\mathrm{Pf}\left(\epsilon\sigma_1\right)<0$ and $\nu=1$, and it becomes trivial when $\epsilon>0$.
On the other hand, for the case of $\epsilon\equiv0$, the chain is topological when $r>0$ as $\mathrm{Pf}\left(r\sigma_3\right)<0$ and $\nu=1$, and it becomes trivial when $r<0$, as shown in Fig.~\ref{fig:topoclass}~(b).

\ed{This again reveals the duality between the variations $\epsilon$ and $r$ (or, equivalently, between the spring and the mass variations). As we have seen from Fig.~\ref{fig:di_all3D},
the chain with even (odd) number of particles (springs) and spring terminals shows bound states for $\epsilon<0$, and the chain with even (odd) number of springs (particles) and free terminals shows bound states for $r>0$. Similar dual relationships are also found in other chain configurations in Fig.~\ref{fig:di_all3D}.
}
}

\begin{figure}[!htb]
    \centering
\includegraphics[width=3in]{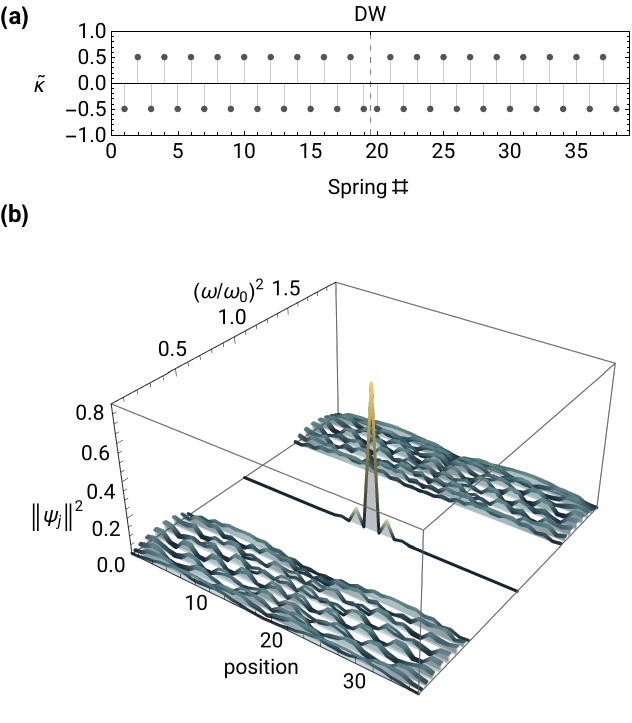}
    \caption{
    (a) Spring constants of the assembled chain connecting two topologically distinct chains with $r=0$ and $\epsilon=\pm0.5$. The domain wall is identified by an anomaly in the alternating spring constant pattern.
    (b) Mode shapes of the assembled chain. One mid-gap bound state appears at the position of the domain wall.
    }
    \label{fig:1DDW}
\end{figure}

These results match the existence of the topological zero modes in the numerical results shown in Fig.~\ref{fig:di_off_r} and Fig.~\ref{fig:di_ess_e}. However, we should note that $\epsilon>0$ and $\epsilon<0$ (and similarly $r>0$ and $r<0$) describe the same bulk periodic chain, only with their choices of unit cells displaced by half a lattice constant. Similar to charge polarization in a lattice, the value $\nu$ of the diatomic chain is not well defined, as it depends on the choice of the unit cell, but the change in $\nu$ is. Under a fixed reference frame (choice of unit cell), tuning the lattice from $\epsilon=-\epsilon_0$ to $+\epsilon_0$ must change the topological invariant from $\nu$ to $\nu'=(\nu + 1)\! \mod{2}$. The assembled chain with two chains having distinct $\nu$ values will then have a ``skipping'' in the pattern and a localized topological state at the connection. Fig.~\ref{fig:1DDW} shows the zero bound state located at the dislocation interface.
On the other hand, if we always take the first two particles from the left as the reference unit cell, adding a particle from the left also switches $\nu$, which creates or annihilate the Majorana zero mode; this latter comment is also well aligned with previous arguments based on paired or isolated Majoranas at the terminal.

\section{Conclusions}
\label{sec:kitaevconcl}

In this study, we investigated a simple yet topologically nontrivial classical mechanical system, that is a 1D spring-mass chain. Using a combination of theoretical and numerical methodologies, we showed that in certain ranges of the parameters the system resembles the
electronic Kitaev chain, hence giving rise to topological states that are reminiscent of Majorana bound states.

We developed a mathematical description applicable to classical mechanical systems that closely resembles the second quantization formalism in quantum mechanics and that efficiently handles many-body problems.
\label{ParaA3}
{\ed{
The second quantization formalism is extensively used in solid state physics to describe the behavior of electrons in crystals. Despite its different physical foundation, the proposed classical description shares a highly correlated mathematical structure that allows a deeper understanding of the classical system and of the parameters that contribute to its global behavior. The proposed method does not apply only to spring-mass systems but also to continuous systems such as networks of coupled acoustic resonators or elastic phononic lattices. In addition to providing an alternative tool for analyzing complex classical systems, this approach also offers a new way to synthesize and correlate nontrivial classical systems to their quantum mechanical counterparts.
}}
Under such description, the Hamiltonian function of the 1D mechanical chain is shown to take a form analogous to the Hamiltonian operator of the 1D electronic superconducting Kitaev's model (rather than the semiconducting forms characteristic of either the SSH or the Rice-Mele models) due to the emergence of Cooper-pair-like terms. 

Notwithstanding, the quantum and classical systems are fundamentally different and follow distinct sets of governing equations, hence resulting in different system matrices.
Although the classical system does not possess a BdG-Hamiltonian-like system matrix,
physics analogous to those of systems with BdG-Hamiltonians can still be replicated by introducing additional symmetries.
We show that when the mechanical chain is dimerized (with alternating mass and spring constants within the equivalent particle system), the synthetic particle-hole symmetry (a key element for the emergence of Majorana-like zero modes) is reformulated by involving the inversion and chiral (sublattice) symmetry of the chain, and the squared eigenfrequency spectrum exhibits symmetry about a reference frequency level.

It was shown that, within certain ranges of parameters and symmetry constraints, zero-frequency (about the reference frequency level) bound states appear at either the terminals of finite dimerized chains, or at the domain walls connecting distinct chains.
\label{ParaB6}
{
\ed{The emergence of the local bound states can be interpreted by unpaired terms in the Hamiltonian function, which well aligns with the Kitaev's model of Majorana zero modes.
We also note that, although prior research based on a conventional dynamical matrix approach has commonly referred to the classical dimerized chain as a system analog to the SSH chain, our results are not in conflict with them. Indeed, our analysis provides a more in-depth characterization of the system that uncovers additional features of this classical mechanical analog system.}}

To further confirm the topological origin of these localized bound states, we carried out a topological band structure analysis by transforming the system matrix into $k$-space. The sign change in the Pfaffian of the system matrix at $k=\pi$, as the band gap closes and reopens, indicated the topological phase transition. On the other hand, the integral representation, namely, the Zak phase and the winding number approaches also confirmed the same transition. The topological invariant $\nu=0,1$ identified the distinct topological phases for the dimerized chains consistently with the topological protection of the bound states observed in real space.


\ed{
The classical mechanical chain considered in this study belongs to the BDI symmetry class \cite{chiu2016classification}, and the Majorana-like bound states do not possess anyonic characteristics, which implies that non-Abelian braiding of these bound states may not be possible.
}
\label{ParaC5}
{
\ed{
Nonetheless, the 0D topological bound states at the terminals of our topologically nontrivial chain do serve as robust locally resonant states. These states effectively trap mechanical energy at selected spatial locations and at prescribed frequencies and could prove effective for those applications relying on energy extraction, such has vibration control or energy harvesting. Indeed the ability to avoid back-scattering at the energy extraction location could result in almost ideal levels (i.e. close to $100\%$) of energy extraction.
These states can also be replicated in micro-fabricated quasi-1D piezoelectric surface acoustic wave devices or stacked film bulk acoustic wave devices, hence making them suitable for building acoustic filters for telecommunication.
The resonant frequency of the topological bound state is determined by the unpaired terminal oscillators, and can be easily adjusted by altering the effective mass and stiffness. It is also possible to envision that, by measuring frequency changes, these systems could be utilized in sensing applications to identify and locate fluctuations in the surrounding environment, such as temperature, pressure, or mass loading.
}}

\begin{acknowledgments}
The authors gratefully acknowledge the financial support of the Office of Naval Research under the project N00014-20-1-2608.
\end{acknowledgments}

\appendix
\section{Proof of canonical second quantized coordinates}\label{sec:pf}
This section shows that the new set of coordinates  according to Eq.~\ref{eq:ajbj},
\begin{equation*}
    \mathbf{A}'\equiv \left(
         a^+_1,b^+_1, \dots , a^+_N,b^+_N;
         a^-_1,b^-_1, \dots , a^-_N,b^-_N
     \right)^\intercal
\end{equation*}
    is canonical.
    For simplicity, we use this notation before introducing the sublattice degree of freedom,
\begin{equation*}
    \mathbf{A}\equiv \left(
         a^+_1, \dots , a^+_{2N};
         a^-_1, \dots , a^-_{2N}
     \right)^\intercal
\end{equation*}
It results only in a different nomenclature and will not affect the results,
\begin{equation}
    \begin{pmatrix}
    a_1^+\\a_2^+\\a_3^+\\ \vdots\\
    \hline
    a_1^-\\a_2^-\\a_3^-\\  \vdots
    \end{pmatrix}
    \rightarrow
    \begin{pmatrix}
    a_1^+\\b_1^+\\a_2^+\\ \vdots\\
    \hline
    a_1^-\\b_1^-\\a_2^-\\ \vdots
    \end{pmatrix}.
\end{equation}

    The coordinates $\mathbf{A}$ are canonical if and only if the Hamilton's equations remain the same in $\mathbf{A}$, as they do in $\mathbf{X}$. That is,
    \begin{equation}
        \dot{\mathbf{A}} = \mathbf{E}\frac{\partial H}{\partial \mathbf{A}}.
    \end{equation}
    Substitute $\mathbf{A}=\mathbf{J}\mathbf{X}$, where $J_{ij}=\partial a_j/\partial X_j$ is the Jacobian matrix, then
    \begin{equation}
        \dot{\mathbf{A}}=\frac{\dd}{\dd t}
        \left(\mathbf{J}\mathbf{X}\right) =
        \mathbf{J}\dot{\mathbf{X}}=
        \mathbf{J}\mathbf{E}\frac{\partial H}{\partial \mathbf{X}}=
        \mathbf{J}\mathbf{E}\mathbf{J}^\intercal \frac{\partial H}{\partial \mathbf{A}}.
    \end{equation}
    Comparing the right-hand sides of the above two equations, the Hamilton's equations hold if and only if $\mathbf{J}\mathbf{E}\mathbf{J}^\intercal = \mathbf{E}$. Based on Eq.~\ref{eq:ajbj}, the $4N\times 4N$ Jacobian matrix $\mathbf{J}$ reads
    \begin{equation}
        \mathbf{J}=
        \frac{1}{\sqrt{\ii \omega_0}}
        \frac{1}{\sqrt{2}}
        \left(
        \begin{array}{ccc|ccc}
        \diagdown & & & \diagdown & & \\
         & m_0^{-\frac{1}{2}} & & & -\ii \sqrt{2\kappa_0} &\\
         &  & \diagdown & &  & \diagdown\\
         \hline
         \diagdown & & & \diagdown & & \\
         & m_0^{-\frac{1}{2}} & & & \ii \sqrt{2\kappa_0}&\\
         &  & \diagdown & &  & \diagdown
        \end{array}\right),
    \end{equation}
    and
    \begin{align}
    \begin{split}
        \mathbf{J}\mathbf{E}\mathbf{J}^\intercal &=
        \frac{1}{2\ii\omega_0}
        \left(
        \begin{array}{ccc|ccc}
         & & & \diagdown & & \\
         & \mathbf{0} & & & -2\ii\left(\frac{2\kappa_0}{m_0}\right)^\frac{1}{2} &\\
         &  &  & &  & \diagdown\\
         \hline
         \diagdown & & &  & & \\
         & 2\ii\left(\frac{2\kappa_0}{m_0}\right)^\frac{1}{2} & & & \mathbf{0} &\\
          &  & \diagdown & &  & 
        \end{array}\right)
        \\
        &= \begin{pmatrix}
        \mathbf{0} & -\mathbf{1}\\
        \mathbf{1} & \mathbf{0}
        \end{pmatrix} =\mathbf{E}.
    \end{split}
    \end{align}
So the coordinate system $\mathbf{A}$
    is canonical and the Hamilton's equations hold, and the scaling factor $1/\sqrt{\ii \omega_0}$ is necessary.

\section{Properties of classical analogue second quantized variables}\label{sec:pr}
This section summarizes the properties of both the quantum and the proposed classical second quantized variables.

Let $\hat{b}$ and $\hat{c}$ be the bosonic and fermionic annihilation operators,
    \begin{subequations}
    \begin{align}
        \left[
        \hat{b}_j,\hat{b}_l
        \right]&=
        \left[
        \hat{b}_j^\dagger,\hat{b}_l^\dagger\right] = 0,\\
        \left[
        \hat{b}_j,\hat{b}_l^\dagger\right] &= \delta_{jl},\\
        \left\{
        \hat{c}_j,\hat{c}_l
        \right\}&=
        \left\{
        \hat{c}_j^\dagger,\hat{c}_l^\dagger\right\} = 0,\\
        \left\{
        \hat{c}_j,\hat{c}_l^\dagger\right\} &= \delta_{jl}.
    \end{align}
    \end{subequations}
    where
    \begin{subequations}
    \begin{align}
        \left[
        \hat{\alpha},\hat{\beta}
        \right]&\equiv \hat{\alpha}\hat{\beta}-\hat{\beta}\hat{\alpha}\;\;\mathrm{(commutator)},\\
        \left\{
        \hat{\alpha},\hat{\beta}
        \right\}&\equiv \hat{\alpha}\hat{\beta}+\hat{\beta}\hat{\alpha}\;\;\mathrm{(anticommutator)},
    \end{align}
    \end{subequations}

On the other hand, the variables of the classical mechanical chain, $a_j^\pm$ ($b_j^\pm$), are complex scalars rather than operators, and they always commute in multiplication,
\begin{subequations}\label{eq:comm}
\begin{align}
    [a_j^+,a_l^-]
    &\equiv a_j^+a_l^- - a_l^-a_j^+
    =0,\label{eq:comma}\\
    [a_j^+,a_l^+]
    &= [a_j^-,a_l^-]
    =0,\label{eq:commb}
\end{align}
\end{subequations}
The commensurate relations reminiscent of those fermionic anticommutation relations are the fundamental Poisson brackets applied to the coordinate variables themselves. Recall the definition of the Poisson brackets of two functions $F$ and $G$ of the canonical coordinates $(\mathbf{a}^+;\mathbf{a}^-)$,
\begin{equation}
\{F,G\}_\mathrm{Poisson}\equiv\sum_j{\left(
    \frac{\partial F}{\partial a_j^-}
    \frac{\partial G}{\partial a_j^+}-
    \frac{\partial F}{\partial a_j^+}
    \frac{\partial G}{\partial a_j^-}
\right)},
\end{equation}
then we have
\begin{subequations}\label{eq:fpb}
\begin{align}
    \{a_j^+,a_l^-\}_\mathrm{Poisson} &=
    -\{a_l^-,a_j^+\}_\mathrm{Poisson} = \delta_{jl},\\
    \{a_j^+,a_l^+\}_\mathrm{Poisson} &=
    \{a_j^-,a_l^-\}_\mathrm{Poisson} = 0.
\end{align}
\end{subequations}

\section{Dispersion of the 1D dimerized mechanical lattice}\label{sec:disperion}

The $\omega$-$k$ dispersion of the dimerized chain can be obtained by either following the classical dynamical matrix approach or the proposed second quantized formalism. Here, we calculate the dispersion using the former approach, which is identical to the results shown in Sec.~\ref{sec:kspace}.

Contemporary textbooks on solid state physics \cite{kittel1996introduction,simon2013oxford} usually employ a diatomic mechanical lattice as the first toy model to introduce the concepts of band gap and reciprocal space. These diatomic lattices can either exhibit alternating (different) masses for two particles \cite{kittel1996introduction} or alternating spring constants \cite{simon2013oxford}, but not both simultaneously. Here, we demonstrate a diatomic lattice with particle masses ($m_{1,2}$) and spring constants ($\kappa_{1,2}$),
both allowed to
change and repeat every other unit
as shown in Fig.~\ref{fig:di}. 

Given two particles in a unit cell, let the displacement of the first and second particles of the $j$-th cell be $x_j$ and $y_j$, respectively. The equations of motion are obtained as,
\begin{equation} \label{eq:di}
    \begin{cases}
        m_1 \ddot{x}_j=\kappa_1(y_j-x_j)+\kappa_2(v_{n-1}-x_j)\\
        m_2 \ddot{y}_j=\kappa_1(x_j-y_j)+\kappa_2(u_{n+1}-y_j)
    \end{cases},
    \text{\;$j \in \mathbb{Z}$}.
\end{equation}
We let $a$ be the lattice constant, and substitute the following ansatz into Eq.~(\ref{eq:di}),
\begin{equation} \label{eq:ansdi}
    \begin{cases}
        x_j = A \ee^{\ii (\omega t - k j a)}\\
        y_j = B \ee^{\ii (\omega t - k j a)}
    \end{cases}
    , \; A,B \in \mathbb{C}.
\end{equation}
Upon simplification, $y_{j-1}$ and $x_{j+1}$ are eliminated, and we obtain two equations with variables $x_j$ and $y_j$ only. Written in matrix form, we get
\begin{multline} \label{eq:dimatrix}
    \begin{pmatrix}
        \kappa_1 +\kappa_2 & -\kappa_1-\kappa_2 \ee^{\ii k a} \\
        -\kappa_1 -\kappa_2 \ee^{-\ii k a} & \kappa_1 +\kappa_2 \\
    \end{pmatrix}
    \begin{pmatrix}
        A\\
        B\\
    \end{pmatrix}
    \\=
    \omega^2
    \begin{pmatrix}
        m_1 & 0\\
        0 & m_2\\
    \end{pmatrix}
    \begin{pmatrix}
        A\\
        B\\
    \end{pmatrix}
    ,
\end{multline}
which forms a generalized eigenvalue problem $\left(\mathbf{K} -\omega^2 \mathbf{M}\right) \mathbf{u}  = \mathbf{0}$.
Solving the characteristic equation yields the eigenfrequencies $\omega$ in terms of a given wavenumber $k$,
\begin{widetext}
\begin{subequations}
\begin{align}\label{eq:k2w}
\omega &=
\sqrt{\frac{(m_1+m_2) (\kappa_1 +\kappa_2)\pm\sqrt{8 \kappa_1 
   \kappa_2 m_1 m_2 (\cos{ 
   ka}-1)+(\kappa_1 +\kappa_2)^2
   (m_1+m_2)^2}}{2m_1 m_2}}\\
   &=
   \omega_0\sqrt{1\pm \sqrt{1+(1-r^2)(1-\epsilon^2)(\cos{ka}-1)/2}},
\end{align}
\end{subequations}
\end{widetext}
where Eqs.~(\ref{eq:mk12},\ref{eq:mk0}, \ref{eq:o0}) were used to obtain the second equality.

\section{Numerical examples of spectra and mode shapes of finite dimerized chains}\label{sec:examples}
\subsection{Duality between spring and mass dimerization}
\begin{figure*}[!htb]
\centering
\includegraphics[width=.9\textwidth]{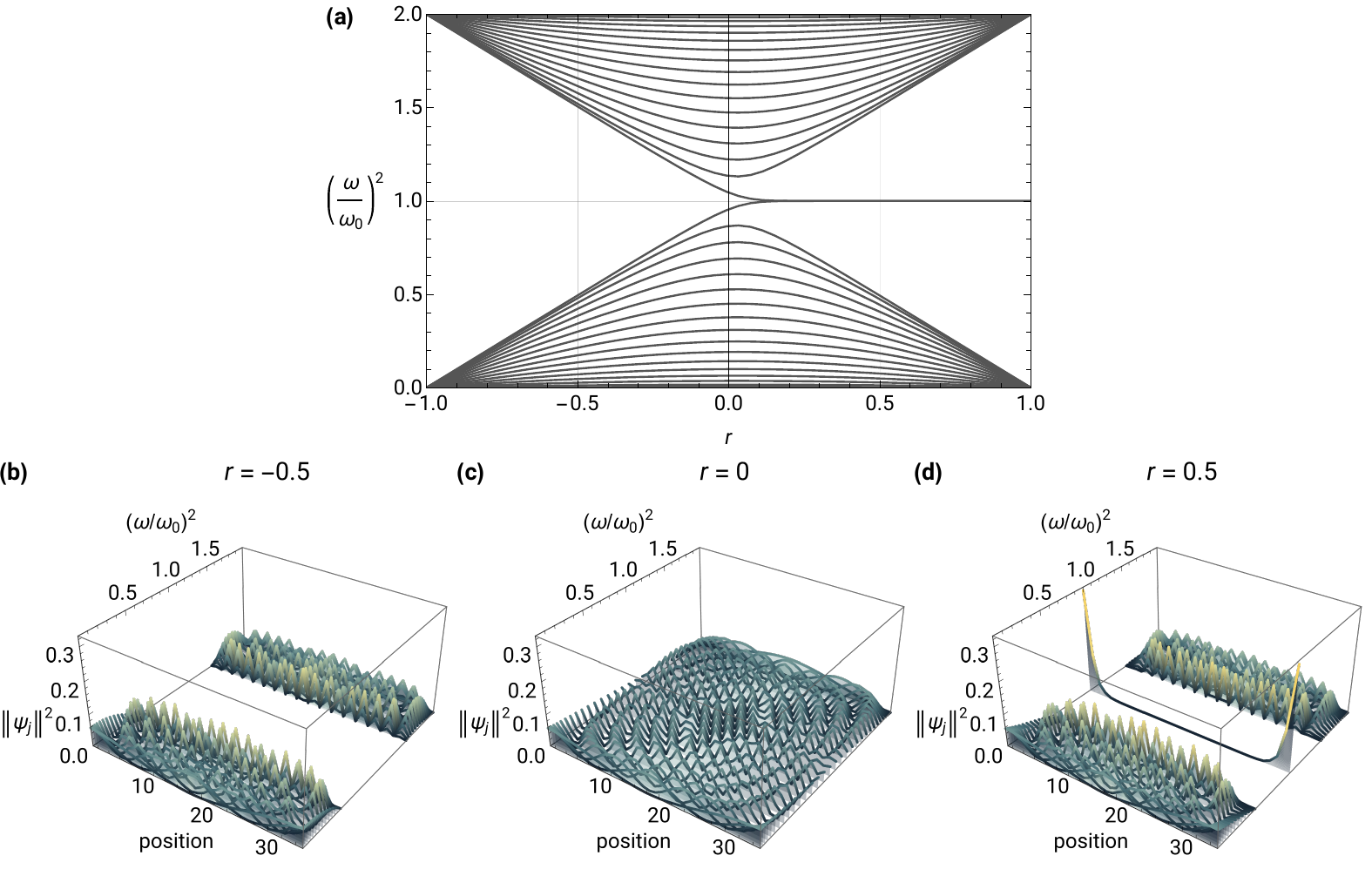}
\caption{\label{fig:di_off_r}
(a) Spectrum and (b-d) mode shapes of the diatomic chain with 35 (odd) particles (17.5 cells), both ends free, $\epsilon=0$, and varying $r$ values.
When $r>0$, Majorana-like bound states with $\omega=\omega_0$ appear at the open ends of the mechanical chain. Symbolic representations of the chains for $r<0$:
$\bullet-\circ-\cdots-\circ-\bullet$,
and for $r>0$:
$\circ-\bullet-\cdots-\bullet-\circ$.
}
\end{figure*}

Fig.~\ref{fig:di_off_r} shows the spectrum and the mode shapes (i.e. the eigenvalues and the eigenstates) of the diatomic chain with odd number of particles (assumed as 35, which correspond to $N=17.5$ unit cells) and free ends. $\epsilon$ is fixed at 0, with $r$ varying from $-1$ to $1$. When $r>0$, topological zero bound states appear at both ends of the chain. Together with Fig.~\ref{fig:di_ess_e}, they show the duality between staggered spring constants and particle masses.

\subsection{Single Majorana-like zero modes in the diatomic chain}
\begin{figure*}[!htb]
\centering
\includegraphics[width=.9\textwidth]{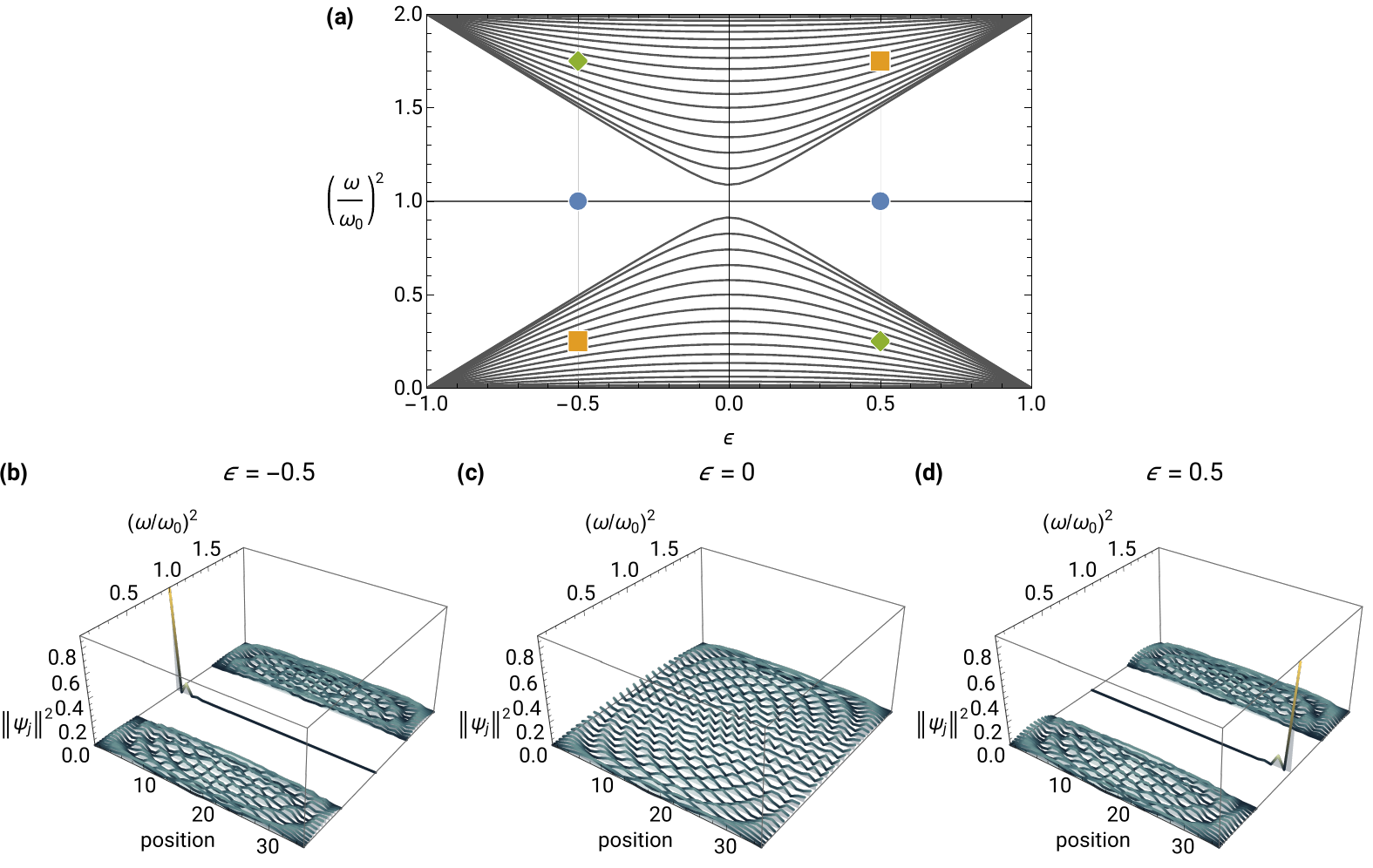}
\caption{\label{fig:di_oss_e}
(a) Spectrum and (b-d) mode shapes of the diatomic chain with 35 (odd) particles (17.5 cells), spring-terminated ends, $r=0$, and varying $\epsilon$ values.
Symbolic representations of the chains for $\epsilon<0$:
$\lvert-\bullet=\bullet\cdots\bullet-\bullet=\rvert$,
and $\epsilon>0$:
$\lvert=\bullet-\bullet\cdots\bullet=\bullet-\rvert$.
Pairs of identical markers in (a) label the counterparts of $\mathbf{P}_\mathrm{odd}$ operation.  
}
\end{figure*}

It is possible to construct a mechanical chain that supports a single Majorana-like zero mode at only one of its terminals.
First, we let either $r=0$ or $\epsilon=0$ and obtain a bulk lattice with inversion symmetry, having the bulk spectrum in agreement with the Kitaev chain.
A finite chain can lose inversion symmetry due to boundary conditions. For example, a chain with an odd number of particles, spring-spring terminals and $r=0$, reading $\lvert-\bullet=\bullet-\bullet=\rvert$, does not have inversion symmetry (the two ends appear to be distinct viewed from each side) despite its periodic extension does.
Fig.~\ref{fig:di_oss_e} plots the $\omega^2$ spectrum and mode shapes with varying $\epsilon$.
For positive and negative $\epsilon$ values, one zero mode appears at either the left or the right ends, respectively. It has a constant zero frequency ($\tilde{\omega}^2=0$) with respect to the reference level, and a symmetric spectrum. Given that there is an odd number of branches in the spectrum, one would then ask: what is the synthetic-particle-hole-exchanged state for this unpaired zero mode? The answer can be found in Eq.~\ref{eq:phsodd}, that shows how for such odd, spring-spring chain, the synthetic PH counterparts of the eigenmodes exist in a different chain with opposite dimerization parameters; this latter chain is the space-inverted image of the original chain. In the spectrum shown in Fig.~\ref{fig:di_oss_e}, they correspond to modes in the opposite quadrant (with the opposite signs of $\epsilon$ and $\tilde{\omega}^2$) of the spectrum. Three pairs of markers in Fig.~\ref{fig:di_oss_e} label the synthetic PH pairs in the spectrum, for the zero modes and the bulk modes.

%


\section{Fourier transform of the classical second quantized variables}\label{sec:FT}
Consider the Fourier transform relations,
\begin{equation}\label{eq:ftaj2ak_ap}
    \begin{cases}
			\;a_k^\pm &= \displaystyle\frac{1}{\sqrt{N}} \sum_{j=1}^N \ee^{-\ii k j} a_j^\pm,\\
            \;a_j^\pm &= \displaystyle\frac{1}{\sqrt{N}} \sum_{q=1}^N \ee^{+\ii k j} a_k^\pm,
	\end{cases}\;\;
	\begin{cases}
			\;b_k^\pm &= \displaystyle\frac{1}{\sqrt{N}} \sum_{j=1}^N \ee^{-\ii k j} b_j^\pm,\\
            \;b_j^\pm &= \displaystyle\frac{1}{\sqrt{N}} \sum_{q=1}^N \ee^{+\ii k j} b_k^\pm,
	\end{cases}
\end{equation}
where
\begin{equation}\label{eq:kk}
    k=2\pi \frac{l}{N},\; l=1,2,\dots,N,
\end{equation}
are the $N$ discrete wavenumbers allowed in a finite chain with $N$ concatenated cells. The resolution in $k$-space increases as the chain gets longer, as $\Delta k = 2\pi/N$. The maximum allowed $k=\pi/a$ is limited by the lattice constant $a$ (spatial sampling period). Here, we use a discrete formulation with $j$ being the cell number, hence it is equivalent to $a=1$, and $ka=k\in (0,2\pi]$. This is in agreement with the fact that $k$ (or $ka$) acts as the phase difference across a unit cell, i.e., $k$ shows in the imaginary exponent $\ee^{\ii k \times j}$, where $j$ is an integer. Hence, replacing $k$ by $k\pm2n\pi$ makes no difference. It is a common convention to move the subdomain $k\in (\pi, 2\pi]$ to $k\in (-\pi, 0]$ and make the $k$-space symmetric, $k\in (-\pi,\pi]$, also known as the first Brillouin zone of a 1D lattice. This is equivalent to take, in Eq.~(\ref{eq:kk}),
\begin{equation}
    \begin{cases}
        \displaystyle\;l = -\frac{N-2}{2},\dots,0,\dots,+\frac{N}{2}, & N \in \mathrm{even\,integers},\\[6pt]
        \displaystyle\;l = -\frac{N-1}{2},\dots,0,\dots,+\frac{N-1}{2}, & N \in \mathrm{\,odd\,integers}.
    \end{cases}
\end{equation}
Note that in a monoatomic (uniform) chain, the highest-spatial-frequency wave is limited by the maximum allowed wavenumber $\pi/a$ corresponding to a wavelength of twice the lattice constant (at least two particles are needed to show a non-constant waveform).
In a diatomic chain (or in chains with multiple-atom unit cells, or even in continuous periodic structures), there can be waves with higher spatial frequencies (wavelengths smaller than twice the lattice constant). Given that inside a unit cell the mechanical properties are not uniform (e.g., particle spacing from $A$ to $B$ and $B$ to the next $A$), it is not possible to precisely define the wavenumber above $\pi/a$. Instead, we adopt the Bloch wavenumber that always lies within $ka\in (-\pi,\pi]$, obtained by the phase shift $-\pi<\phi<\pi$ divided by $a$, measured from one point at $x$ to another at $x+a$. Those higher-frequency waves appearing in this range can be understood as aliased, or higher-order, modes of the same Bloch wavenumber.

Some caution should be taken with the notation. Recall that the real-space quantities $a_j^+$ and $a_j^-$ form a complex conjugate pair, $\left(a_j^+\right)^\ast=a_j^-$ and vice versa. However, this is not the case for the $k$-space counterparts, $\left(a_k^+\right)^\ast\neq a_k^-$.
Given the Fourier transform defined in Eqs.~(\ref{eq:ftaj2ak}), the wavenumber $k$ should also switch signs under complex conjugation,
\begin{subequations}
\begin{align}
    \left(a_k^\pm\right)^* = \frac{1}{\sqrt{N}} \sum_{j=1}^N e^{+i k j} a_j^\mp = a_{-k}^\mp,\\
    \left(b_k^\pm\right)^* = \frac{1}{\sqrt{N}} \sum_{j=1}^N e^{+i k j} b_j^\mp = b_{-k}^\mp,
\end{align}
\end{subequations}
We should not confuse our notation with slightly different conventions typically used in quantum mechanics. While they also start with the real-space Hermitian conjugate pairs $\hat{c}_j$ and $\hat{c}_j^\dagger$ in $k$-space, only $\hat{c}_k$ is obtained as the Fourier transform of $\hat{c}_j$, and $\hat{c}_k^\dagger$ is the Hermitian conjugate of $\hat{c}_k$.

The Fourier transform relations shown in Eqs.~\ref{eq:ftaj2ak} are linear transforms, so they can be expressed in matrix form as
$\mathbf{a}_k = \mathbf{J} \mathbf{a}$, with $\mathbf{J}=\partial\mathbf{a}_k/\partial\mathbf{a}$, or,
\begin{equation}
    \left(
    \begin{array}{c}
         a_{k_1}^+ \\
         b_{k_1}^+ \\
         \vdots \\
         \hline
         a_{k_1}^- \\
         b_{k_1}^- \\
         \vdots
    \end{array}
    \right)
    =
    \left(
    \begin{array}{c|c}
          & \\
         \mathrm{F.T.\; for\;} a_j^+,b_j^+ & \mathbf{0}\\
          & \\
         \hline
          & \\
         \mathbf{0} & \mathrm{F.T.\; for\;} a_j^-,b_j^- \\
          & \\
    \end{array}
    \right)
    \left(
    \begin{array}{c}
         a_{1}^+ \\
         b_{1}^+ \\
         \vdots \\
         \hline
         a_{1}^- \\
         b_{1}^- \\
         \vdots
    \end{array}
    \right).
\end{equation}
Apparently, the matrix $\mathbf{J}$ contains two identical blocks of $2N\times 2N$ matrices (let us call them $\mathbf{F}$), that can be written as $\mathbf{J}=\mathbf{1}\otimes\mathbf{F}$. Below we derive the explicit expression for $\mathbf{F}$.

The $k$-space coordinates in real space are sinusoidal functions for some wavenumber $k$. For brevity, define the corresponding column vector
\begin{equation}
    \ket{k} = \frac{1}{\sqrt{N}} \sum_{j=1}^N e^{i k j} \ket{j},
\end{equation}
or
$\ket{k} =\frac{1}{\sqrt{N}}\left(\ee^{\ii k\times 1}, \ee^{\ii k\times 2},\dots, {\ee^{\ii k N}}\right)^\intercal$.
Note that $k$ can take different values, $k_l=2\pi\times l/N$, so explicitly, 
$\ket{k_l} =\frac{1}{\sqrt{N}}\left(\ee^{\ii 2\pi l\times 1/N}, \ee^{\ii 2\pi l\times 2/N},\dots, \cancelto{1}{\ee^{\ii 2\pi l\times N/N}}\;\;\; \right)^\intercal$.

If we only needed to transform $a_j^+$ to $a_{k_l}^+$, the Fourier transform matrix will simply be given by stacking these column vectors from $\ket{k_1}$ to $\ket{k_N}$, i.e., $a_{k_l}^+=F'_{lj}a_j^+$, with $F'_{lj}=\frac{1}{\sqrt{N}}\ee^{\ii 2\pi l\times j/N}$, or
\begin{equation}
    \left(
    \begin{array}{c}
         a_{k_1}^+ \\
         \vdots \\ \vdots \\
         a_{k_N}^+ \\
    \end{array}
    \right)
    =\frac{1}{\sqrt{N}}
    \begin{pmatrix}
    \ee^{\ii2\pi\times 1/N} & \ee^{\ii2\pi 2\times 1/N} & \dots & 1\\
    \ee^{\ii2\pi\times 2/N} & \ee^{\ii2\pi 2\times 2/N} & \dots & 1\\
    \vdots & \vdots & \ddots & \vdots\\
    1 & 1 & \dots & 1\\
    \end{pmatrix}
    \left(
    \begin{array}{c}
         a_{1}^+ \\
         \vdots \\ \vdots \\
         a_{N}^+ \\
    \end{array}
    \right)
\end{equation}
However, our basis is composed of interlaced $a_j^+$ and $b_j^+$ quantities. So the transform matrix becomes $\mathbf{F}=\mathbf{F}'_{N\times N}\otimes\mathbf{1}$, having twice the dimensions and reading
\begin{equation}\small
    \frac{1}{\sqrt{N}}
    \begin{pmatrix}
    \ee^{\ii2\pi\times 1/N} & 0 & \ee^{\ii2\pi 2\times 1/N} & 0 & \dots & 1 & 0\\
    0 & \ee^{\ii2\pi\times 1/N} & 0 & \ee^{\ii2\pi 2\times 1/N} & \dots & 0 & 1\\
    \ee^{\ii2\pi\times 2/N} & 0 & \ee^{\ii2\pi 2\times 2/N} & 0 & \dots & 1 & 0\\
    0 & \ee^{\ii2\pi\times 2/N} & 0 & \ee^{\ii2\pi 2\times 2/N} & \dots & 0 & 1\\
    \vdots & \vdots & \vdots & \vdots & \ddots & \vdots & \vdots\\
    1 & 0 & 1 & 0 & \dots & 1 & 0\\
    0 & 1 & 0 & 1 & \dots & 0 & 1\\
    \end{pmatrix}
    .
    \normalsize
\end{equation}
We arrive at $\mathbf{J}=\mathbf{1}\otimes\mathbf{F}'\otimes\mathbf{1}$, with $F'_{lj}=\frac{1}{\sqrt{N}}\ee^{\ii 2\pi l\times j/N}$.
Given that $\mathbf{F}'$ is unitary ($\mathbf{F}'^\dagger\mathbf{F}'=\mathbf{1}_{N\times N}$, as a result of the orthogonality of the sinusoidal functions), $\mathbf{J}$ is also unitary, $\mathbf{J}^{-1}=\mathbf{J}^\dagger$ ($\mathbf{J}$ is not Hermitian).

The equations of motion in $k$-space can then be obtained as follows,
\begin{align}
    \begin{split}
        &\tilde{\mathsf{H}}^2 \mathbf{X} = \tilde{\omega}^2 \mathbf{X} \\ \Rightarrow
        &\mathbf{J}\tilde{\mathsf{H}}^2 \mathbf{X}= \tilde{\omega}^2 \mathbf{J}\mathbf{X}\\
        \Rightarrow
        & \mathbf{J}\tilde{\mathsf{H}}^2 \mathbf{J}^\dagger\mathbf{J} \mathbf{X}= \tilde{\omega}^2 \mathbf{J}\mathbf{X}\\
        \Rightarrow
        &\tilde{\mathsf{H}}^2_k  \mathbf{X}_k= \tilde{\omega}^2 \mathbf{X}_k,
    \end{split}
\end{align}
where $\tilde{\mathsf{H}}^2_k=\mathbf{J}\tilde{\mathsf{H}}^2 \mathbf{J}^\dagger$ is the $k$-space system matrix and $\mathbf{X}_k=\mathbf{J}\mathbf{X}$ is the $k$-space eigenvector, both under $(a^+_{k_1},b^+_{k_1},\dots;a^-_{k_1},b^-_{k_1},\dots)$-basis representation.

\begin{figure*}[!htb]
\centering
\includegraphics[width=\textwidth]{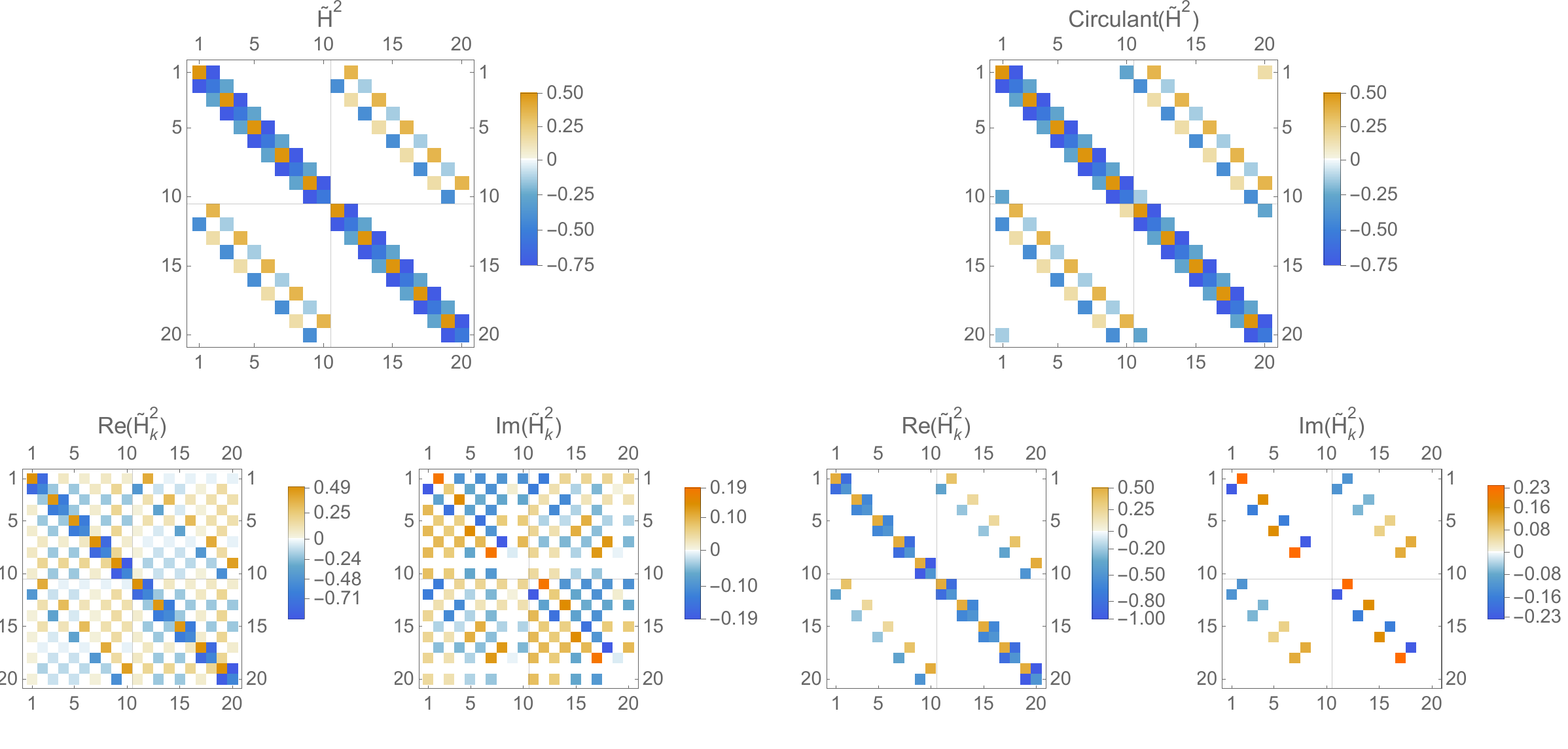}
\caption{\label{fig:circmat}
Left: system matrices of a finite diatomic chain in real space (top) and in $k$-space (bottom), which are not block-diagonalized.
Right: system matrices of a ring of diatomic chain in real space (top) and in $k$-space (bottom). The real-space matrix is circulant and the $k$-space matrix is block-diagonalized by discrete Fourier transform.
}
\end{figure*}

Getting the equations of motion in $k$-space is the first step. In order to obtain the band structure of a chain, that is, to find the eigenfrequency and eigenmode for a given $k$, we must ensure that any two equations involving different $k$'s are fully decoupled. In other words, each of the four $2N\times 2N$ blocks in matrix $\tilde{\mathsf{H}}^2_k$ should be $2\times2$-block-diagonalized.
However, given any finite chain with two terminals, this is not possible. In other terms, the eigenmodes in a finite chain are composed of mixed sinusoidal functions of wavenumbers $k_l$.
The left-hand side of Fig.~\ref{fig:circmat} visualizes the real-space system matrix $\tilde{\mathsf{H}}^2$ (top), and the $k$-space system matrix $\tilde{\mathsf{H}}_k^2$ (bottom), of a 5-cell (10-particle), spring-spring chain with $r=\epsilon=0.5$. Nonzero elements appear everywhere in the matrix $\tilde{\mathsf{H}}_k^2$.

Remember that a {circulant} matrix can always be diagonalized by discrete Fourier transform (in ordinary cases it would be by the $\mathbf{F}'$ matrix).
In our case, the Fourier transform is performed with $\mathbf{J}$, and the above statement can be adapted as follows.
Each of the four $2N\times2N$ blocks in $\tilde{\mathsf{H}}^2$ can be diagonalized by $\mathbf{J}$ if those $2N\times2N$ blocks are circulant with a step of two elements (and rows/columns).
The matrix $\tilde{\mathsf{H}}^2$ of a finite chain is almost circulant with a step of two elements except for the first and the last columns/rows. This is due to the fact that the periodic dimerization pattern stops at the terminals. The matrix can be made 2-element-circulant by connecting the two terminals, which turns the chain into a ring. The corresponding circulant $\tilde{\mathsf{H}}^2$ is shown in the upper-right of Fig.~\ref{fig:circmat}. The only differences from the original matrix are the $(1,2N)$ and $(2N,1)$ elements of the four $2N\times 2N$ blocks, which are now nonzero, as the extension of the original tridiagonal pattern penetrates the block boundary and appears on the other side.
The lower-right shows the $k$-space matrix, $\tilde{\mathsf{H}}_k^2$, and each of the four block is $2\times 2$-block diagonalized, meaning all wavenumbers $k_l$ are decoupled in the equations.

We can now select any $k_l$ and pick the corresponding $2\times 2$ blocks from the four $2N\times 2N$ blocks, assemble them and get the $4\times 4$ matrix equation for the particular $k_l$. Furthermore, as $N\rightarrow\infty$, $k_l=2\pi l/N$, $l=1,2,\dots,N$ becomes continuous.
For all $k\in(0,2\pi]$, the equations read
\begin{equation}
    \tilde{\mathsf{H}}_k^2(k)
    \left(
    \begin{array}{c}
         a_{k}^+ \\
         b_{k}^+ \\
         a_{k}^- \\
         b_{k}^- \\
    \end{array}
    \right)
    =
    \tilde{\omega}^2_{k_l}
    \left(
    \begin{array}{c}
         a_{k}^+ \\
         b_{k}^+ \\
         a_{k}^- \\
         b_{k}^- \\
    \end{array}
    \right),
\end{equation}
in which the system matrix reads
\begin{widetext}
\begin{equation}
    \tilde{\mathsf{H}}_k^2(k)=\left(
    \begin{array}{cc|cc}
         r & -\frac{1}{2} \left(\ee^{-\ii k} (1-\epsilon )+\epsilon +1\right) & 0 & \frac{1}{2} r \left(\ee^{-\ii k} (1-\epsilon )+\epsilon +1\right) \\
         -\frac{1}{2} \left(\ee^{\ii k} (1-\epsilon )+\epsilon +1\right) & -r & -\frac{1}{2} r \left(\ee^{\ii k} (1-\epsilon )+\epsilon +1\right) & 0 \\
         \hline
         0 & \frac{1}{2} r \left(\ee^{-\ii k} (1-\epsilon )+\epsilon +1\right) & r & -\frac{1}{2} \left(\ee^{-\ii k} (1-\epsilon )+\epsilon +1\right) \\
         -\frac{1}{2} r \left(\ee^{\ii k} (1-\epsilon )+\epsilon +1\right) & 0 & -\frac{1}{2} \left(\ee^{\ii k} (1-\epsilon )+\epsilon +1\right) & -r \\
    \end{array}
    \right).
\end{equation}
\end{widetext}

\section{Topological invariant, integral representation}\label{sec:topoint}
It has been shown that the $\mathbb{Z}_2$ invariant obtained based on the Pfaffian of the system matrix is equivalent to the invariant represented by the quantized Zak-Berry phase
\cite{budich2013equivalent}. Below we provide the calculation for the diatomic chain.

The Berry phase in a 1D domain is also known as the Zak phase. In 1D $k$-space, the Zak phase is the integral of the Berry connection in the domain $k\in (-\pi,\pi]$, or equivalently, $k\in (0,2\pi]$,
\begin{equation}\label{eq:zakp}
    \phi_Z=
    \int_0^{2\pi} \mathcal{A}(k) \dd k,
\end{equation}
where
\begin{equation}
    \mathcal{A}(k)
    =-\ii\bra{X}\partial_k\ket{X}
    =-\ii \mathbf{X}(k)^\dagger \cdot \left(\frac{\dd}{\dd k} \mathbf{X}(k)\right)
\end{equation}
is the Berry connection measuring the differential phase change in the eigenvector as it evolves along $k$. The integral in Eq.~\ref{eq:zakp} can be considered as a loop integral but, differently from the Berry phase in 2D manifolds, the integral cannot be converted into a surface integral (of Berry curvature). The Zak phase of the chain is gauge invariant (unchanged under a smooth gauge transformation $\mathbf{X}'(k)=\ee^{\ii \chi(k)} \mathbf{X}(k)$) and is quantized to an integer multiple of $\pi$. In the following, we use the chain with $r=0$ as the example to calculate the Zak phase.

The system matrix for the $r=0$ chain can be obtained from Eq.~\ref{eq:hk2t} as
\begin{align}
    &\tilde{\mathsf{H}}_k^2\rvert_{r=0}(k) =
    \mathbf{1}\otimes\mathbf{A},\\
    &\mathbf{A} = \frac{-1}{2}
    \begin{pmatrix}
    0 &  \ee^{-\ii k} (1-\epsilon)+1+\epsilon\\
    \ee^{\ii k} (1-\epsilon)+1+\epsilon & 0
    \end{pmatrix},
\end{align}
which is in block-diagonal form of identical blocks $\mathbf{A}$, and $\mathbf{A}$ is reminiscent of the $2\times 2$ Hamiltonian of the SSH model. The eigenvalues and normalized eigenvectors are calculated and shown below.
\begin{widetext}
    \begin{equation}
\renewcommand{\arraystretch}{1.25}
    \begin{array}{|c|c|}
        \hline
    \mathrm{Eigenvalues} & \mathrm{Eigenvectors}\\
    \hline\hline
        \tilde{\omega}_1^2 = -\dfrac{\sqrt{(1-\epsilon ^2) \cos k + (1+\epsilon ^2)}}{\sqrt{2}}
        &
        \begin{array}{c}
            \mathbf{X}_{1+} = \left(+\frac{1}{\sqrt{2}},\frac{\ee^{\ii k} (1-\epsilon)+(1+\epsilon)}{2 \sqrt{(1-\epsilon ^2) \cos k + (1+\epsilon ^2)}},0,0\right)^\intercal \\
            \mathbf{X}_{1-} = \left(0,0,+\frac{1}{\sqrt{2}},\frac{\ee^{\ii k} (1-\epsilon)+(1+\epsilon)}{2 \sqrt{(1-\epsilon ^2) \cos k + (1+\epsilon ^2)}}\right)^\intercal 
        \end{array}
        \\
    \hline
        \tilde{\omega}_2^2 = +\dfrac{\sqrt{(1-\epsilon ^2) \cos k + (1+\epsilon ^2)}}{\sqrt{2}}
        &
        \begin{array}{c}
        \mathbf{X}_{2+} = \left(-\frac{1}{\sqrt{2}},\frac{\ee^{\ii k} (1-\epsilon)+(1+\epsilon)}{2 \sqrt{(1-\epsilon ^2) \cos k + (1+\epsilon ^2)}},0,0\right)^\intercal\\
        \mathbf{X}_{2-} = \left(0,0,-\frac{1}{\sqrt{2}},\frac{\ee^{\ii k} (1-\epsilon)+(1+\epsilon)}{2 \sqrt{(1-\epsilon ^2) \cos k + (1+\epsilon ^2)}}\right)^\intercal
        \end{array}
        \\
    \hline 
    \end{array}
    \renewcommand{\arraystretch}{1}
\end{equation}
\end{widetext}

\begin{figure*}[!htb]
\centering
\includegraphics[width=.8\textwidth]{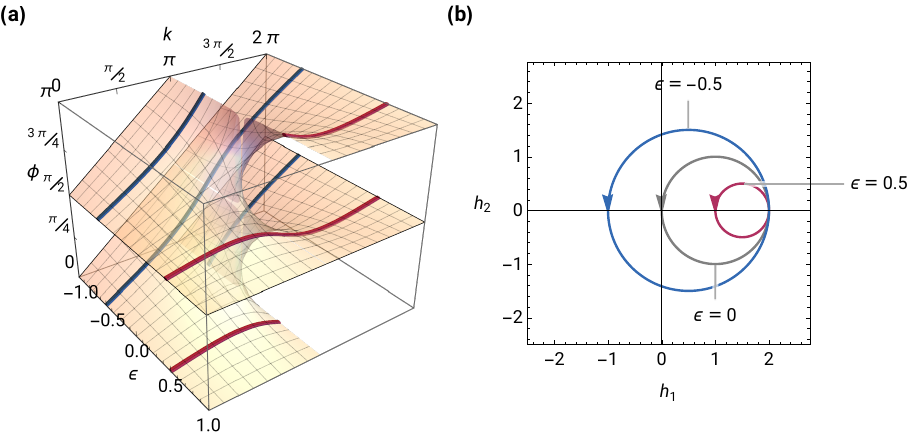}
\caption{\label{fig:zak}
(a) Surfaces of multiple branches of the antiderivative of the Berry connection. Magenta and blue curves show $\epsilon=+0.5$ and $\epsilon=-0.5$ section curves, respectively.
For the former (later), the Zak phase integral returns zero ($\pi$), indicating a trivial (topological) phase.
(b) Loci of $(h_1,h_2)$ components of the system matrix for $k\in[0.2\pi]$, for $\epsilon=+0.5$ (magenta), $0$ (gray), and $+0.5$ (blue), respectively. Winding numbers 0 and 1 indicate trivial and topological phases, respectively. 
}
\end{figure*}

Note that given $(a_k^+,b_k^+)$ and $(a_k^-,b_k^-)$ are decoupled, the two eigenvectors of a degenerate eigenvalue can be separated accordingly. They are composed of identical expressions only at different components, therefore resulting in the same Berry connection.
The Berry connection for the lower bands are found to be
\begin{multline}
\mathcal{A}_{1\pm}(k)
    =-\ii\bra{X_{1\pm}}\partial_k\ket{X_{1\pm}}\\
    =\frac{1-\epsilon}{4}
    \frac{(1+\epsilon) \cos k+(1-\epsilon)}{(1-\epsilon ^2) \cos k + (1+\epsilon ^2)}.
\end{multline}
the Zak phase $\phi_Z$ is the definite integral of $\mathcal{A}(k)$ over the interval $k\in[0,2\pi]$,
\begin{subequations}
\begin{align}
\phi_Z&=\int_0^{2\pi}
\mathcal{A}_{1\pm}(k) \dd k
= \phi(2\pi) - \phi(0), \label{eq:zakint}\\
\phi(k)&=\frac{1}{4} \left[k-2 \tan
   ^{-1}\left(\epsilon  \tan
   \frac{k}{2}\right)\right].
\end{align}
\end{subequations}
$\phi(k)$ is the antiderivative (indefinite integral) function of $\mathcal{A}(k)$, which contains the arctangent function of multiple branches. The Zak phase integral (Eq.~\ref{eq:zakint}) should follow a continuous path that may connect different branches. Fig.~\ref{fig:zak} (a) shows two neighboring branches of $\phi(k,\epsilon)$ (staking one above the other), and the blue and magenta curves show $\epsilon=-0.5$ and $\epsilon=+0.5$ sections, respectively. The center curves are continuous for $k\in[0,\pi]$. It can be concluded that,
\begin{equation}
    \phi_Z=
    \begin{cases}
    0, & \epsilon >0\\
    \pi, & \epsilon <0
    \end{cases}.
\end{equation}

Another approach to obtain the Zak phase is through the {winding number}, which avoids complicated integrals. The matrix $\mathbf{A}$ is first decomposed into Pauli matrices $\mathbf{A} = \mathbf{h}\cdot \bm\sigma = h_i\sigma_i$, and the Zak-Berry phase can be obtained as half of the solid angle enclosed by the loop $\mathbf{h}(k)$, $k\in[0,2\pi]$, viewed from the origin in $\mathbf{h}$-space (where the degeneracy locates).
Particularly $\mathbf{A}$ has no $\sigma_3$-component, $h_3=0$, so the loop of $\mathbf{h}(k)$ is coplanar with the degenerate point and the Zak-Berry phase can only be 0 or $\pi$, depending on whether the origin is encircled by the path. In other words, let $\nu=0,1$ be the winding number of the path $\mathbf{h}(k)$ rounding the origin, then the Zak phase is $\phi_Z=\nu \pi$. For the matrix $\mathbf{A}$, we have
\begin{subequations}
\begin{align}
    h_1 &= (1-\epsilon)\cos{k}+(1+\epsilon),\\
    h_2 &= (1-\epsilon)\sin{k},\\
    h_3 &= 0,
\end{align}
\end{subequations}
which is a counter-clockwise circular path on $h_1h_2$-plane, with a radius $(1-\epsilon)$ and centered at the point $(1+\epsilon,0)$. Fig.~\ref{fig:zak} (b) shows three paths with $\epsilon=-0.5$ (blue), 0 (gray), and $+0.5$ (magenta).
Clearly for $\epsilon>0$, the winding number (and therefore the Zak phase) is 0, corresponding to the trivial state. For $\epsilon<0$, the winding number is 1 and the Zak phase becomes $\pi$, representing the topological state. For $\epsilon=0$, the winding number, the Zak phase, and the topological state are indeterminate.

All the above analysis based on topological band theory in $k$-space confirms the existence of the topological bound states appearing at the terminals of nontrivial chains and the domain walls connecting two distinct chains, are indeed topologically protected.

\bibliography{allbib}

\begin{thebibliography}{74}%
\makeatletter
\providecommand \@ifxundefined [1]{%
 \@ifx{#1\undefined}
}%
\providecommand \@ifnum [1]{%
 \ifnum #1\expandafter \@firstoftwo
 \else \expandafter \@secondoftwo
 \fi
}%
\providecommand \@ifx [1]{%
 \ifx #1\expandafter \@firstoftwo
 \else \expandafter \@secondoftwo
 \fi
}%
\providecommand \natexlab [1]{#1}%
\providecommand \enquote  [1]{``#1''}%
\providecommand \bibnamefont  [1]{#1}%
\providecommand \bibfnamefont [1]{#1}%
\providecommand \citenamefont [1]{#1}%
\providecommand \href@noop [0]{\@secondoftwo}%
\providecommand \href [0]{\begingroup \@sanitize@url \@href}%
\providecommand \@href[1]{\@@startlink{#1}\@@href}%
\providecommand \@@href[1]{\endgroup#1\@@endlink}%
\providecommand \@sanitize@url [0]{\catcode `\\12\catcode `\$12\catcode
  `\&12\catcode `\#12\catcode `\^12\catcode `\_12\catcode `\%12\relax}%
\providecommand \@@startlink[1]{}%
\providecommand \@@endlink[0]{}%
\providecommand \url  [0]{\begingroup\@sanitize@url \@url }%
\providecommand \@url [1]{\endgroup\@href {#1}{\urlprefix }}%
\providecommand \urlprefix  [0]{URL }%
\providecommand \Eprint [0]{\href }%
\providecommand \doibase [0]{https://doi.org/}%
\providecommand \selectlanguage [0]{\@gobble}%
\providecommand \bibinfo  [0]{\@secondoftwo}%
\providecommand \bibfield  [0]{\@secondoftwo}%
\providecommand \translation [1]{[#1]}%
\providecommand \BibitemOpen [0]{}%
\providecommand \bibitemStop [0]{}%
\providecommand \bibitemNoStop [0]{.\EOS\space}%
\providecommand \EOS [0]{\spacefactor3000\relax}%
\providecommand \BibitemShut  [1]{\csname bibitem#1\endcsname}%
\let\auto@bib@innerbib\@empty
\bibitem [{\citenamefont {Liang}\ \emph {et~al.}(2009)\citenamefont {Liang},
  \citenamefont {Yuan},\ and\ \citenamefont {Cheng}}]{liang2009acoustic}%
  \BibitemOpen
  \bibfield  {author} {\bibinfo {author} {\bibfnamefont {B.}~\bibnamefont
  {Liang}}, \bibinfo {author} {\bibfnamefont {B.}~\bibnamefont {Yuan}},\ and\
  \bibinfo {author} {\bibfnamefont {J.-c.}\ \bibnamefont {Cheng}},\ }\href@noop
  {} {\bibfield  {journal} {\bibinfo  {journal} {Physical review letters}\
  }\textbf {\bibinfo {volume} {103}},\ \bibinfo {pages} {104301} (\bibinfo
  {year} {2009})}\BibitemShut {NoStop}%
\bibitem [{\citenamefont {Fleury}\ \emph {et~al.}(2014)\citenamefont {Fleury},
  \citenamefont {Sounas}, \citenamefont {Sieck}, \citenamefont {Haberman},\
  and\ \citenamefont {Al{\`u}}}]{fleury2014sound}%
  \BibitemOpen
  \bibfield  {author} {\bibinfo {author} {\bibfnamefont {R.}~\bibnamefont
  {Fleury}}, \bibinfo {author} {\bibfnamefont {D.~L.}\ \bibnamefont {Sounas}},
  \bibinfo {author} {\bibfnamefont {C.~F.}\ \bibnamefont {Sieck}}, \bibinfo
  {author} {\bibfnamefont {M.~R.}\ \bibnamefont {Haberman}},\ and\ \bibinfo
  {author} {\bibfnamefont {A.}~\bibnamefont {Al{\`u}}},\ }\href@noop {}
  {\bibfield  {journal} {\bibinfo  {journal} {Science}\ }\textbf {\bibinfo
  {volume} {343}},\ \bibinfo {pages} {516} (\bibinfo {year}
  {2014})}\BibitemShut {NoStop}%
\bibitem [{\citenamefont {Li}\ \emph {et~al.}(2014)\citenamefont {Li},
  \citenamefont {Anzel}, \citenamefont {Yang}, \citenamefont {Kevrekidis},\
  and\ \citenamefont {Daraio}}]{li2014granular}%
  \BibitemOpen
  \bibfield  {author} {\bibinfo {author} {\bibfnamefont {F.}~\bibnamefont
  {Li}}, \bibinfo {author} {\bibfnamefont {P.}~\bibnamefont {Anzel}}, \bibinfo
  {author} {\bibfnamefont {J.}~\bibnamefont {Yang}}, \bibinfo {author}
  {\bibfnamefont {P.~G.}\ \bibnamefont {Kevrekidis}},\ and\ \bibinfo {author}
  {\bibfnamefont {C.}~\bibnamefont {Daraio}},\ }\href@noop {} {\bibfield
  {journal} {\bibinfo  {journal} {Nature communications}\ }\textbf {\bibinfo
  {volume} {5}},\ \bibinfo {pages} {1} (\bibinfo {year} {2014})}\BibitemShut
  {NoStop}%
\bibitem [{\citenamefont {Yu}\ \emph {et~al.}(2018)\citenamefont {Yu},
  \citenamefont {He}, \citenamefont {Wang}, \citenamefont {Liu}, \citenamefont
  {Sun}, \citenamefont {Li}, \citenamefont {Lu}, \citenamefont {Lu},
  \citenamefont {Liu},\ and\ \citenamefont {Chen}}]{yu2018elastic}%
  \BibitemOpen
  \bibfield  {author} {\bibinfo {author} {\bibfnamefont {S.-Y.}\ \bibnamefont
  {Yu}}, \bibinfo {author} {\bibfnamefont {C.}~\bibnamefont {He}}, \bibinfo
  {author} {\bibfnamefont {Z.}~\bibnamefont {Wang}}, \bibinfo {author}
  {\bibfnamefont {F.-K.}\ \bibnamefont {Liu}}, \bibinfo {author} {\bibfnamefont
  {X.-C.}\ \bibnamefont {Sun}}, \bibinfo {author} {\bibfnamefont
  {Z.}~\bibnamefont {Li}}, \bibinfo {author} {\bibfnamefont {H.-Z.}\
  \bibnamefont {Lu}}, \bibinfo {author} {\bibfnamefont {M.-H.}\ \bibnamefont
  {Lu}}, \bibinfo {author} {\bibfnamefont {X.-P.}\ \bibnamefont {Liu}},\ and\
  \bibinfo {author} {\bibfnamefont {Y.-F.}\ \bibnamefont {Chen}},\ }\href@noop
  {} {\bibfield  {journal} {\bibinfo  {journal} {Nature communications}\
  }\textbf {\bibinfo {volume} {9}},\ \bibinfo {pages} {1} (\bibinfo {year}
  {2018})}\BibitemShut {NoStop}%
\bibitem [{\citenamefont {Ma}\ \emph {et~al.}(2019)\citenamefont {Ma},
  \citenamefont {Sun},\ and\ \citenamefont {Gonella}}]{ma2019valley}%
  \BibitemOpen
  \bibfield  {author} {\bibinfo {author} {\bibfnamefont {J.}~\bibnamefont
  {Ma}}, \bibinfo {author} {\bibfnamefont {K.}~\bibnamefont {Sun}},\ and\
  \bibinfo {author} {\bibfnamefont {S.}~\bibnamefont {Gonella}},\ }\href@noop
  {} {\bibfield  {journal} {\bibinfo  {journal} {Physical Review Applied}\
  }\textbf {\bibinfo {volume} {12}},\ \bibinfo {pages} {044015} (\bibinfo
  {year} {2019})}\BibitemShut {NoStop}%
\bibitem [{\citenamefont {El~Helou}\ \emph {et~al.}(2022)\citenamefont
  {El~Helou}, \citenamefont {Grossmann}, \citenamefont {Tabor}, \citenamefont
  {Buskohl},\ and\ \citenamefont {Harne}}]{el2022mechanical}%
  \BibitemOpen
  \bibfield  {author} {\bibinfo {author} {\bibfnamefont {C.}~\bibnamefont
  {El~Helou}}, \bibinfo {author} {\bibfnamefont {B.}~\bibnamefont {Grossmann}},
  \bibinfo {author} {\bibfnamefont {C.~E.}\ \bibnamefont {Tabor}}, \bibinfo
  {author} {\bibfnamefont {P.~R.}\ \bibnamefont {Buskohl}},\ and\ \bibinfo
  {author} {\bibfnamefont {R.~L.}\ \bibnamefont {Harne}},\ }\href@noop {}
  {\bibfield  {journal} {\bibinfo  {journal} {Nature}\ }\textbf {\bibinfo
  {volume} {608}},\ \bibinfo {pages} {699} (\bibinfo {year}
  {2022})}\BibitemShut {NoStop}%
\bibitem [{\citenamefont {Zivari}\ \emph {et~al.}(2022)\citenamefont {Zivari},
  \citenamefont {Stockill}, \citenamefont {Fiaschi},\ and\ \citenamefont
  {Gr{\"o}blacher}}]{zivari2022non}%
  \BibitemOpen
  \bibfield  {author} {\bibinfo {author} {\bibfnamefont {A.}~\bibnamefont
  {Zivari}}, \bibinfo {author} {\bibfnamefont {R.}~\bibnamefont {Stockill}},
  \bibinfo {author} {\bibfnamefont {N.}~\bibnamefont {Fiaschi}},\ and\ \bibinfo
  {author} {\bibfnamefont {S.}~\bibnamefont {Gr{\"o}blacher}},\ }\href@noop {}
  {\bibfield  {journal} {\bibinfo  {journal} {Nature Physics}\ ,\ \bibinfo
  {pages} {1}} (\bibinfo {year} {2022})}\BibitemShut {NoStop}%
\bibitem [{\citenamefont {Chen}\ \emph {et~al.}(2019)\citenamefont {Chen},
  \citenamefont {Lera}, \citenamefont {Chaunsali}, \citenamefont {Torrent},
  \citenamefont {Alvarez}, \citenamefont {Yang}, \citenamefont {San-Jose},\
  and\ \citenamefont {Christensen}}]{chen2019mechanical}%
  \BibitemOpen
  \bibfield  {author} {\bibinfo {author} {\bibfnamefont {C.-W.}\ \bibnamefont
  {Chen}}, \bibinfo {author} {\bibfnamefont {N.}~\bibnamefont {Lera}}, \bibinfo
  {author} {\bibfnamefont {R.}~\bibnamefont {Chaunsali}}, \bibinfo {author}
  {\bibfnamefont {D.}~\bibnamefont {Torrent}}, \bibinfo {author} {\bibfnamefont
  {J.~V.}\ \bibnamefont {Alvarez}}, \bibinfo {author} {\bibfnamefont
  {J.}~\bibnamefont {Yang}}, \bibinfo {author} {\bibfnamefont {P.}~\bibnamefont
  {San-Jose}},\ and\ \bibinfo {author} {\bibfnamefont {J.}~\bibnamefont
  {Christensen}},\ }\href@noop {} {\bibfield  {journal} {\bibinfo  {journal}
  {Advanced Materials}\ }\textbf {\bibinfo {volume} {31}},\ \bibinfo {pages}
  {1904386} (\bibinfo {year} {2019})}\BibitemShut {NoStop}%
\bibitem [{\citenamefont {Barlas}\ and\ \citenamefont
  {Prodan}(2020)}]{barlas2020topological}%
  \BibitemOpen
  \bibfield  {author} {\bibinfo {author} {\bibfnamefont {Y.}~\bibnamefont
  {Barlas}}\ and\ \bibinfo {author} {\bibfnamefont {E.}~\bibnamefont
  {Prodan}},\ }\href@noop {} {\bibfield  {journal} {\bibinfo  {journal}
  {Physical Review Letters}\ }\textbf {\bibinfo {volume} {124}},\ \bibinfo
  {pages} {146801} (\bibinfo {year} {2020})}\BibitemShut {NoStop}%
\bibitem [{\citenamefont {Qian}\ \emph {et~al.}(2022)\citenamefont {Qian},
  \citenamefont {Apigo}, \citenamefont {Padavi{\'c}}, \citenamefont {Ahn},
  \citenamefont {Vishveshwara},\ and\ \citenamefont
  {Prodan}}]{qian2022observation}%
  \BibitemOpen
  \bibfield  {author} {\bibinfo {author} {\bibfnamefont {K.}~\bibnamefont
  {Qian}}, \bibinfo {author} {\bibfnamefont {D.~J.}\ \bibnamefont {Apigo}},
  \bibinfo {author} {\bibfnamefont {K.}~\bibnamefont {Padavi{\'c}}}, \bibinfo
  {author} {\bibfnamefont {K.~H.}\ \bibnamefont {Ahn}}, \bibinfo {author}
  {\bibfnamefont {S.}~\bibnamefont {Vishveshwara}},\ and\ \bibinfo {author}
  {\bibfnamefont {C.}~\bibnamefont {Prodan}},\ }\href@noop {} {\bibfield
  {journal} {\bibinfo  {journal} {arXiv preprint arXiv:2201.12377}\ } (\bibinfo
  {year} {2022})}\BibitemShut {NoStop}%
\bibitem [{\citenamefont {Allein}\ \emph {et~al.}(2022)\citenamefont {Allein},
  \citenamefont {Chaunsali}, \citenamefont {Anastasiadis}, \citenamefont
  {Frankel}, \citenamefont {Boechler}, \citenamefont {Diakonos},\ and\
  \citenamefont {Theocharis}}]{allein2022duality}%
  \BibitemOpen
  \bibfield  {author} {\bibinfo {author} {\bibfnamefont {F.}~\bibnamefont
  {Allein}}, \bibinfo {author} {\bibfnamefont {R.}~\bibnamefont {Chaunsali}},
  \bibinfo {author} {\bibfnamefont {A.}~\bibnamefont {Anastasiadis}}, \bibinfo
  {author} {\bibfnamefont {I.}~\bibnamefont {Frankel}}, \bibinfo {author}
  {\bibfnamefont {N.}~\bibnamefont {Boechler}}, \bibinfo {author}
  {\bibfnamefont {F.~K.}\ \bibnamefont {Diakonos}},\ and\ \bibinfo {author}
  {\bibfnamefont {G.}~\bibnamefont {Theocharis}},\ }\href@noop {} {\bibfield
  {journal} {\bibinfo  {journal} {arXiv preprint arXiv:2203.10311}\ } (\bibinfo
  {year} {2022})}\BibitemShut {NoStop}%
\bibitem [{\citenamefont {Qian}\ \emph {et~al.}(2023)\citenamefont {Qian},
  \citenamefont {Apigo}, \citenamefont {Padavi{\'c}}, \citenamefont {Ahn},
  \citenamefont {Vishveshwara},\ and\ \citenamefont
  {Prodan}}]{qian2023observation}%
  \BibitemOpen
  \bibfield  {author} {\bibinfo {author} {\bibfnamefont {K.}~\bibnamefont
  {Qian}}, \bibinfo {author} {\bibfnamefont {D.~J.}\ \bibnamefont {Apigo}},
  \bibinfo {author} {\bibfnamefont {K.}~\bibnamefont {Padavi{\'c}}}, \bibinfo
  {author} {\bibfnamefont {K.~H.}\ \bibnamefont {Ahn}}, \bibinfo {author}
  {\bibfnamefont {S.}~\bibnamefont {Vishveshwara}},\ and\ \bibinfo {author}
  {\bibfnamefont {C.}~\bibnamefont {Prodan}},\ }\href@noop {} {\bibfield
  {journal} {\bibinfo  {journal} {Physical Review Research}\ }\textbf {\bibinfo
  {volume} {5}},\ \bibinfo {pages} {L012012} (\bibinfo {year}
  {2023})}\BibitemShut {NoStop}%
\bibitem [{\citenamefont {Majorana}(1937)}]{majorana1937teoria}%
  \BibitemOpen
  \bibfield  {author} {\bibinfo {author} {\bibfnamefont {E.}~\bibnamefont
  {Majorana}},\ }\href@noop {} {\bibfield  {journal} {\bibinfo  {journal} {Il
  Nuovo Cimento (1924-1942)}\ }\textbf {\bibinfo {volume} {14}},\ \bibinfo
  {pages} {171} (\bibinfo {year} {1937})}\BibitemShut {NoStop}%
\bibitem [{\citenamefont {Kitaev}(2001)}]{kitaev2001unpaired}%
  \BibitemOpen
  \bibfield  {author} {\bibinfo {author} {\bibfnamefont {A.~Y.}\ \bibnamefont
  {Kitaev}},\ }\href@noop {} {\bibfield  {journal} {\bibinfo  {journal}
  {Physics-uspekhi}\ }\textbf {\bibinfo {volume} {44}},\ \bibinfo {pages} {131}
  (\bibinfo {year} {2001})}\BibitemShut {NoStop}%
\bibitem [{\citenamefont {Kitaev}(2003)}]{kitaev2003fault}%
  \BibitemOpen
  \bibfield  {author} {\bibinfo {author} {\bibfnamefont {A.~Y.}\ \bibnamefont
  {Kitaev}},\ }\href@noop {} {\bibfield  {journal} {\bibinfo  {journal} {Annals
  of Physics}\ }\textbf {\bibinfo {volume} {303}},\ \bibinfo {pages} {2}
  (\bibinfo {year} {2003})}\BibitemShut {NoStop}%
\bibitem [{\citenamefont {Lutchyn}\ \emph {et~al.}(2010)\citenamefont
  {Lutchyn}, \citenamefont {Sau},\ and\ \citenamefont
  {Sarma}}]{lutchyn2010majorana}%
  \BibitemOpen
  \bibfield  {author} {\bibinfo {author} {\bibfnamefont {R.~M.}\ \bibnamefont
  {Lutchyn}}, \bibinfo {author} {\bibfnamefont {J.~D.}\ \bibnamefont {Sau}},\
  and\ \bibinfo {author} {\bibfnamefont {S.~D.}\ \bibnamefont {Sarma}},\
  }\href@noop {} {\bibfield  {journal} {\bibinfo  {journal} {Physical review
  letters}\ }\textbf {\bibinfo {volume} {105}},\ \bibinfo {pages} {077001}
  (\bibinfo {year} {2010})}\BibitemShut {NoStop}%
\bibitem [{\citenamefont {Alicea}(2010)}]{alicea2010majorana}%
  \BibitemOpen
  \bibfield  {author} {\bibinfo {author} {\bibfnamefont {J.}~\bibnamefont
  {Alicea}},\ }\href@noop {} {\bibfield  {journal} {\bibinfo  {journal}
  {Physical Review B}\ }\textbf {\bibinfo {volume} {81}},\ \bibinfo {pages}
  {125318} (\bibinfo {year} {2010})}\BibitemShut {NoStop}%
\bibitem [{\citenamefont {Sau}\ \emph {et~al.}(2010)\citenamefont {Sau},
  \citenamefont {Lutchyn}, \citenamefont {Tewari},\ and\ \citenamefont
  {Sarma}}]{sau2010generic}%
  \BibitemOpen
  \bibfield  {author} {\bibinfo {author} {\bibfnamefont {J.~D.}\ \bibnamefont
  {Sau}}, \bibinfo {author} {\bibfnamefont {R.~M.}\ \bibnamefont {Lutchyn}},
  \bibinfo {author} {\bibfnamefont {S.}~\bibnamefont {Tewari}},\ and\ \bibinfo
  {author} {\bibfnamefont {S.~D.}\ \bibnamefont {Sarma}},\ }\href@noop {}
  {\bibfield  {journal} {\bibinfo  {journal} {Physical review letters}\
  }\textbf {\bibinfo {volume} {104}},\ \bibinfo {pages} {040502} (\bibinfo
  {year} {2010})}\BibitemShut {NoStop}%
\bibitem [{\citenamefont {Alicea}\ \emph {et~al.}(2011)\citenamefont {Alicea},
  \citenamefont {Oreg}, \citenamefont {Refael}, \citenamefont {Von~Oppen},\
  and\ \citenamefont {Fisher}}]{alicea2011non}%
  \BibitemOpen
  \bibfield  {author} {\bibinfo {author} {\bibfnamefont {J.}~\bibnamefont
  {Alicea}}, \bibinfo {author} {\bibfnamefont {Y.}~\bibnamefont {Oreg}},
  \bibinfo {author} {\bibfnamefont {G.}~\bibnamefont {Refael}}, \bibinfo
  {author} {\bibfnamefont {F.}~\bibnamefont {Von~Oppen}},\ and\ \bibinfo
  {author} {\bibfnamefont {M.}~\bibnamefont {Fisher}},\ }\href@noop {}
  {\bibfield  {journal} {\bibinfo  {journal} {Nature Physics}\ }\textbf
  {\bibinfo {volume} {7}},\ \bibinfo {pages} {412} (\bibinfo {year}
  {2011})}\BibitemShut {NoStop}%
\bibitem [{\citenamefont {Chien}\ \emph {et~al.}(2018)\citenamefont {Chien},
  \citenamefont {Velizhanin}, \citenamefont {Dubi}, \citenamefont {Ilic},\ and\
  \citenamefont {Zwolak}}]{chien2018topological}%
  \BibitemOpen
  \bibfield  {author} {\bibinfo {author} {\bibfnamefont {C.-C.}\ \bibnamefont
  {Chien}}, \bibinfo {author} {\bibfnamefont {K.~A.}\ \bibnamefont
  {Velizhanin}}, \bibinfo {author} {\bibfnamefont {Y.}~\bibnamefont {Dubi}},
  \bibinfo {author} {\bibfnamefont {B.~R.}\ \bibnamefont {Ilic}},\ and\
  \bibinfo {author} {\bibfnamefont {M.}~\bibnamefont {Zwolak}},\ }\href@noop {}
  {\bibfield  {journal} {\bibinfo  {journal} {Physical Review B}\ }\textbf
  {\bibinfo {volume} {97}},\ \bibinfo {pages} {125425} (\bibinfo {year}
  {2018})}\BibitemShut {NoStop}%
\bibitem [{\citenamefont {Chien}\ \emph {et~al.}(2017)\citenamefont {Chien},
  \citenamefont {Kouachi}, \citenamefont {Velizhanin}, \citenamefont {Dubi},\
  and\ \citenamefont {Zwolak}}]{chien2017thermal}%
  \BibitemOpen
  \bibfield  {author} {\bibinfo {author} {\bibfnamefont {C.-C.}\ \bibnamefont
  {Chien}}, \bibinfo {author} {\bibfnamefont {S.}~\bibnamefont {Kouachi}},
  \bibinfo {author} {\bibfnamefont {K.~A.}\ \bibnamefont {Velizhanin}},
  \bibinfo {author} {\bibfnamefont {Y.}~\bibnamefont {Dubi}},\ and\ \bibinfo
  {author} {\bibfnamefont {M.}~\bibnamefont {Zwolak}},\ }\href@noop {}
  {\bibfield  {journal} {\bibinfo  {journal} {Physical Review E}\ }\textbf
  {\bibinfo {volume} {95}},\ \bibinfo {pages} {012137} (\bibinfo {year}
  {2017})}\BibitemShut {NoStop}%
\bibitem [{\citenamefont {Greiter}\ \emph {et~al.}(2014)\citenamefont
  {Greiter}, \citenamefont {Schnells},\ and\ \citenamefont
  {Thomale}}]{greiter20141d}%
  \BibitemOpen
  \bibfield  {author} {\bibinfo {author} {\bibfnamefont {M.}~\bibnamefont
  {Greiter}}, \bibinfo {author} {\bibfnamefont {V.}~\bibnamefont {Schnells}},\
  and\ \bibinfo {author} {\bibfnamefont {R.}~\bibnamefont {Thomale}},\
  }\href@noop {} {\bibfield  {journal} {\bibinfo  {journal} {Annals of
  Physics}\ }\textbf {\bibinfo {volume} {351}},\ \bibinfo {pages} {1026}
  (\bibinfo {year} {2014})}\BibitemShut {NoStop}%
\bibitem [{\citenamefont {Nadj-Perge}\ \emph {et~al.}(2014)\citenamefont
  {Nadj-Perge}, \citenamefont {Drozdov}, \citenamefont {Li}, \citenamefont
  {Chen}, \citenamefont {Jeon}, \citenamefont {Seo}, \citenamefont {MacDonald},
  \citenamefont {Bernevig},\ and\ \citenamefont
  {Yazdani}}]{nadj2014observation}%
  \BibitemOpen
  \bibfield  {author} {\bibinfo {author} {\bibfnamefont {S.}~\bibnamefont
  {Nadj-Perge}}, \bibinfo {author} {\bibfnamefont {I.~K.}\ \bibnamefont
  {Drozdov}}, \bibinfo {author} {\bibfnamefont {J.}~\bibnamefont {Li}},
  \bibinfo {author} {\bibfnamefont {H.}~\bibnamefont {Chen}}, \bibinfo {author}
  {\bibfnamefont {S.}~\bibnamefont {Jeon}}, \bibinfo {author} {\bibfnamefont
  {J.}~\bibnamefont {Seo}}, \bibinfo {author} {\bibfnamefont {A.~H.}\
  \bibnamefont {MacDonald}}, \bibinfo {author} {\bibfnamefont {B.~A.}\
  \bibnamefont {Bernevig}},\ and\ \bibinfo {author} {\bibfnamefont
  {A.}~\bibnamefont {Yazdani}},\ }\href@noop {} {\bibfield  {journal} {\bibinfo
   {journal} {Science}\ }\textbf {\bibinfo {volume} {346}},\ \bibinfo {pages}
  {602} (\bibinfo {year} {2014})}\BibitemShut {NoStop}%
\bibitem [{\citenamefont {Kim}\ \emph {et~al.}(2018)\citenamefont {Kim},
  \citenamefont {Palacio-Morales}, \citenamefont {Posske}, \citenamefont
  {R{\'o}zsa}, \citenamefont {Palot{\'a}s}, \citenamefont {Szunyogh},
  \citenamefont {Thorwart},\ and\ \citenamefont
  {Wiesendanger}}]{kim2018toward}%
  \BibitemOpen
  \bibfield  {author} {\bibinfo {author} {\bibfnamefont {H.}~\bibnamefont
  {Kim}}, \bibinfo {author} {\bibfnamefont {A.}~\bibnamefont
  {Palacio-Morales}}, \bibinfo {author} {\bibfnamefont {T.}~\bibnamefont
  {Posske}}, \bibinfo {author} {\bibfnamefont {L.}~\bibnamefont {R{\'o}zsa}},
  \bibinfo {author} {\bibfnamefont {K.}~\bibnamefont {Palot{\'a}s}}, \bibinfo
  {author} {\bibfnamefont {L.}~\bibnamefont {Szunyogh}}, \bibinfo {author}
  {\bibfnamefont {M.}~\bibnamefont {Thorwart}},\ and\ \bibinfo {author}
  {\bibfnamefont {R.}~\bibnamefont {Wiesendanger}},\ }\href@noop {} {\bibfield
  {journal} {\bibinfo  {journal} {Science Advances}\ }\textbf {\bibinfo
  {volume} {4}},\ \bibinfo {pages} {eaar5251} (\bibinfo {year}
  {2018})}\BibitemShut {NoStop}%
\bibitem [{\citenamefont {Attig}\ \emph {et~al.}(2019)\citenamefont {Attig},
  \citenamefont {Roychowdhury}, \citenamefont {Lawler},\ and\ \citenamefont
  {Trebst}}]{attig2019topological}%
  \BibitemOpen
  \bibfield  {author} {\bibinfo {author} {\bibfnamefont {J.}~\bibnamefont
  {Attig}}, \bibinfo {author} {\bibfnamefont {K.}~\bibnamefont {Roychowdhury}},
  \bibinfo {author} {\bibfnamefont {M.~J.}\ \bibnamefont {Lawler}},\ and\
  \bibinfo {author} {\bibfnamefont {S.}~\bibnamefont {Trebst}},\ }\href@noop {}
  {\bibfield  {journal} {\bibinfo  {journal} {Physical Review Research}\
  }\textbf {\bibinfo {volume} {1}},\ \bibinfo {pages} {032047} (\bibinfo {year}
  {2019})}\BibitemShut {NoStop}%
\bibitem [{\citenamefont {Zhang}\ and\ \citenamefont
  {Zhou}(2017)}]{zhang2017two}%
  \BibitemOpen
  \bibfield  {author} {\bibinfo {author} {\bibfnamefont {S.-L.}\ \bibnamefont
  {Zhang}}\ and\ \bibinfo {author} {\bibfnamefont {Q.}~\bibnamefont {Zhou}},\
  }\href@noop {} {\bibfield  {journal} {\bibinfo  {journal} {Physical Review
  A}\ }\textbf {\bibinfo {volume} {95}},\ \bibinfo {pages} {061601} (\bibinfo
  {year} {2017})}\BibitemShut {NoStop}%
\bibitem [{\citenamefont {Ezawa}(2019)}]{ezawa2019braiding}%
  \BibitemOpen
  \bibfield  {author} {\bibinfo {author} {\bibfnamefont {M.}~\bibnamefont
  {Ezawa}},\ }\href@noop {} {\bibfield  {journal} {\bibinfo  {journal}
  {Physical Review B}\ }\textbf {\bibinfo {volume} {100}},\ \bibinfo {pages}
  {045407} (\bibinfo {year} {2019})}\BibitemShut {NoStop}%
\bibitem [{\citenamefont {Ezawa}(2020)}]{ezawa2020non}%
  \BibitemOpen
  \bibfield  {author} {\bibinfo {author} {\bibfnamefont {M.}~\bibnamefont
  {Ezawa}},\ }\href@noop {} {\bibfield  {journal} {\bibinfo  {journal}
  {Physical Review B}\ }\textbf {\bibinfo {volume} {102}},\ \bibinfo {pages}
  {075424} (\bibinfo {year} {2020})}\BibitemShut {NoStop}%
\bibitem [{\citenamefont {Gao}\ \emph {et~al.}(2019)\citenamefont {Gao},
  \citenamefont {Torrent}, \citenamefont {Cervera}, \citenamefont {San-Jose},
  \citenamefont {S{\'a}nchez-Dehesa},\ and\ \citenamefont
  {Christensen}}]{gao2019majorana}%
  \BibitemOpen
  \bibfield  {author} {\bibinfo {author} {\bibfnamefont {P.}~\bibnamefont
  {Gao}}, \bibinfo {author} {\bibfnamefont {D.}~\bibnamefont {Torrent}},
  \bibinfo {author} {\bibfnamefont {F.}~\bibnamefont {Cervera}}, \bibinfo
  {author} {\bibfnamefont {P.}~\bibnamefont {San-Jose}}, \bibinfo {author}
  {\bibfnamefont {J.}~\bibnamefont {S{\'a}nchez-Dehesa}},\ and\ \bibinfo
  {author} {\bibfnamefont {J.}~\bibnamefont {Christensen}},\ }\href@noop {}
  {\bibfield  {journal} {\bibinfo  {journal} {Physical review letters}\
  }\textbf {\bibinfo {volume} {123}},\ \bibinfo {pages} {196601} (\bibinfo
  {year} {2019})}\BibitemShut {NoStop}%
\bibitem [{\citenamefont {Raghu}\ and\ \citenamefont
  {Haldane}(2008)}]{raghu2008analogs}%
  \BibitemOpen
  \bibfield  {author} {\bibinfo {author} {\bibfnamefont {S.}~\bibnamefont
  {Raghu}}\ and\ \bibinfo {author} {\bibfnamefont {F.~D.~M.}\ \bibnamefont
  {Haldane}},\ }\href@noop {} {\bibfield  {journal} {\bibinfo  {journal}
  {Physical Review A}\ }\textbf {\bibinfo {volume} {78}},\ \bibinfo {pages}
  {033834} (\bibinfo {year} {2008})}\BibitemShut {NoStop}%
\bibitem [{\citenamefont {Yang}\ \emph {et~al.}(2015)\citenamefont {Yang},
  \citenamefont {Gao}, \citenamefont {Shi}, \citenamefont {Lin}, \citenamefont
  {Gao}, \citenamefont {Chong},\ and\ \citenamefont
  {Zhang}}]{yang2015topological}%
  \BibitemOpen
  \bibfield  {author} {\bibinfo {author} {\bibfnamefont {Z.}~\bibnamefont
  {Yang}}, \bibinfo {author} {\bibfnamefont {F.}~\bibnamefont {Gao}}, \bibinfo
  {author} {\bibfnamefont {X.}~\bibnamefont {Shi}}, \bibinfo {author}
  {\bibfnamefont {X.}~\bibnamefont {Lin}}, \bibinfo {author} {\bibfnamefont
  {Z.}~\bibnamefont {Gao}}, \bibinfo {author} {\bibfnamefont {Y.}~\bibnamefont
  {Chong}},\ and\ \bibinfo {author} {\bibfnamefont {B.}~\bibnamefont {Zhang}},\
  }\href@noop {} {\bibfield  {journal} {\bibinfo  {journal} {Physical review
  letters}\ }\textbf {\bibinfo {volume} {114}},\ \bibinfo {pages} {114301}
  (\bibinfo {year} {2015})}\BibitemShut {NoStop}%
\bibitem [{\citenamefont {Ni}\ \emph {et~al.}(2015)\citenamefont {Ni},
  \citenamefont {He}, \citenamefont {Sun}, \citenamefont {Liu}, \citenamefont
  {Lu}, \citenamefont {Feng},\ and\ \citenamefont
  {Chen}}]{ni2015topologically}%
  \BibitemOpen
  \bibfield  {author} {\bibinfo {author} {\bibfnamefont {X.}~\bibnamefont
  {Ni}}, \bibinfo {author} {\bibfnamefont {C.}~\bibnamefont {He}}, \bibinfo
  {author} {\bibfnamefont {X.-C.}\ \bibnamefont {Sun}}, \bibinfo {author}
  {\bibfnamefont {X.-p.}\ \bibnamefont {Liu}}, \bibinfo {author} {\bibfnamefont
  {M.-H.}\ \bibnamefont {Lu}}, \bibinfo {author} {\bibfnamefont
  {L.}~\bibnamefont {Feng}},\ and\ \bibinfo {author} {\bibfnamefont {Y.-F.}\
  \bibnamefont {Chen}},\ }\href@noop {} {\bibfield  {journal} {\bibinfo
  {journal} {New Journal of Physics}\ }\textbf {\bibinfo {volume} {17}},\
  \bibinfo {pages} {053016} (\bibinfo {year} {2015})}\BibitemShut {NoStop}%
\bibitem [{\citenamefont {Khanikaev}\ \emph {et~al.}(2015)\citenamefont
  {Khanikaev}, \citenamefont {Fleury}, \citenamefont {Mousavi},\ and\
  \citenamefont {Alu}}]{khanikaev2015topologically}%
  \BibitemOpen
  \bibfield  {author} {\bibinfo {author} {\bibfnamefont {A.~B.}\ \bibnamefont
  {Khanikaev}}, \bibinfo {author} {\bibfnamefont {R.}~\bibnamefont {Fleury}},
  \bibinfo {author} {\bibfnamefont {S.~H.}\ \bibnamefont {Mousavi}},\ and\
  \bibinfo {author} {\bibfnamefont {A.}~\bibnamefont {Alu}},\ }\href@noop {}
  {\bibfield  {journal} {\bibinfo  {journal} {Nature communications}\ }\textbf
  {\bibinfo {volume} {6}},\ \bibinfo {pages} {1} (\bibinfo {year}
  {2015})}\BibitemShut {NoStop}%
\bibitem [{\citenamefont {Wang}\ \emph {et~al.}(2015)\citenamefont {Wang},
  \citenamefont {Lu},\ and\ \citenamefont {Bertoldi}}]{wang2015topological}%
  \BibitemOpen
  \bibfield  {author} {\bibinfo {author} {\bibfnamefont {P.}~\bibnamefont
  {Wang}}, \bibinfo {author} {\bibfnamefont {L.}~\bibnamefont {Lu}},\ and\
  \bibinfo {author} {\bibfnamefont {K.}~\bibnamefont {Bertoldi}},\ }\href@noop
  {} {\bibfield  {journal} {\bibinfo  {journal} {Physical review letters}\
  }\textbf {\bibinfo {volume} {115}},\ \bibinfo {pages} {104302} (\bibinfo
  {year} {2015})}\BibitemShut {NoStop}%
\bibitem [{\citenamefont {Nash}\ \emph {et~al.}(2015)\citenamefont {Nash},
  \citenamefont {Kleckner}, \citenamefont {Read}, \citenamefont {Vitelli},
  \citenamefont {Turner},\ and\ \citenamefont {Irvine}}]{nash2015topological}%
  \BibitemOpen
  \bibfield  {author} {\bibinfo {author} {\bibfnamefont {L.~M.}\ \bibnamefont
  {Nash}}, \bibinfo {author} {\bibfnamefont {D.}~\bibnamefont {Kleckner}},
  \bibinfo {author} {\bibfnamefont {A.}~\bibnamefont {Read}}, \bibinfo {author}
  {\bibfnamefont {V.}~\bibnamefont {Vitelli}}, \bibinfo {author} {\bibfnamefont
  {A.~M.}\ \bibnamefont {Turner}},\ and\ \bibinfo {author} {\bibfnamefont
  {W.~T.}\ \bibnamefont {Irvine}},\ }\href@noop {} {\bibfield  {journal}
  {\bibinfo  {journal} {Proceedings of the National Academy of Sciences}\
  }\textbf {\bibinfo {volume} {112}},\ \bibinfo {pages} {14495} (\bibinfo
  {year} {2015})}\BibitemShut {NoStop}%
\bibitem [{\citenamefont {Mousavi}\ \emph {et~al.}(2015)\citenamefont
  {Mousavi}, \citenamefont {Khanikaev},\ and\ \citenamefont
  {Wang}}]{mousavi2015topologically}%
  \BibitemOpen
  \bibfield  {author} {\bibinfo {author} {\bibfnamefont {S.~H.}\ \bibnamefont
  {Mousavi}}, \bibinfo {author} {\bibfnamefont {A.~B.}\ \bibnamefont
  {Khanikaev}},\ and\ \bibinfo {author} {\bibfnamefont {Z.}~\bibnamefont
  {Wang}},\ }\href@noop {} {\bibfield  {journal} {\bibinfo  {journal} {Nature
  communications}\ }\textbf {\bibinfo {volume} {6}},\ \bibinfo {pages} {1}
  (\bibinfo {year} {2015})}\BibitemShut {NoStop}%
\bibitem [{\citenamefont {Miniaci}\ \emph {et~al.}(2018)\citenamefont
  {Miniaci}, \citenamefont {Pal}, \citenamefont {Morvan},\ and\ \citenamefont
  {Ruzzene}}]{miniaci2018experimental}%
  \BibitemOpen
  \bibfield  {author} {\bibinfo {author} {\bibfnamefont {M.}~\bibnamefont
  {Miniaci}}, \bibinfo {author} {\bibfnamefont {R.}~\bibnamefont {Pal}},
  \bibinfo {author} {\bibfnamefont {B.}~\bibnamefont {Morvan}},\ and\ \bibinfo
  {author} {\bibfnamefont {M.}~\bibnamefont {Ruzzene}},\ }\href@noop {}
  {\bibfield  {journal} {\bibinfo  {journal} {Physical Review X}\ }\textbf
  {\bibinfo {volume} {8}},\ \bibinfo {pages} {031074} (\bibinfo {year}
  {2018})}\BibitemShut {NoStop}%
\bibitem [{\citenamefont {He}\ \emph {et~al.}(2016)\citenamefont {He},
  \citenamefont {Ni}, \citenamefont {Ge}, \citenamefont {Sun}, \citenamefont
  {Chen}, \citenamefont {Lu}, \citenamefont {Liu},\ and\ \citenamefont
  {Chen}}]{he2016acoustic}%
  \BibitemOpen
  \bibfield  {author} {\bibinfo {author} {\bibfnamefont {C.}~\bibnamefont
  {He}}, \bibinfo {author} {\bibfnamefont {X.}~\bibnamefont {Ni}}, \bibinfo
  {author} {\bibfnamefont {H.}~\bibnamefont {Ge}}, \bibinfo {author}
  {\bibfnamefont {X.-C.}\ \bibnamefont {Sun}}, \bibinfo {author} {\bibfnamefont
  {Y.-B.}\ \bibnamefont {Chen}}, \bibinfo {author} {\bibfnamefont {M.-H.}\
  \bibnamefont {Lu}}, \bibinfo {author} {\bibfnamefont {X.-P.}\ \bibnamefont
  {Liu}},\ and\ \bibinfo {author} {\bibfnamefont {Y.-F.}\ \bibnamefont
  {Chen}},\ }\href@noop {} {\bibfield  {journal} {\bibinfo  {journal} {Nature
  physics}\ }\textbf {\bibinfo {volume} {12}},\ \bibinfo {pages} {1124}
  (\bibinfo {year} {2016})}\BibitemShut {NoStop}%
\bibitem [{\citenamefont {S{\"u}sstrunk}\ and\ \citenamefont
  {Huber}(2015)}]{susstrunk2015observation}%
  \BibitemOpen
  \bibfield  {author} {\bibinfo {author} {\bibfnamefont {R.}~\bibnamefont
  {S{\"u}sstrunk}}\ and\ \bibinfo {author} {\bibfnamefont {S.~D.}\ \bibnamefont
  {Huber}},\ }\href@noop {} {\bibfield  {journal} {\bibinfo  {journal}
  {Science}\ }\textbf {\bibinfo {volume} {349}},\ \bibinfo {pages} {47}
  (\bibinfo {year} {2015})}\BibitemShut {NoStop}%
\bibitem [{\citenamefont {Wu}\ and\ \citenamefont {Hu}(2015)}]{wu2015scheme}%
  \BibitemOpen
  \bibfield  {author} {\bibinfo {author} {\bibfnamefont {L.-H.}\ \bibnamefont
  {Wu}}\ and\ \bibinfo {author} {\bibfnamefont {X.}~\bibnamefont {Hu}},\
  }\href@noop {} {\bibfield  {journal} {\bibinfo  {journal} {Physical review
  letters}\ }\textbf {\bibinfo {volume} {114}},\ \bibinfo {pages} {223901}
  (\bibinfo {year} {2015})}\BibitemShut {NoStop}%
\bibitem [{\citenamefont {Yang}\ \emph {et~al.}(2018)\citenamefont {Yang},
  \citenamefont {Xu}, \citenamefont {Xu}, \citenamefont {Wang}, \citenamefont
  {Jiang}, \citenamefont {Hu},\ and\ \citenamefont
  {Hang}}]{yang2018visualization}%
  \BibitemOpen
  \bibfield  {author} {\bibinfo {author} {\bibfnamefont {Y.}~\bibnamefont
  {Yang}}, \bibinfo {author} {\bibfnamefont {Y.~F.}\ \bibnamefont {Xu}},
  \bibinfo {author} {\bibfnamefont {T.}~\bibnamefont {Xu}}, \bibinfo {author}
  {\bibfnamefont {H.-X.}\ \bibnamefont {Wang}}, \bibinfo {author}
  {\bibfnamefont {J.-H.}\ \bibnamefont {Jiang}}, \bibinfo {author}
  {\bibfnamefont {X.}~\bibnamefont {Hu}},\ and\ \bibinfo {author}
  {\bibfnamefont {Z.~H.}\ \bibnamefont {Hang}},\ }\href@noop {} {\bibfield
  {journal} {\bibinfo  {journal} {Physical review letters}\ }\textbf {\bibinfo
  {volume} {120}},\ \bibinfo {pages} {217401} (\bibinfo {year}
  {2018})}\BibitemShut {NoStop}%
\bibitem [{\citenamefont {Xia}\ \emph {et~al.}(2017)\citenamefont {Xia},
  \citenamefont {Liu}, \citenamefont {Huang}, \citenamefont {Dai},
  \citenamefont {Jiao}, \citenamefont {Zang}, \citenamefont {Yu}, \citenamefont
  {Zheng},\ and\ \citenamefont {Liu}}]{xia2017topological}%
  \BibitemOpen
  \bibfield  {author} {\bibinfo {author} {\bibfnamefont {B.-Z.}\ \bibnamefont
  {Xia}}, \bibinfo {author} {\bibfnamefont {T.-T.}\ \bibnamefont {Liu}},
  \bibinfo {author} {\bibfnamefont {G.-L.}\ \bibnamefont {Huang}}, \bibinfo
  {author} {\bibfnamefont {H.-Q.}\ \bibnamefont {Dai}}, \bibinfo {author}
  {\bibfnamefont {J.-R.}\ \bibnamefont {Jiao}}, \bibinfo {author}
  {\bibfnamefont {X.-G.}\ \bibnamefont {Zang}}, \bibinfo {author}
  {\bibfnamefont {D.-J.}\ \bibnamefont {Yu}}, \bibinfo {author} {\bibfnamefont
  {S.-J.}\ \bibnamefont {Zheng}},\ and\ \bibinfo {author} {\bibfnamefont
  {J.}~\bibnamefont {Liu}},\ }\href@noop {} {\bibfield  {journal} {\bibinfo
  {journal} {Physical Review B}\ }\textbf {\bibinfo {volume} {96}},\ \bibinfo
  {pages} {094106} (\bibinfo {year} {2017})}\BibitemShut {NoStop}%
\bibitem [{\citenamefont {Deng}\ \emph {et~al.}(2017)\citenamefont {Deng},
  \citenamefont {Ge}, \citenamefont {Tian}, \citenamefont {Lu},\ and\
  \citenamefont {Jing}}]{deng2017observation}%
  \BibitemOpen
  \bibfield  {author} {\bibinfo {author} {\bibfnamefont {Y.}~\bibnamefont
  {Deng}}, \bibinfo {author} {\bibfnamefont {H.}~\bibnamefont {Ge}}, \bibinfo
  {author} {\bibfnamefont {Y.}~\bibnamefont {Tian}}, \bibinfo {author}
  {\bibfnamefont {M.}~\bibnamefont {Lu}},\ and\ \bibinfo {author}
  {\bibfnamefont {Y.}~\bibnamefont {Jing}},\ }\href@noop {} {\bibfield
  {journal} {\bibinfo  {journal} {Physical Review B}\ }\textbf {\bibinfo
  {volume} {96}},\ \bibinfo {pages} {184305} (\bibinfo {year}
  {2017})}\BibitemShut {NoStop}%
\bibitem [{\citenamefont {Chaunsali}\ \emph {et~al.}(2018)\citenamefont
  {Chaunsali}, \citenamefont {Chen},\ and\ \citenamefont
  {Yang}}]{chaunsali2018subwavelength}%
  \BibitemOpen
  \bibfield  {author} {\bibinfo {author} {\bibfnamefont {R.}~\bibnamefont
  {Chaunsali}}, \bibinfo {author} {\bibfnamefont {C.-W.}\ \bibnamefont
  {Chen}},\ and\ \bibinfo {author} {\bibfnamefont {J.}~\bibnamefont {Yang}},\
  }\href@noop {} {\bibfield  {journal} {\bibinfo  {journal} {Physical Review
  B}\ }\textbf {\bibinfo {volume} {97}},\ \bibinfo {pages} {054307} (\bibinfo
  {year} {2018})}\BibitemShut {NoStop}%
\bibitem [{\citenamefont {Liu}\ and\ \citenamefont
  {Semperlotti}(2020)}]{liu2020robust}%
  \BibitemOpen
  \bibfield  {author} {\bibinfo {author} {\bibfnamefont {T.-W.}\ \bibnamefont
  {Liu}}\ and\ \bibinfo {author} {\bibfnamefont {F.}~\bibnamefont
  {Semperlotti}},\ }\href@noop {} {\bibfield  {journal} {\bibinfo  {journal}
  {Bulletin of the American Physical Society}\ }\textbf {\bibinfo {volume}
  {65}} (\bibinfo {year} {2020})}\BibitemShut {NoStop}%
\bibitem [{\citenamefont {Liu}\ and\ \citenamefont
  {Semperlotti}(2021)}]{liu2021synthetic}%
  \BibitemOpen
  \bibfield  {author} {\bibinfo {author} {\bibfnamefont {T.-W.}\ \bibnamefont
  {Liu}}\ and\ \bibinfo {author} {\bibfnamefont {F.}~\bibnamefont
  {Semperlotti}},\ }\href@noop {} {\bibfield  {journal} {\bibinfo  {journal}
  {Advanced Materials}\ }\textbf {\bibinfo {volume} {33}},\ \bibinfo {pages}
  {2005160} (\bibinfo {year} {2021})}\BibitemShut {NoStop}%
\bibitem [{\citenamefont {Lu}\ \emph {et~al.}(2017)\citenamefont {Lu},
  \citenamefont {Qiu}, \citenamefont {Ye}, \citenamefont {Fan}, \citenamefont
  {Ke}, \citenamefont {Zhang},\ and\ \citenamefont {Liu}}]{lu2017observation}%
  \BibitemOpen
  \bibfield  {author} {\bibinfo {author} {\bibfnamefont {J.}~\bibnamefont
  {Lu}}, \bibinfo {author} {\bibfnamefont {C.}~\bibnamefont {Qiu}}, \bibinfo
  {author} {\bibfnamefont {L.}~\bibnamefont {Ye}}, \bibinfo {author}
  {\bibfnamefont {X.}~\bibnamefont {Fan}}, \bibinfo {author} {\bibfnamefont
  {M.}~\bibnamefont {Ke}}, \bibinfo {author} {\bibfnamefont {F.}~\bibnamefont
  {Zhang}},\ and\ \bibinfo {author} {\bibfnamefont {Z.}~\bibnamefont {Liu}},\
  }\href@noop {} {\bibfield  {journal} {\bibinfo  {journal} {Nature Physics}\
  }\textbf {\bibinfo {volume} {13}},\ \bibinfo {pages} {369} (\bibinfo {year}
  {2017})}\BibitemShut {NoStop}%
\bibitem [{\citenamefont {Pal}\ and\ \citenamefont
  {Ruzzene}(2017)}]{pal2017edge}%
  \BibitemOpen
  \bibfield  {author} {\bibinfo {author} {\bibfnamefont {R.~K.}\ \bibnamefont
  {Pal}}\ and\ \bibinfo {author} {\bibfnamefont {M.}~\bibnamefont {Ruzzene}},\
  }\href@noop {} {\bibfield  {journal} {\bibinfo  {journal} {New Journal of
  Physics}\ }\textbf {\bibinfo {volume} {19}},\ \bibinfo {pages} {025001}
  (\bibinfo {year} {2017})}\BibitemShut {NoStop}%
\bibitem [{\citenamefont {Vila}\ \emph {et~al.}(2017)\citenamefont {Vila},
  \citenamefont {Pal},\ and\ \citenamefont {Ruzzene}}]{vila2017observation}%
  \BibitemOpen
  \bibfield  {author} {\bibinfo {author} {\bibfnamefont {J.}~\bibnamefont
  {Vila}}, \bibinfo {author} {\bibfnamefont {R.~K.}\ \bibnamefont {Pal}},\ and\
  \bibinfo {author} {\bibfnamefont {M.}~\bibnamefont {Ruzzene}},\ }\href@noop
  {} {\bibfield  {journal} {\bibinfo  {journal} {Physical Review B}\ }\textbf
  {\bibinfo {volume} {96}},\ \bibinfo {pages} {134307} (\bibinfo {year}
  {2017})}\BibitemShut {NoStop}%
\bibitem [{\citenamefont {Liu}\ and\ \citenamefont
  {Semperlotti}(2018)}]{liu2018tunable}%
  \BibitemOpen
  \bibfield  {author} {\bibinfo {author} {\bibfnamefont {T.-W.}\ \bibnamefont
  {Liu}}\ and\ \bibinfo {author} {\bibfnamefont {F.}~\bibnamefont
  {Semperlotti}},\ }\href {https://doi.org/10.1103/PhysRevApplied.9.014001}
  {\bibfield  {journal} {\bibinfo  {journal} {Phys. Rev. Applied}\ }\textbf
  {\bibinfo {volume} {9}},\ \bibinfo {pages} {014001} (\bibinfo {year}
  {2018})}\BibitemShut {NoStop}%
\bibitem [{\citenamefont {Zhu}\ \emph {et~al.}(2018)\citenamefont {Zhu},
  \citenamefont {Liu},\ and\ \citenamefont {Semperlotti}}]{zhu2018design}%
  \BibitemOpen
  \bibfield  {author} {\bibinfo {author} {\bibfnamefont {H.}~\bibnamefont
  {Zhu}}, \bibinfo {author} {\bibfnamefont {T.-W.}\ \bibnamefont {Liu}},\ and\
  \bibinfo {author} {\bibfnamefont {F.}~\bibnamefont {Semperlotti}},\
  }\href@noop {} {\bibfield  {journal} {\bibinfo  {journal} {Physical Review
  B}\ }\textbf {\bibinfo {volume} {97}},\ \bibinfo {pages} {174301} (\bibinfo
  {year} {2018})}\BibitemShut {NoStop}%
\bibitem [{\citenamefont {Liu}\ and\ \citenamefont
  {Semperlotti}(2019)}]{liu2019experimental}%
  \BibitemOpen
  \bibfield  {author} {\bibinfo {author} {\bibfnamefont {T.-W.}\ \bibnamefont
  {Liu}}\ and\ \bibinfo {author} {\bibfnamefont {F.}~\bibnamefont
  {Semperlotti}},\ }\href@noop {} {\bibfield  {journal} {\bibinfo  {journal}
  {Physical Review Applied}\ }\textbf {\bibinfo {volume} {11}},\ \bibinfo
  {pages} {014040} (\bibinfo {year} {2019})}\BibitemShut {NoStop}%
\bibitem [{\citenamefont {Ganti}\ \emph
  {et~al.}(2020{\natexlab{a}})\citenamefont {Ganti}, \citenamefont {Liu},\ and\
  \citenamefont {Semperlotti}}]{ganti2020weyl}%
  \BibitemOpen
  \bibfield  {author} {\bibinfo {author} {\bibfnamefont {S.~S.}\ \bibnamefont
  {Ganti}}, \bibinfo {author} {\bibfnamefont {T.-W.}\ \bibnamefont {Liu}},\
  and\ \bibinfo {author} {\bibfnamefont {F.}~\bibnamefont {Semperlotti}},\
  }\href@noop {} {\bibfield  {journal} {\bibinfo  {journal} {New Journal of
  Physics}\ }\textbf {\bibinfo {volume} {22}},\ \bibinfo {pages} {083001}
  (\bibinfo {year} {2020}{\natexlab{a}})}\BibitemShut {NoStop}%
\bibitem [{\citenamefont {Ganti}\ \emph
  {et~al.}(2020{\natexlab{b}})\citenamefont {Ganti}, \citenamefont {Liu},\ and\
  \citenamefont {Semperlotti}}]{ganti2020topological}%
  \BibitemOpen
  \bibfield  {author} {\bibinfo {author} {\bibfnamefont {S.~S.}\ \bibnamefont
  {Ganti}}, \bibinfo {author} {\bibfnamefont {T.-W.}\ \bibnamefont {Liu}},\
  and\ \bibinfo {author} {\bibfnamefont {F.}~\bibnamefont {Semperlotti}},\
  }\href@noop {} {\bibfield  {journal} {\bibinfo  {journal} {Journal of Sound
  and Vibration}\ }\textbf {\bibinfo {volume} {466}},\ \bibinfo {pages}
  {115060} (\bibinfo {year} {2020}{\natexlab{b}})}\BibitemShut {NoStop}%
\bibitem [{\citenamefont {Xiao}\ \emph {et~al.}(2015)\citenamefont {Xiao},
  \citenamefont {Ma}, \citenamefont {Yang}, \citenamefont {Sheng},
  \citenamefont {Zhang},\ and\ \citenamefont {Chan}}]{xiao2015geometric}%
  \BibitemOpen
  \bibfield  {author} {\bibinfo {author} {\bibfnamefont {M.}~\bibnamefont
  {Xiao}}, \bibinfo {author} {\bibfnamefont {G.}~\bibnamefont {Ma}}, \bibinfo
  {author} {\bibfnamefont {Z.}~\bibnamefont {Yang}}, \bibinfo {author}
  {\bibfnamefont {P.}~\bibnamefont {Sheng}}, \bibinfo {author} {\bibfnamefont
  {Z.}~\bibnamefont {Zhang}},\ and\ \bibinfo {author} {\bibfnamefont {C.~T.}\
  \bibnamefont {Chan}},\ }\href@noop {} {\bibfield  {journal} {\bibinfo
  {journal} {Nature Physics}\ }\textbf {\bibinfo {volume} {11}},\ \bibinfo
  {pages} {240} (\bibinfo {year} {2015})}\BibitemShut {NoStop}%
\bibitem [{\citenamefont {Xiao}\ \emph {et~al.}(2017)\citenamefont {Xiao},
  \citenamefont {Ma}, \citenamefont {Zhang},\ and\ \citenamefont
  {Chan}}]{xiao2017topological}%
  \BibitemOpen
  \bibfield  {author} {\bibinfo {author} {\bibfnamefont {Y.-X.}\ \bibnamefont
  {Xiao}}, \bibinfo {author} {\bibfnamefont {G.}~\bibnamefont {Ma}}, \bibinfo
  {author} {\bibfnamefont {Z.-Q.}\ \bibnamefont {Zhang}},\ and\ \bibinfo
  {author} {\bibfnamefont {C.~T.}\ \bibnamefont {Chan}},\ }\href@noop {}
  {\bibfield  {journal} {\bibinfo  {journal} {Physical review letters}\
  }\textbf {\bibinfo {volume} {118}},\ \bibinfo {pages} {166803} (\bibinfo
  {year} {2017})}\BibitemShut {NoStop}%
\bibitem [{\citenamefont {Chaunsali}\ \emph {et~al.}(2017)\citenamefont
  {Chaunsali}, \citenamefont {Kim}, \citenamefont {Thakkar}, \citenamefont
  {Kevrekidis},\ and\ \citenamefont {Yang}}]{chaunsali2017demonstrating}%
  \BibitemOpen
  \bibfield  {author} {\bibinfo {author} {\bibfnamefont {R.}~\bibnamefont
  {Chaunsali}}, \bibinfo {author} {\bibfnamefont {E.}~\bibnamefont {Kim}},
  \bibinfo {author} {\bibfnamefont {A.}~\bibnamefont {Thakkar}}, \bibinfo
  {author} {\bibfnamefont {P.~G.}\ \bibnamefont {Kevrekidis}},\ and\ \bibinfo
  {author} {\bibfnamefont {J.}~\bibnamefont {Yang}},\ }\href@noop {} {\bibfield
   {journal} {\bibinfo  {journal} {Physical review letters}\ }\textbf {\bibinfo
  {volume} {119}},\ \bibinfo {pages} {024301} (\bibinfo {year}
  {2017})}\BibitemShut {NoStop}%
\bibitem [{\citenamefont {Chen}\ \emph {et~al.}(2018)\citenamefont {Chen},
  \citenamefont {Nassar},\ and\ \citenamefont {Huang}}]{chen2018study}%
  \BibitemOpen
  \bibfield  {author} {\bibinfo {author} {\bibfnamefont {H.}~\bibnamefont
  {Chen}}, \bibinfo {author} {\bibfnamefont {H.}~\bibnamefont {Nassar}},\ and\
  \bibinfo {author} {\bibfnamefont {G.}~\bibnamefont {Huang}},\ }\href@noop {}
  {\bibfield  {journal} {\bibinfo  {journal} {Journal of the Mechanics and
  Physics of Solids}\ }\textbf {\bibinfo {volume} {117}},\ \bibinfo {pages}
  {22} (\bibinfo {year} {2018})}\BibitemShut {NoStop}%
\bibitem [{\citenamefont {Ding}\ \emph {et~al.}(2016)\citenamefont {Ding},
  \citenamefont {Ma}, \citenamefont {Xiao}, \citenamefont {Zhang},\ and\
  \citenamefont {Chan}}]{ding2016emergence}%
  \BibitemOpen
  \bibfield  {author} {\bibinfo {author} {\bibfnamefont {K.}~\bibnamefont
  {Ding}}, \bibinfo {author} {\bibfnamefont {G.}~\bibnamefont {Ma}}, \bibinfo
  {author} {\bibfnamefont {M.}~\bibnamefont {Xiao}}, \bibinfo {author}
  {\bibfnamefont {Z.}~\bibnamefont {Zhang}},\ and\ \bibinfo {author}
  {\bibfnamefont {C.~T.}\ \bibnamefont {Chan}},\ }\href@noop {} {\bibfield
  {journal} {\bibinfo  {journal} {Physical Review X}\ }\textbf {\bibinfo
  {volume} {6}},\ \bibinfo {pages} {021007} (\bibinfo {year}
  {2016})}\BibitemShut {NoStop}%
\bibitem [{\citenamefont {Tang}\ \emph {et~al.}(2020)\citenamefont {Tang},
  \citenamefont {Jiang}, \citenamefont {Ding}, \citenamefont {Xiao},
  \citenamefont {Zhang}, \citenamefont {Chan},\ and\ \citenamefont
  {Ma}}]{tang2020exceptional}%
  \BibitemOpen
  \bibfield  {author} {\bibinfo {author} {\bibfnamefont {W.}~\bibnamefont
  {Tang}}, \bibinfo {author} {\bibfnamefont {X.}~\bibnamefont {Jiang}},
  \bibinfo {author} {\bibfnamefont {K.}~\bibnamefont {Ding}}, \bibinfo {author}
  {\bibfnamefont {Y.-X.}\ \bibnamefont {Xiao}}, \bibinfo {author}
  {\bibfnamefont {Z.-Q.}\ \bibnamefont {Zhang}}, \bibinfo {author}
  {\bibfnamefont {C.~T.}\ \bibnamefont {Chan}},\ and\ \bibinfo {author}
  {\bibfnamefont {G.}~\bibnamefont {Ma}},\ }\href@noop {} {\bibfield  {journal}
  {\bibinfo  {journal} {Science}\ }\textbf {\bibinfo {volume} {370}},\ \bibinfo
  {pages} {1077} (\bibinfo {year} {2020})}\BibitemShut {NoStop}%
\bibitem [{\citenamefont {Dom{\'i}nguez-Rocha}\ \emph
  {et~al.}(2020)\citenamefont {Dom{\'i}nguez-Rocha}, \citenamefont
  {Thevamaran}, \citenamefont {Ellis},\ and\ \citenamefont
  {Kottos}}]{dominguez2020environmentally}%
  \BibitemOpen
  \bibfield  {author} {\bibinfo {author} {\bibfnamefont {V.}~\bibnamefont
  {Dom{\'i}nguez-Rocha}}, \bibinfo {author} {\bibfnamefont {R.}~\bibnamefont
  {Thevamaran}}, \bibinfo {author} {\bibfnamefont {F.}~\bibnamefont {Ellis}},\
  and\ \bibinfo {author} {\bibfnamefont {T.}~\bibnamefont {Kottos}},\
  }\href@noop {} {\bibfield  {journal} {\bibinfo  {journal} {Physical Review
  Applied}\ }\textbf {\bibinfo {volume} {13}},\ \bibinfo {pages} {014060}
  (\bibinfo {year} {2020})}\BibitemShut {NoStop}%
\bibitem [{\citenamefont {Liao}\ \emph {et~al.}(2022)\citenamefont {Liao},
  \citenamefont {Zhang}, \citenamefont {Cheng},\ and\ \citenamefont
  {Liu}}]{liao2022engineering}%
  \BibitemOpen
  \bibfield  {author} {\bibinfo {author} {\bibfnamefont {D.}~\bibnamefont
  {Liao}}, \bibinfo {author} {\bibfnamefont {Z.}~\bibnamefont {Zhang}},
  \bibinfo {author} {\bibfnamefont {Y.}~\bibnamefont {Cheng}},\ and\ \bibinfo
  {author} {\bibfnamefont {X.}~\bibnamefont {Liu}},\ }\href@noop {} {\bibfield
  {journal} {\bibinfo  {journal} {Physical Review B}\ }\textbf {\bibinfo
  {volume} {105}},\ \bibinfo {pages} {184108} (\bibinfo {year}
  {2022})}\BibitemShut {NoStop}%
\bibitem [{\citenamefont {Ye}\ \emph {et~al.}(2022)\citenamefont {Ye},
  \citenamefont {Qiu}, \citenamefont {Xiao}, \citenamefont {Li}, \citenamefont
  {Du}, \citenamefont {Ke},\ and\ \citenamefont {Liu}}]{ye2022topological}%
  \BibitemOpen
  \bibfield  {author} {\bibinfo {author} {\bibfnamefont {L.}~\bibnamefont
  {Ye}}, \bibinfo {author} {\bibfnamefont {C.}~\bibnamefont {Qiu}}, \bibinfo
  {author} {\bibfnamefont {M.}~\bibnamefont {Xiao}}, \bibinfo {author}
  {\bibfnamefont {T.}~\bibnamefont {Li}}, \bibinfo {author} {\bibfnamefont
  {J.}~\bibnamefont {Du}}, \bibinfo {author} {\bibfnamefont {M.}~\bibnamefont
  {Ke}},\ and\ \bibinfo {author} {\bibfnamefont {Z.}~\bibnamefont {Liu}},\
  }\href@noop {} {\bibfield  {journal} {\bibinfo  {journal} {Nature
  Communications}\ }\textbf {\bibinfo {volume} {13}},\ \bibinfo {pages} {508}
  (\bibinfo {year} {2022})}\BibitemShut {NoStop}%
\bibitem [{\citenamefont {Su}\ \emph {et~al.}(1979)\citenamefont {Su},
  \citenamefont {Schrieffer},\ and\ \citenamefont {Heeger}}]{su1979solitons}%
  \BibitemOpen
  \bibfield  {author} {\bibinfo {author} {\bibfnamefont {W.}~\bibnamefont
  {Su}}, \bibinfo {author} {\bibfnamefont {J.}~\bibnamefont {Schrieffer}},\
  and\ \bibinfo {author} {\bibfnamefont {A.~J.}\ \bibnamefont {Heeger}},\
  }\href@noop {} {\bibfield  {journal} {\bibinfo  {journal} {Physical review
  letters}\ }\textbf {\bibinfo {volume} {42}},\ \bibinfo {pages} {1698}
  (\bibinfo {year} {1979})}\BibitemShut {NoStop}%
\bibitem [{\citenamefont {Yin}\ \emph {et~al.}(2018)\citenamefont {Yin},
  \citenamefont {Ruzzene}, \citenamefont {Wen}, \citenamefont {Yu},
  \citenamefont {Cai},\ and\ \citenamefont {Yue}}]{yin2018band}%
  \BibitemOpen
  \bibfield  {author} {\bibinfo {author} {\bibfnamefont {J.}~\bibnamefont
  {Yin}}, \bibinfo {author} {\bibfnamefont {M.}~\bibnamefont {Ruzzene}},
  \bibinfo {author} {\bibfnamefont {J.}~\bibnamefont {Wen}}, \bibinfo {author}
  {\bibfnamefont {D.}~\bibnamefont {Yu}}, \bibinfo {author} {\bibfnamefont
  {L.}~\bibnamefont {Cai}},\ and\ \bibinfo {author} {\bibfnamefont
  {L.}~\bibnamefont {Yue}},\ }\href@noop {} {\bibfield  {journal} {\bibinfo
  {journal} {Scientific reports}\ }\textbf {\bibinfo {volume} {8}},\ \bibinfo
  {pages} {1} (\bibinfo {year} {2018})}\BibitemShut {NoStop}%
\bibitem [{\citenamefont {Vila}\ \emph {et~al.}(2019)\citenamefont {Vila},
  \citenamefont {Paulino},\ and\ \citenamefont {Ruzzene}}]{vila2019role}%
  \BibitemOpen
  \bibfield  {author} {\bibinfo {author} {\bibfnamefont {J.}~\bibnamefont
  {Vila}}, \bibinfo {author} {\bibfnamefont {G.~H.}\ \bibnamefont {Paulino}},\
  and\ \bibinfo {author} {\bibfnamefont {M.}~\bibnamefont {Ruzzene}},\
  }\href@noop {} {\bibfield  {journal} {\bibinfo  {journal} {Physical Review
  B}\ }\textbf {\bibinfo {volume} {99}},\ \bibinfo {pages} {125116} (\bibinfo
  {year} {2019})}\BibitemShut {NoStop}%
\bibitem [{\citenamefont {Shi}\ \emph {et~al.}(2021)\citenamefont {Shi},
  \citenamefont {Kiorpelidis}, \citenamefont {Chaunsali}, \citenamefont
  {Achilleos}, \citenamefont {Theocharis},\ and\ \citenamefont
  {Yang}}]{shi2021disorder}%
  \BibitemOpen
  \bibfield  {author} {\bibinfo {author} {\bibfnamefont {X.}~\bibnamefont
  {Shi}}, \bibinfo {author} {\bibfnamefont {I.}~\bibnamefont {Kiorpelidis}},
  \bibinfo {author} {\bibfnamefont {R.}~\bibnamefont {Chaunsali}}, \bibinfo
  {author} {\bibfnamefont {V.}~\bibnamefont {Achilleos}}, \bibinfo {author}
  {\bibfnamefont {G.}~\bibnamefont {Theocharis}},\ and\ \bibinfo {author}
  {\bibfnamefont {J.}~\bibnamefont {Yang}},\ }\href@noop {} {\bibfield
  {journal} {\bibinfo  {journal} {Physical Review Research}\ }\textbf {\bibinfo
  {volume} {3}},\ \bibinfo {pages} {033012} (\bibinfo {year}
  {2021})}\BibitemShut {NoStop}%
\bibitem [{\citenamefont {Dulock}\ and\ \citenamefont
  {McIntosh}(1965)}]{dulock1965degeneracy}%
  \BibitemOpen
  \bibfield  {author} {\bibinfo {author} {\bibfnamefont {V.~A.}\ \bibnamefont
  {Dulock}}\ and\ \bibinfo {author} {\bibfnamefont {H.~V.}\ \bibnamefont
  {McIntosh}},\ }\href@noop {} {\bibfield  {journal} {\bibinfo  {journal}
  {American Journal of Physics}\ }\textbf {\bibinfo {volume} {33}},\ \bibinfo
  {pages} {109} (\bibinfo {year} {1965})}\BibitemShut {NoStop}%
\bibitem [{\citenamefont {Rice}\ and\ \citenamefont
  {Mele}(1982)}]{rice1982elementary}%
  \BibitemOpen
  \bibfield  {author} {\bibinfo {author} {\bibfnamefont {M.}~\bibnamefont
  {Rice}}\ and\ \bibinfo {author} {\bibfnamefont {E.}~\bibnamefont {Mele}},\
  }\href@noop {} {\bibfield  {journal} {\bibinfo  {journal} {Physical Review
  Letters}\ }\textbf {\bibinfo {volume} {49}},\ \bibinfo {pages} {1455}
  (\bibinfo {year} {1982})}\BibitemShut {NoStop}%
\bibitem [{\citenamefont {De~Gennes}\ and\ \citenamefont
  {Pincus}(1966)}]{de1966superconductivity}%
  \BibitemOpen
  \bibfield  {author} {\bibinfo {author} {\bibfnamefont {P.-G.}\ \bibnamefont
  {De~Gennes}}\ and\ \bibinfo {author} {\bibfnamefont {P.~A.}\ \bibnamefont
  {Pincus}},\ }\href@noop {} {\emph {\bibinfo {title} {Superconductivity of
  metals and alloys}}}\ (\bibinfo  {publisher} {CRC Press},\ \bibinfo {year}
  {1966})\BibitemShut {NoStop}%
\bibitem [{\citenamefont {Chiu}\ \emph {et~al.}(2016)\citenamefont {Chiu},
  \citenamefont {Teo}, \citenamefont {Schnyder},\ and\ \citenamefont
  {Ryu}}]{chiu2016classification}%
  \BibitemOpen
  \bibfield  {author} {\bibinfo {author} {\bibfnamefont {C.-K.}\ \bibnamefont
  {Chiu}}, \bibinfo {author} {\bibfnamefont {J.~C.}\ \bibnamefont {Teo}},
  \bibinfo {author} {\bibfnamefont {A.~P.}\ \bibnamefont {Schnyder}},\ and\
  \bibinfo {author} {\bibfnamefont {S.}~\bibnamefont {Ryu}},\ }\href@noop {}
  {\bibfield  {journal} {\bibinfo  {journal} {Reviews of Modern Physics}\
  }\textbf {\bibinfo {volume} {88}},\ \bibinfo {pages} {035005} (\bibinfo
  {year} {2016})}\BibitemShut {NoStop}%
\bibitem [{\citenamefont {Kittel}\ \emph {et~al.}(1996)\citenamefont {Kittel},
  \citenamefont {McEuen},\ and\ \citenamefont
  {McEuen}}]{kittel1996introduction}%
  \BibitemOpen
  \bibfield  {author} {\bibinfo {author} {\bibfnamefont {C.}~\bibnamefont
  {Kittel}}, \bibinfo {author} {\bibfnamefont {P.}~\bibnamefont {McEuen}},\
  and\ \bibinfo {author} {\bibfnamefont {P.}~\bibnamefont {McEuen}},\
  }\href@noop {} {\emph {\bibinfo {title} {Introduction to solid state
  physics}}},\ \bibinfo {edition} {8th}\ ed.\ (\bibinfo  {publisher} {Wiley New
  York},\ \bibinfo {year} {1996})\BibitemShut {NoStop}%
\bibitem [{\citenamefont {Simon}(2013)}]{simon2013oxford}%
  \BibitemOpen
  \bibfield  {author} {\bibinfo {author} {\bibfnamefont {S.~H.}\ \bibnamefont
  {Simon}},\ }\href@noop {} {\emph {\bibinfo {title} {The Oxford solid state
  basics}}}\ (\bibinfo  {publisher} {OUP Oxford},\ \bibinfo {year}
  {2013})\BibitemShut {NoStop}%
\bibitem [{\citenamefont {Budich}\ and\ \citenamefont
  {Ardonne}(2013)}]{budich2013equivalent}%
  \BibitemOpen
  \bibfield  {author} {\bibinfo {author} {\bibfnamefont {J.~C.}\ \bibnamefont
  {Budich}}\ and\ \bibinfo {author} {\bibfnamefont {E.}~\bibnamefont
  {Ardonne}},\ }\href@noop {} {\bibfield  {journal} {\bibinfo  {journal}
  {Physical Review B}\ }\textbf {\bibinfo {volume} {88}},\ \bibinfo {pages}
  {075419} (\bibinfo {year} {2013})}\BibitemShut {NoStop}%
\end{thebibliography}%

\end{document}